\documentclass[usenatbib]{mnras}
\usepackage{graphicx}
\usepackage{amsmath}
\usepackage{breqn}
\usepackage{amssymb}
\usepackage{hyperref}
\usepackage{xcolor}

\newcommand{\descendantseedmass}{M_{\mathrm{seed}}^{\mathrm{ESD}}}

\newcommand{\seedmass}{M_{\mathrm{seed}}^{\mathrm{DGB}}}

\newcommand{\mh}{\tilde{M}_{\mathrm{h}}}
\newcommand{\msfmp}{\tilde{M}_{\mathrm{sfmp}}}

\newcommand{\massembly}{M^{\mathrm{galaxy}}_{\mathrm{total}}}
\newcommand{\massemblyFOF}{M^{\mathrm{halo}}_{\mathrm{total}}}
\newcommand{\environmentseedprobability}
{P_{\mathrm{seed}}^{\mathrm{env}}}

\title[Low mass seeds in cosmological simulations]{Representing low mass black hole seeds in cosmological simulations: A new sub-grid stochastic seed model}
\author[Bhowmick et al.]{Aklant K. Bhowmick$^{1}$,
Laura Blecha$^{1}$,
Paul Torrey$^{1,2}$,
Rainer Weinberger$^{3}$,\newauthor
Luke Zoltan Kelley$^{4}$, 
Mark Vogelsberger$^{5}$,
Lars Hernquist$^{6}$,
Rachel S. Somerville$^{7,8}$
\\
$^{1}$Department of Physics, University of Florida, Gainesville, FL 32611, USA\\
$^{2}$Department of Astronomy, University of Virginia, 530 McCormick Road, Charlottesville, VA 22903, USA\\
$^{3}$Leibniz Institute for Astrophysics, An der Sternwarte 16, 14482 Potsdam, Germany\\
$^{4}$Department of Astronomy, University of California at Berkeley, 501 Campbell Hall, Berkeley, CA 94720, USA\\
$^{5}$Department of Physics, Kavli Institute for Astrophysics and Space Research, Massachusetts Institute of Technology, Cambridge, MA 02139, USA \\
$^{6}$Harvard-Smithsonian Center for Astrophysics, 60 Garden Street, Cambridge, MA 02138, USA\\
$^{7}$Center for Computational Astrophysics, Flatiron institute, New York, NY 10010, USA\\
$^{8}$Department of Physics and Astronomy, Rutgers University, 136
}
\begin{document}
\maketitle
\begin{abstract}
The nature of the first seeds of supermassive black holes~(SMBHs) is currently unknown, with postulated initial masses ranging from $\sim10^5~M_{\odot}$ to as low as $\sim10^2~M_{\odot}$. However, most existing cosmological hydrodynamical simulations resolve BHs only down to $\sim10^5-10^6~M_{\odot}$. In this work, we introduce a novel sub-grid BH seeding model for cosmological simulations that is directly calibrated from high resolution zoom simulations that can trace the formation and growth of $\sim 10^3~M_{\odot}$ seeds that form in halos with pristine, star-forming gas. We trace the BH growth along galaxy merger trees until their descendants reach masses of $\sim10^4$ or $10^5~M_{\odot}$. The descendants assemble in galaxies with a broad range of properties~(e.g., halo masses ranging from $\sim10^7-10^9~M_{\odot}$) that evolve with redshift and are also sensitive to seed parameters. The results are used to build a new stochastic seeding model that directly seeds these descendants in lower resolution versions of our zoom region. Remarkably, we find that by seeding the descendants simply based on total galaxy mass, redshift and an environmental richness parameter, we can reproduce the results of the detailed gas based seeding model. The baryonic properties of the host galaxies are well reproduced by the mass-based seeding criterion. The redshift-dependence of the mass-based criterion captures the combined influence of halo growth, star formation and metal enrichment on the formation of $\sim 10^3~M_{\odot}$ seeds. The environment based seeding criterion seeds the descendants in rich environments with higher numbers of neighboring galaxies. This accounts for the impact of unresolved merger dominated growth of BHs, which produces faster growth of descendants in richer environments with more extensive BH merger history. Our new seed model will be useful for representing a variety of low mass seeding channels within next generation larger volume uniform cosmological simulations.  
\end{abstract}


\section{Introduction}
The origin of supermassive black holes~(SMBHs) is a key missing piece in our current understanding of galaxy formation. Several theoretical channels have been proposed for the first ``seeds" of SMBHs, predicting a wide range of postulated initial masses. At the lowest mass end of the initial seed mass function, we have the remnants of the first generation Population III stars, a.k.a. Pop III seeds ~\citep{2001ApJ...550..372F,2001ApJ...551L..27M,2013ApJ...773...83X,2018MNRAS.480.3762S} ranging from $\sim10^2-10^3~M_{\odot}$. Next, we have seeds postulated at the ``intermediate mass" range of $\sim10^3-10^4~M_{\odot}$ that can form via runaway stellar and black hole~(BH) collisions within dense Nuclear Star Clusters, a.k.a NSC seeds ~\citep{2011ApJ...740L..42D,2014MNRAS.442.3616L,2020MNRAS.498.5652K,2021MNRAS.503.1051D,2021MNRAS.tmp.1381D}. Finally, we can have ``high mass seeds" formed via direct isothermal collapse of gas at sufficiently high temperatures~($\gtrsim10^4$ K), a.k.a direct collapse black hole or DCBH seeds~\citep{2003ApJ...596...34B,2006MNRAS.370..289B,2014ApJ...795..137R,2016MNRAS.458..233L,2018MNRAS.476.3523L,2019Natur.566...85W,2020MNRAS.492.4917L,2023arXiv230519081B}. DCBHs masses are traditionally postulated to be ranging within $\sim10^4-10^6~M_{\odot}$, but recent works have suggested that they can also be as massive as $\sim10^8~M_{\odot}$~\citep{2023arXiv230402066M}.  

The growing observed population of luminous quasars at $z\sim6-8$~\citep{2001AJ....122.2833F,2010AJ....139..906W,2011Natur.474..616M,2015MNRAS.453.2259V,2016ApJ...833..222J,2016Banados,2017MNRAS.468.4702R,2018ApJS..237....5M,2018ApJ...869L...9W,2018Natur.553..473B,2019ApJ...872L...2M,2019AJ....157..236Y,2021ApJ...907L...1W} tells us that $\sim10^9-10^{10}~M_{\odot}$ BHs already assembled within the first few hundred million years after the Big Bang. These already pose a serious challenge to models of BH formation as well as BH growth. For example, light seeds may need to sustainably accrete gas at super-Eddington rates to grow by $\sim6-7$ orders of magnitude within such a short time. Alternatively, they can boost their seed mass via mergers, but it is unclear as to how efficiently these seeds sink and merge with each other within the shallow potential wells of high redshift proto-galaxies~\citep{2007ApJ...663L...5V,2021MNRAS.508.1973M}. Heavier seed masses such as DCBHs are substantially more conducive for assembling the high-z quasars, but it is unclear if they form frequently enough to account for the observed number densities~($1~\mathrm{Gpc}^{-3}$).

Due to possible degeneracies in the impact of different BH formation versus BH growth models, it is challenging to constrain seed models solely using observations of luminous high-z quasars. 
To that end, detections of lower mass BH populations at high-z are going to be crucial for constraining seed models as these BHs are more likely to retain the memory of their initial seeds.  The James Webb Space Telescope~\citep[JWST;][]{2006SSRv..123..485G} is pushing the frontiers of SMBH studies by detecting lower luminosity active galactic nuclei~(AGN) at high redshifts. In addition to the first statistical sample of $\sim10^6-10^7~M_{\odot}$ AGN at $z\sim4-7$~\citep{2023arXiv230311946H}, JWST has also produced the first detections at $z\gtrsim8.3$~\citep{2023arXiv230308918L} and $z\sim10.6$~\citep{2023arXiv230512492M}. Moreover, there is an exciting possibility of future detections of BHs as small as $\sim10^5~M_{\odot}$ using JWST, which would potentially enable us to probe the massive end of the seed population for the very first time~\citep{2017ApJ...838..117N,2018ApJ...861..142C,2022ApJ...931L..25I}.

Even with JWST and proposed X-ray facilities like ATHENA~\citep{2017AN....338..153B} and  Axis~\citep{2019BAAS...51g.107M}, low mass seeds $\sim10^2-10^4~M_{\odot}$ are likely to be inaccessible to electromagnetic~(EM) observations at high-z. However, with the new observational window of gravitational waves~(GW) opened for the first time by the Laser Interferometer Gravitational-Wave Observatory~\citep[LIGO;][]{2009RPPh...72g6901A}, we can close this gap. In addition to detecting numerous~($\sim 80$) stellar mass BH mergers, LIGO has also started probing the elusive population of intermediate mass black holes~(IMBH: $\sim10^2-10^5~M_{\odot}$) with GW190521~\citep{2020ApJ...900L..13A} producing a $\sim142~M_{\odot}$ BH remnant. At the other end of BH mass spectrum, the North American Nanohertz Observatory for Gravitational Waves~(NANOGRAV) have also detected the Hellings-Downs correlation expected from a stochastic GW background that most likely originates from populations of merging SMBHs~\citep{2023ApJ...951L...8A}. But the strongest imprints of BH formation will likely be provided by the upcoming Laser Interferometer Space Antenna  %/ LISA.~
\citep[LISA;][]{2019arXiv190706482B}, which is expected to detect GWs from mergers of IMBHs as small as $\sim10^3~M_{\odot}$ up to $z\sim15$~\citep{2017arXiv170200786A}.

Cosmological hydrodynamic simulations~\citep{2012ApJ...745L..29D,2014Natur.509..177V,2015MNRAS.452..575S,2015MNRAS.450.1349K,2015MNRAS.446..521S,2016MNRAS.460.2979V,2016MNRAS.463.3948D,2017MNRAS.467.4739K,2017MNRAS.470.1121T,2019ComAC...6....2N,2020MNRAS.498.2219V} have emerged as powerful tools for testing galaxy formation theories~(see, e.g., the review by \citealt{2020NatRP...2...42V}). However, most such simulations can resolve gas elements only down to $\sim10^5-10^7~M_{\odot}$, depending on the simulation volume. This is particularly true for simulation volumes needed to produce statistical samples of galaxies and BHs that can be directly compared to observations.  Therefore, most cosmological simulations only model BH seeds down to $\sim10^5~M_{\odot}$~\citep[for e.g.][]{2014Natur.509..177V,2015MNRAS.450.1349K,2017MNRAS.470.1121T}. Notably, there are simulations that do attempt to capture seed masses down to $\sim10^4~M_{\odot}$~\citep{2022MNRAS.513..670N} and $\sim10^3~M_{\odot}$~\citep{2014MNRAS.442.2751T,2019MNRAS.483.4640W}, but they do so without explicitly resolving the seed-forming gas to those masses. Overall, directly resolving the low mass seed population~($\sim10^{2}-10^{4}~M_{\odot}$ encompassing Pop III and NSC seeding channels) is completely inaccessible within state of the art cosmological simulations, and pushing beyond current resolution limits will require a substantial advancement in available computing power. 

Given that BH seed formation is primarily governed by properties of the seed-forming gas, the insufficient resolution within cosmological simulations carries the additional liability of having poorly converged gas properties. For instance, Pop III and NSC seeds are supposed to be born out of star-forming and metal poor gas. However, the rates of star formation and metal enrichment may not be well converged in these simulations at their typical gas mass resolutions of $\sim10^5-10^7~M_{\odot}$~(for example, see Figure 19 of \citealt{2021MNRAS.507.2012B}). As a result, many simulations~\citep{2012ApJ...745L..29D,2014Natur.509..177V,2018MNRAS.475..624N,2022MNRAS.513..670N} simply use a host halo mass threshold to seed BHs. Several cosmological simulations have also used local gas properties for seeding~\citep{2014MNRAS.442.2751T,2017MNRAS.470.1121T,2019MNRAS.483.4640W}. These simulations produce seeds directly out of sufficiently dense and metal poor gas cells, which is much more consistent with proposed theoretical seeding channels. But these approaches can lead to stronger resolution dependence in the simulated BH populations~(see Figure 10 of \citealt{2014MNRAS.442.2751T}). In any case, most of these seeding approaches have achieved significant success in generating satisfactory agreement with the observed SMBH populations at $z\sim0$~\citep{2020MNRAS.493..899H}. However, it is important to note that they do not provide definitive discrimination among the potential seeding channels from which the simulated BHs may have originated.

A standard approach to achieve very high resolutions in cosmological simulations is to use the `zoom-in' technique. In our previous work \citep{2021MNRAS.507.2012B,2022MNRAS.510..177B}, we used cosmological zoom-in simulations with gas mass resolutions up to $\sim10^3~M_{\odot}$ to build a new set of gas based seed models that placed seeds down to the lowest masses~($1.56\times10^3~M_{\odot}/h$) within halos containing sufficient amounts of star forming~\&~metal poor gas. We systematically explored these gas based seed models and found that the strongest constraints for seeding are expected within merger rates measurable with LISA. However, the predictions for these zoom simulations are subject to large cosmic variance, as they correspond to biased regions of the large-scale structure. In order to make observationally testable predictions with these gas based seed models, we must find a way to represent them in cosmological simulations despite the lack of sufficient resolution.  

In this work, we build a new sub-grid stochastic seed model that can represent low mass seeds born out of star forming and metal poor gas, within lower-resolution and larger-volume simulations that cannot directly resolve them. To do this, we first run a suite of highest resolution zoom simulations that places $1.56\times10^3~M_{\odot}/h$ seeds within star forming and metal poor gas using %based on 
the gas based seed models from \cite{2021MNRAS.507.2012B}. We then study the growth of $1.56\times10^3~M_{\odot}/h$ seeds and the evolution of their formation environments. We particularly study the halo and galaxy properties wherein these seeds assemble higher mass~($1.25\times10^4~\&~1\times10^5~M_{\odot}/h$) descendants. We then use the results to build our stochastic seed model that directly seeds these descendants within lower resolution versions of the same zoom region. In the process, we determine the key ingredients required for these stochastic seed models to reproduce the results of the gas based seed models in the lower resolution zooms. 

Section \ref{methods} presents the basic methodology, which includes the simulation suite, the underlying galaxy formation model, as well as the BH seed models. Our main results are described in sections \ref{Black hole mass assembly} and \ref{subhalo based stochastic seed model}. In section \ref{Black hole mass assembly}, we present the results for the formation and growth of $1.56\times10^3~M_{\odot}/h$ seeds within our highest resolution zoom simulations. In section \ref{subhalo based stochastic seed model}, we use the results from section \ref{Black hole mass assembly} to build our stochastic seed model. Finally, Section \ref{Summary and Conclusions} summarizes our main results.

%The closer we get to detecting BHs close to their initial seed masses, adds an exciting dimension to our understanding of black hole formation and growth.

\section{Methods}
\label{methods}
\subsection{AREPO cosmological code and the Illustris-TNG model}
\label{AREPO cosmological code and the Illustris-TNG model}
We use the \texttt{AREPO} gravity + magneto-hydrodynamics~(MHD) solver~\citep{2010MNRAS.401..791S,2011MNRAS.418.1392P,2016MNRAS.462.2603P,2020ApJS..248...32W} to run our simulations. The simulations use a $\Lambda$ cold dark matter cosmology with parameters adopted from \cite{2016A&A...594A..13P}: ($\Omega_{\Lambda}=0.6911, \Omega_m=0.3089, \Omega_b=0.0486, H_0=67.74~\mathrm{km}~\mathrm{sec}^{-1}\mathrm{Mpc}^{-1},\sigma_8=0.8159,n_s=0.9667$). The gravity solver uses the PM Tree~\citep{1986Natur.324..446B} method and the MHD solver for gas dynamics uses a quasi-Lagrangian description of the fluid within an unstructured grid generated via a Voronoi tessellation of the domain. Halos are identified using the friends of friends~(FOF) algorithm~\citep{1985ApJ...292..371D} with a linking length of 0.2 times the mean particle separation. Subhalos are computed using the  \texttt{SUBFIND}~\citep{2001MNRAS.328..726S} algorithm for each simulation snapshot. Aside from our BH seed models, our underlying galaxy formation model is the same as the \texttt{IllustrisTNG}~(TNG) simulation suite~\citep{2018MNRAS.475..676S,2018MNRAS.475..648P,2018MNRAS.475..624N,2018MNRAS.477.1206N,2018MNRAS.480.5113M,2019ComAC...6....2N} \citep[see also][]{2018MNRAS.479.4056W,2018MNRAS.474.3976G,2019MNRAS.485.4817D,2019MNRAS.484.5587T,2019MNRAS.483.4140R,2019MNRAS.490.3234N,2019MNRAS.490.3196P,2021MNRAS.500.4597U,2021MNRAS.503.1940H}. The TNG model includes a wide range of sub-grid physics for star formation and evolution, metal enrichment and feedack as detailed in \cite{2018MNRAS.473.4077P} and also summarized in our earlier papers \citep{2021MNRAS.507.2012B,2022MNRAS.510..177B,2022MNRAS.516..138B}. 

\subsection{BH accretion, feedback and dynamics}
BH accretion rates are determined by the Eddington-limited Bondi-Hoyle formalism given by  
\begin{eqnarray}
\dot{M}_{\mathrm{bh}}=\mathrm{min}(\dot{M}_{\mathrm{Bondi}}, \dot{M}_{\mathrm{Edd}})\\
\dot{M}_{\mathrm{Bondi}}=\frac{4 \pi G^2 M_{\mathrm{bh}}^2 \rho}{c_s^3}\\
\dot{M}_{\mathrm{Edd}}=\frac{4\pi G M_{\mathrm{bh}} m_p}{\epsilon_r \sigma_T~c}
\label{bondi_eqn}
\end{eqnarray} 
where $G$ is the gravitational constant, $\rho$ is the local gas density, $M_{\mathrm{bh}}$ is the BH mass, $c_s$ is the local sound speed, $m_p$ is the proton mass, and $\sigma_T$ is the Thompson scattering cross section. Accreting black holes radiate at bolometric luminosities given by, 
\begin{equation}
    L_{\mathrm{bol}}=\epsilon_r \dot{M}_{\mathrm{bh}} c^2,
    \label{bol_lum_eqn}
\end{equation}
where $\epsilon_r=0.2$ is the radiative efficiency.

IllustrisTNG implements a dual mode AGN feedback. `Thermal feedback' is implemented for Eddington ratios~($\eta \equiv \dot{M}_{\mathrm{bh}}/\dot{M}_{\mathrm{edd}}$) higher than a critical value of  $\eta_{\mathrm{crit}}=\mathrm{min}[0.002(M_{\mathrm{BH}}/10^8 M_{\odot})^2,0.1]$. Here, thermal energy is deposited on to the neighboring gas at a rate of $\epsilon_{f,\mathrm{high}} \epsilon_r \dot{M}_{\mathrm{BH}}c^2$ with $\epsilon_{f,\mathrm{high}} \epsilon_r=0.02$ where $\epsilon_{f,\mathrm{high}}$ is the ``high accretion state" coupling efficiency. `Kinetic feedback' is implemented for Eddington ratios lower than the critical value. Here, kinetic energy is injected into the gas in a pulsed fashion whenever sufficient feedback energy is available, which manifests as a `wind' oriented along a randomly chosen direction. The injected rate is $\epsilon_{f,\mathrm{low}}\dot{M}_{\mathrm{BH}}c^2$ where $\epsilon_{f,\mathrm{low}}$ is called the `low accretion state' coupling efficiency~($\epsilon_{f,\mathrm{low}} \lesssim 0.2$). For further details, we direct the interested readers to \cite{2017MNRAS.465.3291W}.

The limited mass resolution hinders our simulations from fully capturing the crucial BH dynamical friction force, especially for low masses. To stabilize the dynamics, BHs are relocated to the nearest potential minimum within their proximity, determined by the closest $10^3$ neighboring gas cells. When one BH enters the neighborhood of another, prompt merger occurs.  

%\subsection{Halo and subhalo identification}
%\label{Halo and subhalo identification}

\subsection{Black hole seed models}
\label{Black hole seed models}
\subsubsection{Gas based seed model}
\label{gas based seeding}

We explore the formation and growth of the lowest mass $1.56\times10^3~M_{\odot}/h$ seeds using the gas based seeding prescriptions developed in \cite{2021MNRAS.507.2012B}. In order to contrast these seeds from those produced by the seed model discussed in the next subsection, we shall hereafter refer to them as  \textit{direct gas based} seeds or DGBs with mass $\seedmass$. These seeding criteria are meant to broadly encompasses popular theoretical channels such as Pop III, NSC and DCBH seeds, that are postulated to form in regions comprised of dense and metal poor gas. We briefly summarize them as follows: 
\begin{itemize}
\item \textit{Star forming \& metal poor gas mass criterion:} We place DGBs in halos with a minimum threshold of dense~($>0.1~\mathrm{cm}^{-3}$) \& metal poor~($Z<10^{-4}~Z_{\odot}$) gas mass, denoted by $\msfmp$~(in the units of $\seedmass$). The values of $\msfmp$ are not constrained, but we expect it to be different for the various seeding channels. In this work, we consider models with $\msfmp=5,50,150~\&~1000$.
\item \textit{Halo mass criterion:} We place DGBs in halos with a total mass exceeding a critical threshold, specified by $\tilde{M}_{h}$ in the units of $\seedmass$. In this work, we consider $\tilde{M}_{h}=3000~\&~10000$. While our seeding prescriptions are meant to be based on the gas properties within halos, we still adopt this criterion to avoid seeding in halos significantly below the atomic cooling threshold. This is because our simulations do not include the necessary physics~(for e.g. $H_2$ cooling) to self-consistently capture the collapse of gas and the formation of stars within these (mini)halos. Additionally, these lowest mass halos are also impacted by the finite simulation resolution, many of which are spuriuosly identified gas clumps with very little DM mass. (Please see Figure \ref{halo_evolution_star_formation} and Appendix \ref{Evolution of star forming metal poor gas in halos} for further discussion about the foregoing points.) Another motivation for this criterion is that NSC seeds are anticipated to grow more efficiently within sufficiently deep gravitational potential wells where runaway BH merger remnants face difficulties escaping the cluster. Deeper gravitational potentials are expected in higher mass halos.
\end{itemize}

Our gas based seed models will therefore contain three parameters, namely $\msfmp$, $\tilde{M}_{\mathrm{h}}$ and $\seedmass$. The simulation suite that will use these seed models will be referred to as \texttt{GAS_BASED}. The individual runs will be labelled as \texttt{SM*_FOF*} where the `*'s correspond to the values of $\msfmp$ and $\tilde{M}_{\mathrm{h}}$ respectively. For example, $\msfmp=5$ and $\tilde{M}_{\mathrm{h}}=3000$ will correspond to \texttt{SM5_FOF3000}. As already mentioned, the seed masses in this suite will be $\seedmass=1.56\times10^3~M_{\odot}/h$.

\subsubsection{Stochastic seed model}
\label{Subhalo-based stochastic seeding}
As we mentioned, the key goal of this work is to build a new approach to represent low mass seeds in larger-volume lower-resolution cosmological simulations that cannot directly resolve them. As we shall see in Section \ref{subhalo based stochastic seed model}, this is achieved via a new stochastic seeding model. The complete details of this seed model are described in Section \ref{subhalo based stochastic seed model}, where we thoroughly discuss their motivation and calibration using the results obtained from the \texttt{GAS_BASED} suite. Here, we briefly summarize key features so that the reader can contrast it against the gas based seed models described in the previous subsection.

Since the simulations here will not fully resolve the $1.56\times10^3~M_{\odot}/h$ DGBs, we will essentially seed their resolvable descendants. To distinguish them from the DGBs, we shall refer to these seeded descendants as \textit{extrapolated seed descendants} or ESDs with masses~(denoted by $\descendantseedmass$) limited to the gas mass resolution of the simulations. In this work, we will largely explore ESD masses $\descendantseedmass=1.25\times10^4~\&~1\times10^5~M_{\odot}/h$, to be used for simulations with gas mass resolutions of $\sim10^4~\&~10^5~M_{\odot}/h$ respectively. 

To seed the ESDs, we identify sites using the FOF algorithm, but with a shorter linking length~(by factor of $\sim1/3$) compared to that used for identifying halos. We shall refer to these short linking length FOFs as ``best-Friends of Friends or bFOFs''. These bFOFs essentially correspond to galaxies or proto-galaxies residing inside the halos. We do this to accommodate the formation of multiple ESDs per halo; this is because even if we seed one DGB per halo in the gas based seed models, subsequent evolution of hierarchical structure naturally leads to halos occupying  multiple higher mass descendants. Notably, one could alternatively seed in subhalos computed by \texttt{SUBFIND}; however, \texttt{SUBFIND} is prohibitively expensive to be called frequently enough for seeding BHs. Hereafter, in most instances, we shall simply refer to these bFOFs as ``galaxies".  %\footnote{Traditionally, galaxies in most simulations have been defined to be baryonic components of subhalos computed using traditional subhalo finders such as \texttt{SUBFIND}. But in this work, they refer to baryonic components of bFOFs. However, we do note that both \texttt{SUBFIND} based as well as bFOF based galaxy catalogs are available for our simulation snapshots}. 
 Their properties are comprehensively studied in Section \ref{bFOF_details_sec}.

%\aklant{In this paper, our primary emphasis is on the utilization of bFOFs as subhalos, while the \texttt{SUBFIND} based subhalos receive comparatively less attention. Henceforth, we will employ the term ``subhalos" to specifically refer to bFOFs. However, in certain specific instances where we do make use of \texttt{SUBFIND} based subhalos, we will refer to them as ``subfind-subhalos" to distinguish them from the bFOFs.}

The ESDs will be stochastically placed in galaxies based on where the descendants of the $1.56\times10^3~M_{\odot}/h$ DGBs end up within the \texttt{GAS_BASED} suite. Below we provide a brief summary of the seeding criteria

\begin{itemize}
\item \textit{Galaxy mass criterion}: We will apply a galaxy mass~(`galaxy mass' hereafter refers to the total mass including dark matter, gas and stars) seeding threshold that will be stochastically drawn from galaxy mass distributions predicted for the assembly of ($1.25\times10^4$ and $10^5~M_{\odot}/h$) BHs that are descendants of $1.56\times10^3~M_{\odot}/h$ DGBs within the \texttt{GAS_BASED} suite. As we explore further, it becomes evident that these distributions vary with redshift and exhibit significant scatter. The redshift dependence will capture the influence of halo growth, star formation, and metal enrichment on seed formation in our gas based seed models.

\item \textit{Galaxy environment criterion}: In the context of a galaxy, we define its \textit{environment} as the count of neighboring halos~($N_{\mathrm{ngb}}$) that exceed the mass of its host halo and are located within a specified distance~(denoted by $D_{\mathrm{ngb}}$) from the host halo. In this study, we determine $N_{\mathrm{ngb}}$ within a range of 5 times the virial radius~($R_{\mathrm{vir}}$) of the host halo, i.e. $D_{\mathrm{ngb}}=5 R_{\mathrm{vir}}$. This choice is suitable for investigating the immediate small-scale external surroundings of the galaxy, extending beyond its host halo. We then apply a seeding probability~(less than unity) to suppress ESD formation in galaxies with $\leq1$ neighboring halos, thereby favoring their formation in richer environments. By doing this, we account for the impact of unresolved hierarchical merger dominated growth from $\seedmass$ to $\descendantseedmass$, as it favors more rapid BH growth within galaxies in richer environments.
\end{itemize}

The simulations that use only the \textit{galaxy mass criterion} will be referred to as the \texttt{STOCHASTIC_MASS_ONLY} suite. For simulations which use both \textit{galaxy mass criterion} and \textit{galaxy environment criterion}, we will refer to them as the \texttt{STOCHASTIC_MASS_ENV} suite. During the course of this paper, we will illustrate that the outcomes of each simulation of a specific region within the \texttt{GAS_BASED} suite, employing a distinct set of gas based seeding parameters, can be reasonably well reproduced in a lower-resolution simulation of the same region within the \texttt{STOCHASTIC_MASS_ENV} suite.

\subsection{Simulation suite}
\label{Simulation Suite}

Our simulation suite consists of zoom runs for the same overdense region as that used in \cite{2021MNRAS.507.2012B}~(referred to as \texttt{ZOOM_REGION_z5}). The region was chosen from a parent uniform volume of $(25~\mathrm{Mpc}/h)^3$, and is targeted to produce a $3.5\times10^{11}~M_{\odot}/h$ halo at $z=5$. The simulations were run from $z=127$ to $z=7$ using the \texttt{MUSIC}~\citep{2011MNRAS.415.2101H} initial condition generator. The background grid's resolution and the resolution of high-resolution zoom regions are determined by two key parameters: $L_{\mathrm{min}}$ (or $\mathrm{levelmin}$) and $L_{\mathrm{max}}$ (or $\mathrm{levelmax}$) respectively. These parameters define the resolution level, denoted as $L$, which is equivalent to the mass resolution produced by $2^L$ number of dark matter (DM) particles per side in a uniform-resolution $(25~\mathrm{Mpc}/h)^3$ box. Specifically, we set $L_{\mathrm{min}}=7$ for the background grid, resulting in a DM mass resolution of $5.3\times10^9~M_{\odot}/h$. For the high-resolution zoom region, we explore $L_{\mathrm{max}}$ values of $10$, $11$ and $12$. In addition, there is a buffer region that consists of DM particles with intermediate resolutions bridging the gap between the background grid and the zoom region. This buffer region serves a crucial purpose of facilitating a smooth transition between the zoom region and the background grid. Our simulation suite is comprised of the following set of resolutions for the zoom regions:
\begin{itemize}
\item In our highest resolution $L_{\mathrm{max}}=12$ runs, we achieve a DM mass resolution of $1.6 \times 10^4~M_{\odot}/h$ and a gas mass resolution of $\sim10^3M_{\odot}/h$~(the gas cell masses are contingent upon the degree of refinement or derefinement of the Voronoi cells, thereby introducing some variability). These runs are used for the \texttt{GAS_BASED} suite that seeds DGBs at $1.56\times10^3~M_{\odot}/h$ using the gas based seed models described in Section \ref{gas based seeding}.

\item For our $L_{\mathrm{max}}=11~\&~10$ runs, we achieve DM mass resolutions of $1.3 \times 10^5~\&~1\times10^6~M_{\odot}/h$ and gas mass resolutions of $\sim10^4~\&~10^5~M_{\odot}/h$ respectively. These runs will be used for the \texttt{STOCHASTIC_MASS_ONLY} and \texttt{STOCHASTIC_MASS_ENV} suite, that will seed ESDs at $1.25\times10^4~\&~1\times10^5~M_{\odot}/h$ for $L_{\mathrm{max}}=11~\&~10$ respectively, using the stochastic seed models described in Section \ref{Subhalo-based stochastic seeding}.
\end{itemize}
Further details of our full simulation suite are summarized in Table \ref{tab:my_label}. It is important to note that our new stochastic seed models will be primarily designed for implementation within larger-volume uniform simulations. However, this paper specifically focuses on zoom simulations. In particular, we are using $L_{\mathrm{max}}=11~\&~10$ zoom simulations for testing the stochastic seed models against the highest resolution $L_{\mathrm{max}}=12$ zooms that use the gas based seed models. In a subsequent paper~(Bhowmick et al in prep), we will be applying the stochastic seed models on uniform volume simulations of the same resolutions as the $L_{\mathrm{max}}=11~\&~10$ zooms.
\begin{table*}
     \centering
    \begin{tabular}{c|c|c|c|c|c|c|}
         $L_{\mathrm{max}}$ & $M_{dm}$~($M_{\odot}/h$) & $M_{gas}$~($M_{\odot}/h$) & $\epsilon~(kpc/h)$ & Black hole neighbors & Seed mass~($M_{\odot}/h$) & Seed model\\
         \hline
         12 & $1.6\times10^4$ & $\sim10^3$ & 0.125 & 256 & $\seedmass=1.56\times10^{3}$ & gas based seeding\\
         11 & $1.3\times10^5$ & $\sim10^4$ & 0.25 & 128 & $\descendantseedmass=1.25\times10^{4}$ & Stochastic seeding  \\
         10 & $1\times10^6$ & $\sim10^5$ & 0.5 & 64 & $\descendantseedmass=1\times10^{5}$ & Stochastic seeding\\
         \hline
    \end{tabular}
    \caption{Spatial and mass resolutions within the zoom region of our simulations for various values of $L_{\mathrm{max}}$~(see Section \ref{Simulation Suite} for the definition). $M_{dm}$ is the mass of a dark matter particle, $M_{gas}$ is the typical mass of a gas cell~(note that gas cells can refine and de-refine depending on the local density), and $\epsilon$ is the gravitational smoothing length. The 4th column represents the number of nearest gas cells that are assigned to be BH neighbors. The 5th and 6th columns correspond to the seed mass and seed model used at the different resolutions.}
    \label{tab:my_label}
\end{table*}

\subsection{Tracing BH growth along merger trees: The \texttt{SUBLINK} algorithm}
\label{Tracing BH growth along merger trees}
We use the \texttt{GAS_BASED} suite to trace the growth of the lowest mass $1.56\times10^3~M_{\odot}/h$ DGBs and study the evolution of their environments~(halo and galaxy properties) as they assemble higher mass BHs. We do this by first constructing subhalo merger trees using the \texttt{SUBLINK} algorithm~\citep{2015MNRAS.449...49R}, which was designed for \texttt{SUBFIND} based subhalos. Note that these \texttt{SUBFIND} based subhalos, like bFOFs, also trace the substructure within halos. Therefore, to avoid confusion, we shall refer to \texttt{SUBFIND} based subhalos as ``subfind-subhalos". It is also very common to interpret the subfind-subhalos as 
``galaxies". As we shall see however, in this work, we only use these subfind-subhalos as an intermediate step to arrive at the FOF and bFOF merger trees. Therefore, there is no further mention of subfind-subhalos after this subsection. On that note, recall again that any mention of ``galaxy" in our paper refers to the bFOFs. 

\texttt{SUBFIND} was run on-the-fly to compute subfind-subhalos within both FOF and bFOF catalogues. Therefore, for obtaining both FOF and bFOF merger trees, we first compute the merger trees of their corresponding subfind-subhalos. Following are the key steps in the construction of the subfind-subhalo merger tree:
\begin{itemize}
\item For each progenitor subfind-subhalo at a given snapshot, \texttt{SUBLINK} determines a set of candidate descendant subfind-subhalos from the next snapshot. Candidate descendants are those subfind-subhalos which have common DM particles with the progenitor. 
\item Next, each candidate descendant is given a score based on the merit function $\chi=\sum_i 1/R_i^{-1}$ where $R_i$ is the binding energy rank of particle $i$ within the progenitor. DM particles with higher binding energy within the progenitor are given a lower rank. $\sum_i$ denotes a sum for all the particles within the candidate descendant.
\item Amongst all the candidate descendants, the final unique descendant is chosen to be the one with the highest score. This essentially ensures that the unique descendant has the highest likelihood of retaining the most bound DM particles that resided within the progenitor.
\end{itemize}

From the subfind-subhalo merger trees, we use the ones that only consist of central subfind-subhalos~(most massive within a FOF or bFOF) and construct the corresponding FOF/ halo merger trees and bFOF/galaxy merger trees. We then trace the growth of BHs along these merger trees, and the outcomes of this analysis are elaborated upon in the subsequent sections.

\section{RESULTS I: Black hole mass assembly in high-resolution zooms}
\label{Black hole mass assembly}
\begin{figure*}
\includegraphics[width=14 cm]{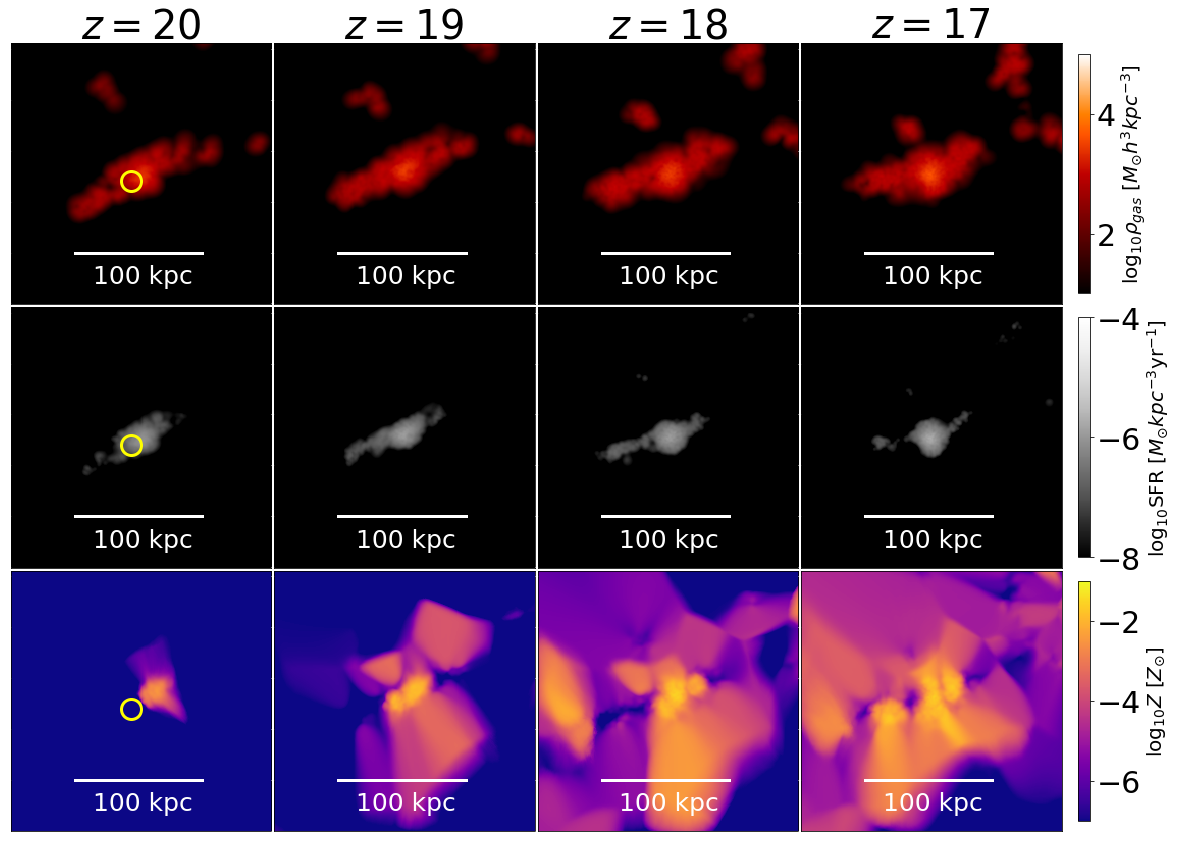}\\
\includegraphics[width=14 cm]{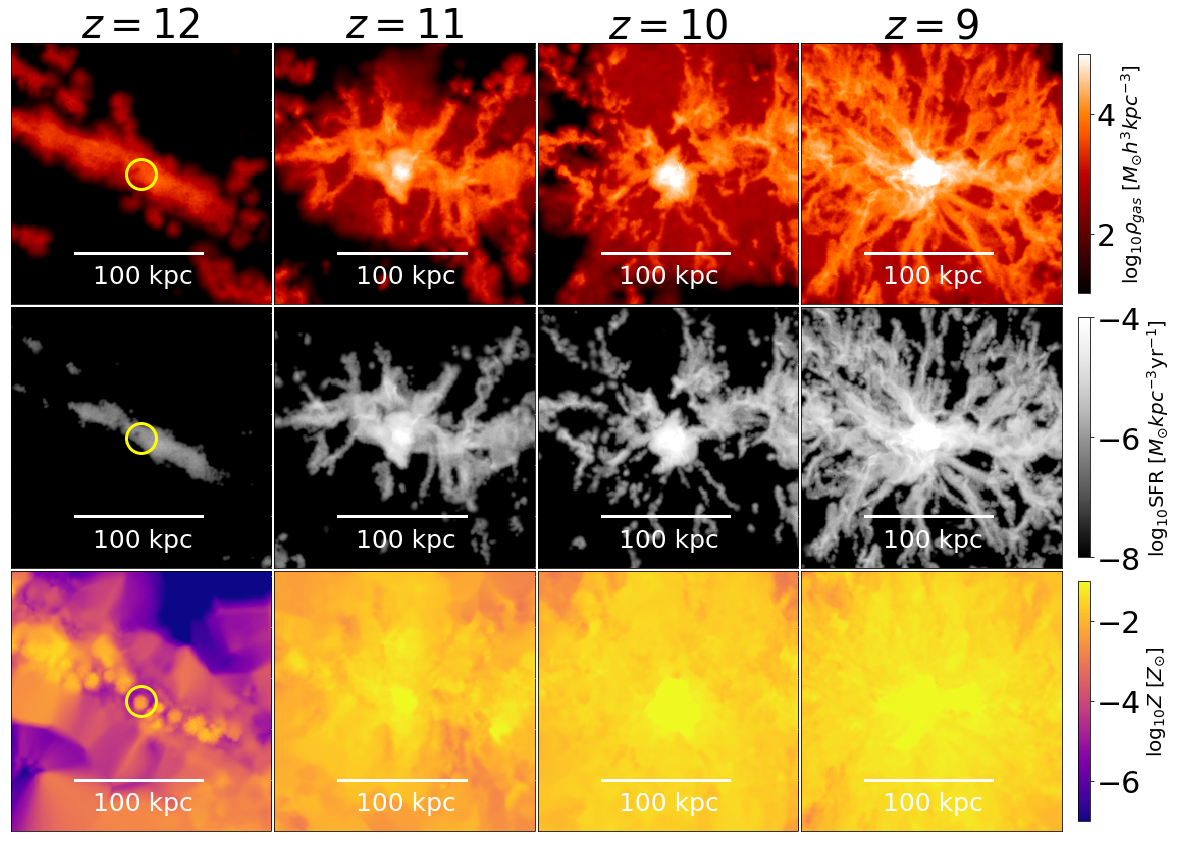}

\caption{Evolution of gas density (red/orange), star formation rate density (grayscale) and gas metallicity (yellow/purple) of various seed forming sites in our zoom simulations that use the gas based seed models described in Section \ref{gas based seeding}. Hereafter, we shall refer to the seeds formed by the gas based seed models as ``Direct Gas Based seeds" or DGBs. The large panels correspond to DGB forming sites from two distinct epochs namely $z=20$~(top) and $z=12$~(bottom). Within each large panel, the leftmost sub-panel corresponds to the snapshot at the time of DGB formation, wherein the yellow circles mark the location of the formation site that contains the star forming~\&~metal poor gas. The remaining subpanels from left to right show the evolution of that formation site along three subsequent snapshots. We can clearly see that at the time of DGB formation, the regions in the immediate vicinity of the formation site have already started the process of metal enrichment. As a result, these regions get completely polluted with metals within a very short time after DGB formation.}
\label{evolution_of_seed_forming_gas}
\end{figure*}

\begin{figure*}
\includegraphics[width=16 cm]{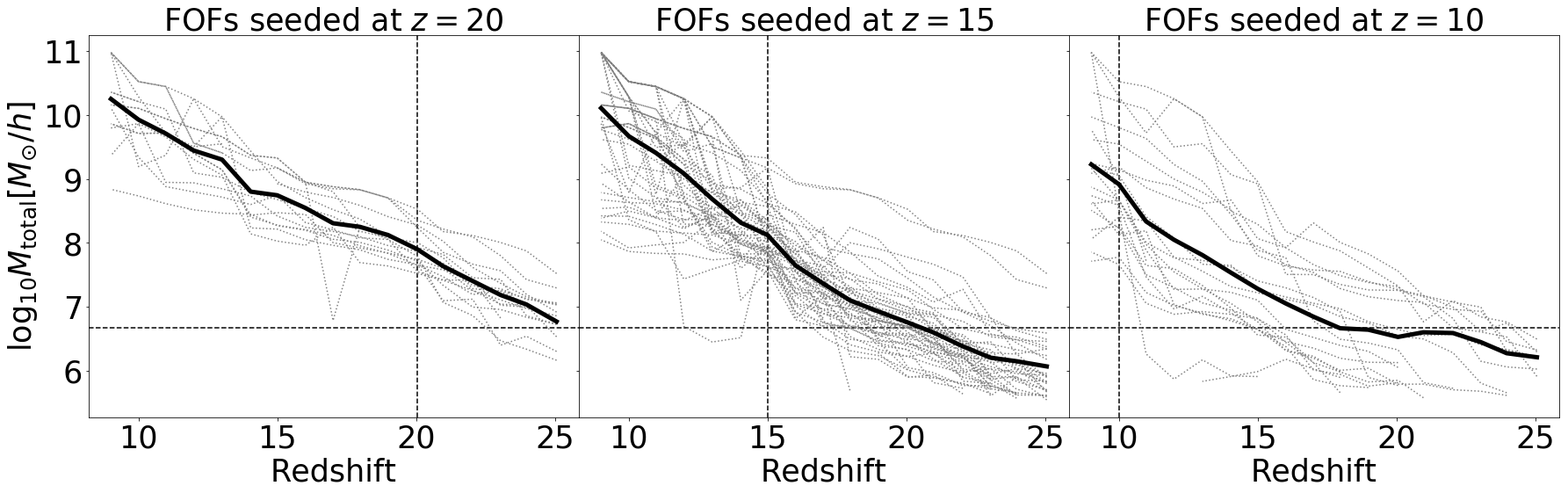}\\
\includegraphics[width=16 cm]{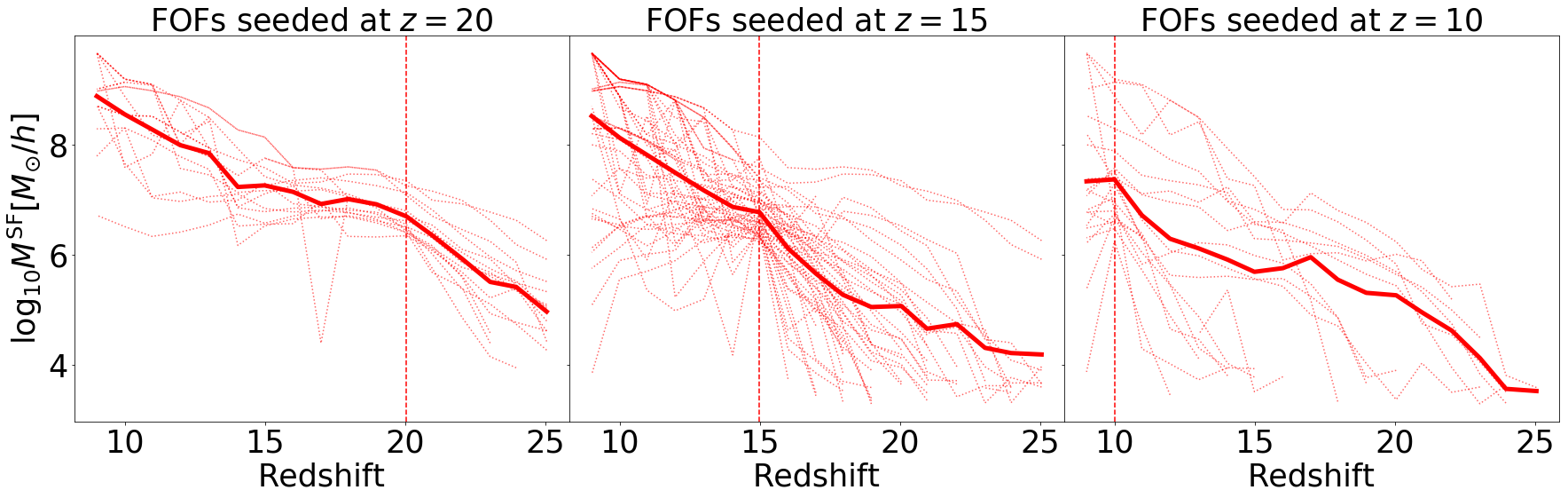}\\
\includegraphics[width=16 cm]{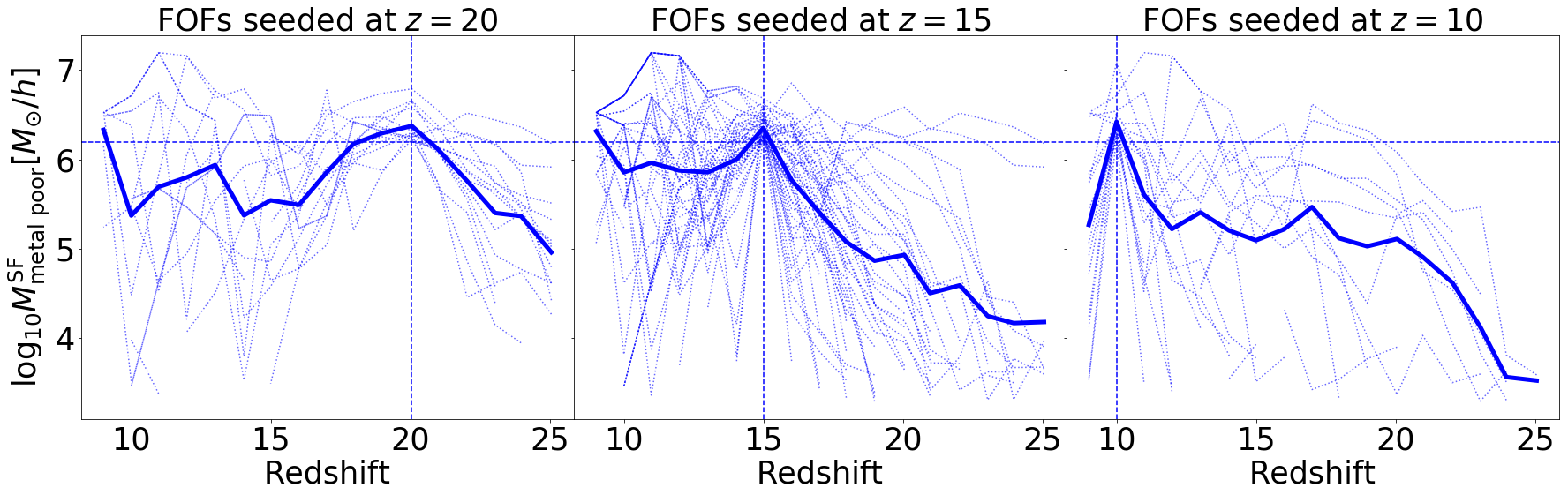}\\
\includegraphics[width=16 cm]{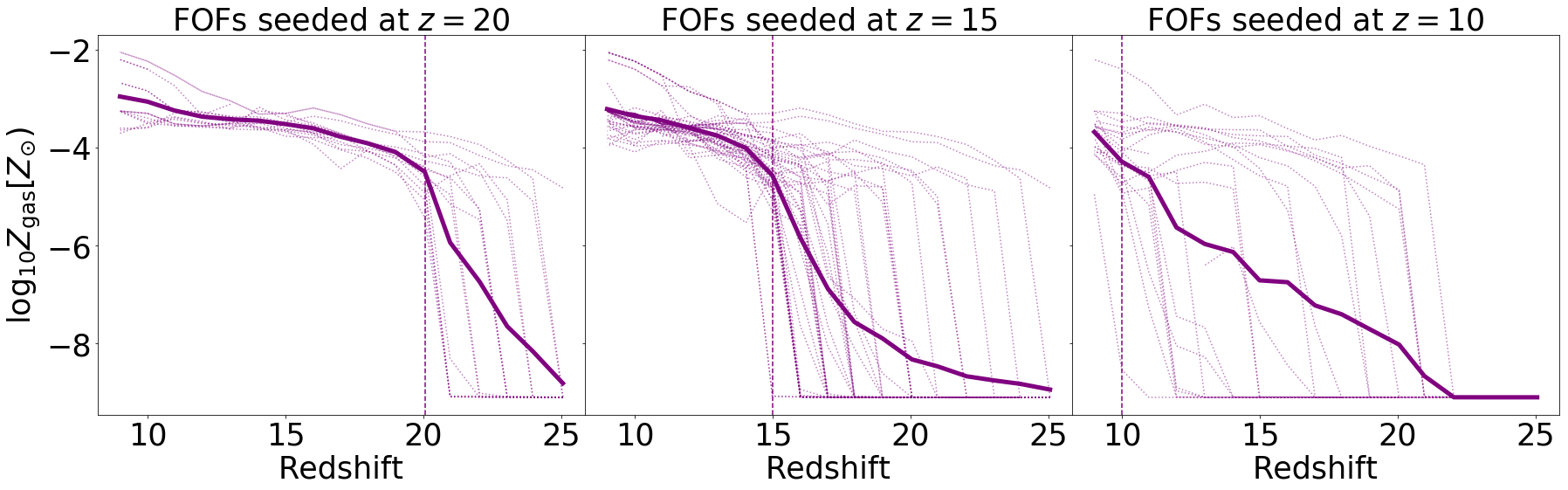}\\
\caption{Assembly history of halos forming $1.56\times10^3~M_{\odot}/h$ DGBs %based on 
using gas based seed models. Top to bottom, the rows %correspond to 
show the evolution of total halo mass~($M_{\mathrm{total}}$), star forming gas mass~($M^{\mathrm{SF}}$), star forming \& metal poor gas mass~($M^{\mathrm{SF}}_{\mathrm{metal \ poor}}$), and gas metallicity~($Z_{\mathrm{gas}}$). Left, middle and right panels show halos seeded at $z=20$, $z=15$ and $z=10$~(vertical dashed lines in each column) respectively, %based on 
using the gas based seeding criterion, $\msfmp = 1000$~(horizontal dashed line in 3rd row) and $\tilde{M}_{\mathrm{h}} = 3000$~(horizontal dashed line in 1st row). The faded dotted lines show the evolution of all DGB-forming halos along their merger trees. The thick solid lines show the mean trend, i.e. logarithmic average of the values of all the faded dotted lines at each redshift. The star forming \& metal poor gas masses tend to sharply drop soon after seeding, independent of the time of seeding. This is because the DGB forming halos have already started to undergo rapid metal enrichment, which is shown in the fourth row by the rapid increase in gas metallicity prior to the seeding event.}
\label{evolution_of_seed_forming_gas_all}
\end{figure*}

\begin{figure*}
\includegraphics[width=16 cm]{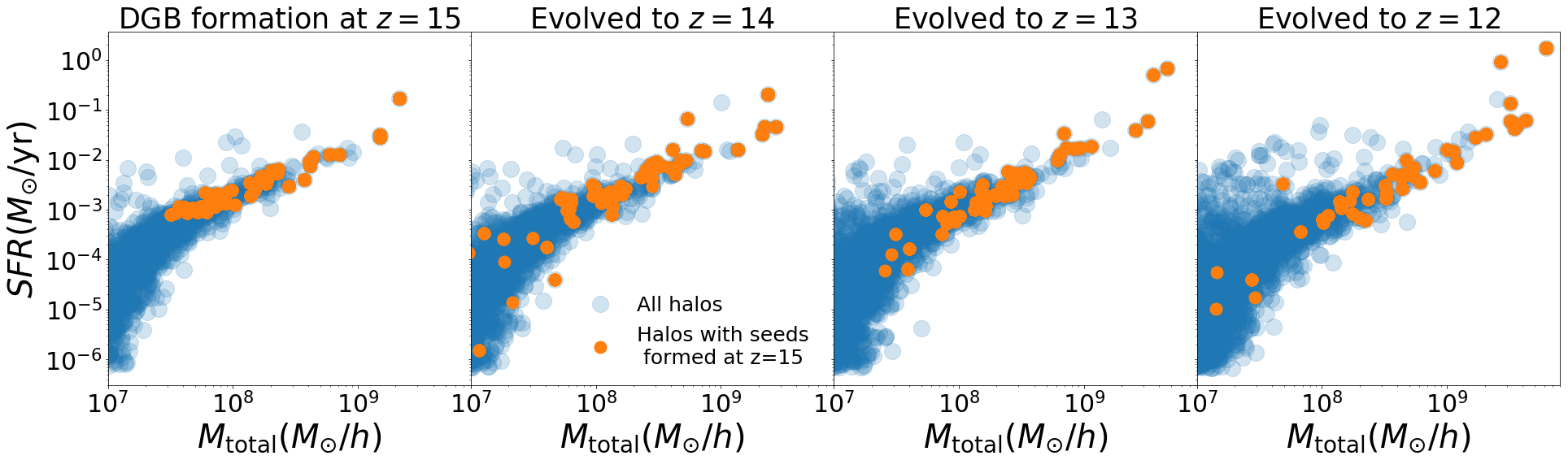}\\
\includegraphics[width=16 cm]{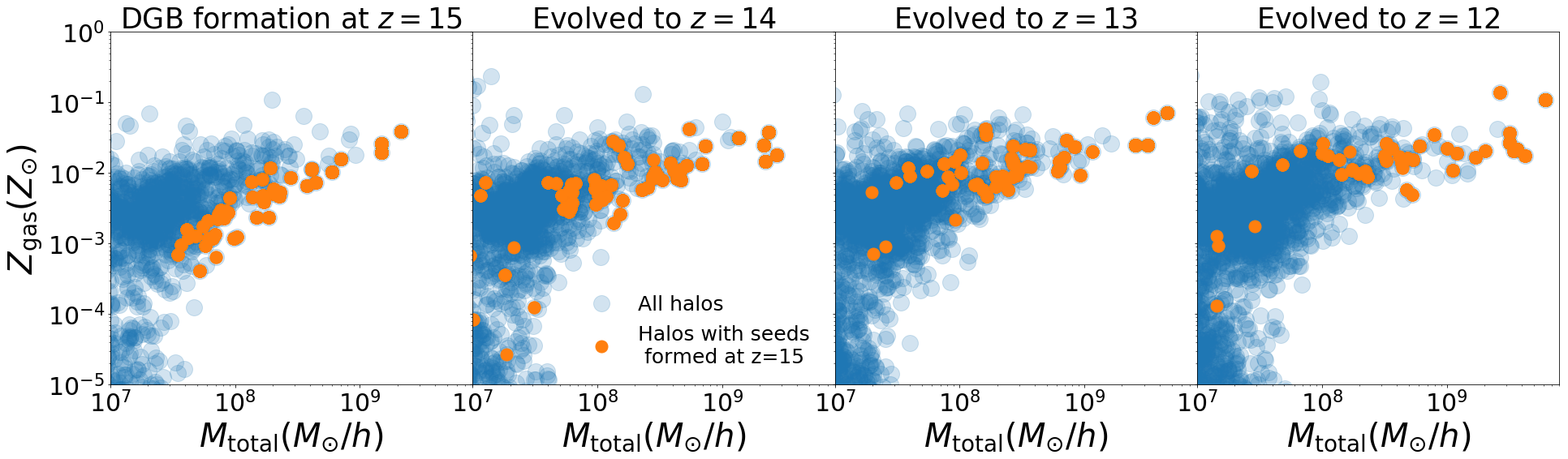}\\
\caption{The evolution of host star formation rates or SFR (top panels) and $Z_{\mathrm{gas}}$ (bottom panels) versus host mass is shown for $1.56\times10^3~M_{\odot}/h$ DGBs formed at $z=15$. In the leftmost panels, the filled orange circles %trace the evolution of 
indicate the halos that form DGBs at $z=15$. The filled orange circles in the subsequent panels (from left to right) show the same host halos at $z=14,13~\&~12$. The full population of halos at each redshift is shown in blue %~(left to right panels) on the $M_{\mathrm{total}}$ vs SFR plane~(top panels) and the $M_{\mathrm{total}}$ vs $Z$ plane~(bottom panels). 
In other words, we select the orange circles at $z=15$ using our gas based seeding criteria $[\mh,\msfmp = 3000,1000]$~(assuming $\seedmass=1.56\times10^3~M_{\odot}/h$ ), and follow their evolution on the halo merger tree. 
Comparing them to the full population of halos at each redshift, we find that even though the DGB forming halos at $z=15$ are biased towards lower gas metallicities at fixed halo mass~(lower left panel), subsequent evolution of these halos to lower redshifts causes them to become more unbiased at $z=14,13~\&~12$. This is due to the rapid metal enrichment of these DGB forming halos depicted in Figure \ref{evolution_of_seed_forming_gas_all}.}
\label{unbiased_halo_at_assembly_fig}
\end{figure*}

We start our analysis by looking at the growth history of $1.5\times10^3~M_{\odot}/h$ DGBs within the \texttt{GAS_BASED} suite. We trace their growth along halo merger trees~(see Section \ref{Tracing BH growth along merger trees}) from the time of their formation to when they assemble higher mass ($1.25\times10^4,1\times10^5~\&~8\times10^5~M_{\odot}/h$) descendant BHs. We choose these descendant BH masses as they encompass the target gas mass resolutions of our lower resolution~ ($L_{\mathrm{max}}=11~\&~10$) zooms. These are also comparable to typical gas mass resolutions of cosmological simulations in the existing literature. For example, the \texttt{TNG100}~\citep{2018MNRAS.475..624N}, \texttt{Illustris}~\citep{2014Natur.509..177V,2014MNRAS.444.1518V}, \texttt{EAGLE}~\citep{2015MNRAS.446..521S}, \texttt{MassiveBlackII}~\citep{2015MNRAS.450.1349K}, \texttt{BlueTides}~\citep{2016MNRAS.455.2778F} and \texttt{HorizonAGN}~\citep{2017MNRAS.467.4739K} simulations have a gas mass resolution of  $\sim10^6~M_{\odot}$ and similar values for the seed masses. The relatively smaller volume cosmological simulations such as \texttt{ROMULUS25}~\citep{2017MNRAS.470.1121T} and TNG50~\citep{2019MNRAS.490.3196P} have a gas mass resolution of $\sim10^5~M_{\odot}$ with a seed mass of $10^6~M_{\odot}$. Recall again that most of these simulations seed BHs simply based on either a constant halo mass threshold, or poorly resolved local gas properties. 
%As we shall see, the results of 
The results presented in this section will be used in Section \ref{subhalo based stochastic seed model} to calibrate the stochastic seed model that will represent the gas based $1.56\times10^3~M_{\odot}/h$ seeds in the lower-resolution zooms without resolving them directly.

\subsection{Evolution of seed forming sites: Rapid metal enrichment after seed formation}
\label{Rapid metal enrichment after seed formation}
Figure \ref{evolution_of_seed_forming_gas} depicts the evolution of gas density, star formation rate~(SFR) density, and gas metallicity at DGB forming sites from two distinct epochs ($z = 20~\&~12$). As dictated by our gas based seed models, for each of the DGB forming sites there exists gas that is simultaneously forming stars but is also metal poor~(marked in yellow circles). However, we also find that metal enrichment has already commenced at the immediate vicinity of these DGB forming sites. In other words, DGB formation occurs in halos where metal enrichment has already begun due to prior star formation and evolution, but it has not polluted the entire halo yet. But soon after DGB formation, i.e. within a few tens of million years, we find that the entirety of the regions becomes polluted with metals.  

The rapid metal enrichment of DGB forming halos is shown much more comprehensively and quantitatively in Figure  \ref{evolution_of_seed_forming_gas_all}. Here we show the evolution of halo mass, star forming gas mass, star forming metal poor gas mass and gas metallicity from $z\sim25-7$ for all DGB forming halos along their respective merger trees~(faded dotted lines). To avoid overcrowding of the plots, we select trees based on the most restrictive seeding criterion of $\msfmp=1000~\&~ \mh=3000$, but our general conclusions hold true for other seeding thresholds as well. Not surprisingly, the halo mass~(1st row) and star forming gas mass~(2nd row) tend to monotonically increase with decreasing redshift on average~(thick solid black lines). Note that for individual trees, the halo mass can occasionally decrease with time due to tidal stripping. On more rare occasions, there may also be a sharp drop in the the halo mass at given snapshot followed by a sharp rise back to being close to the original value. This is likely because the FOF finder ``mistakenly'' splits a larger halo in two at that snapshot. The star forming gas mass can also additionally decrease with time due to the star forming gas being converted to star particles.    

Very importantly, the star forming~\&~metal poor gas mass~(3rd row of Figure \ref{evolution_of_seed_forming_gas_all}) increases initially and peaks at the time of DGB formation, following which it rapidly drops down. This happens independent of the formation redshift, and is due to the rapid metal enrichment depicted in Figure \ref{evolution_of_seed_forming_gas}. The rapid metal enrichment can be quantitatively seen in the average gas metallicity evolution~(4th row of Figure \ref{evolution_of_seed_forming_gas_all}). We can see that even prior to the DGB formation, the average gas metallicities already start to increase from the pre-enrichment values~($\sim10^{-8}~Z_{\odot}$), to $\sim10^{-3}~Z_{\odot}$ at the time of formation. Therefore, even at the time of formation, the average metallicities of halos are already greater than the maximum seeding threshold of $10^{-4}~Z_{\odot}$; however, there are still pockets of star forming gas with metallicities $\leq10^{-4}~Z_{\odot}$, wherein DGBs form.

In Figure \ref{unbiased_halo_at_assembly_fig}, we select halos that form DGBs at $z=15$ %based on 
using gas based seeding parameters $\msfmp=1000~\&~ \mh=3000$, and we show their evolution~(orange circles) to $z=14,13~\&~12$ on the SFR versus halo mass plane~(upper panels) and the gas metallicity versus halo mass plane~(lower panels). We compare them to the full population of halos at their respective redshifts~(blue points). We investigate how biased %are 
these DGB forming halos are compared to typical halos of similar masses. On the SFR versus halo mass plane, the DGB forming halos have similar SFRs compared to halos of similar masses; not surprisingly, this continues to be so as they evolve to lower redshifts. On the metallicity versus halo mass plane, we find that DGB forming halos have significantly lower metallicities compared to halos of similar masses. This is a natural consequence of the requirement that the DGB forming halos have sufficient amounts of metal poor gas. However, due to the rapid metal enrichment of these halos seen in Figures \ref{evolution_of_seed_forming_gas} and \ref{evolution_of_seed_forming_gas_all}, their descendants at $z=14,13~\&~12$ end up having metallicities similar to halos of comparable mass. 

The picture that emerges from Figures \ref{evolution_of_seed_forming_gas} - \ref{unbiased_halo_at_assembly_fig} is one in which DGB-forming halos are generally {\em not} a special subset of halos (in terms of properties that persist to lower redshift), but rather they are fairly typical halos that have  the right conditions for DGB formation at a special moment in {\em time}. In other words, despite our seeding criterion favoring low-metallicity, star-forming halos, their descendants still end up with similar SFRs and metallicities compared to the general population of similar-mass halos. While Figure \ref{unbiased_halo_at_assembly_fig} only shows the evolution of DGB-forming halos at $z=15$, this general conclusion holds true for DGB-forming halos at all redshifts. A key consequence is that the descendants of seed forming halos can be well characterized by their halo mass distributions, largely because they are in this transient phase %in the process 
of rapid metal enrichment at the time of seed formation. 

We utilize this characteristic of our gas based seeding models to develop the new sub-grid seeding model for lower-resolution simulations in Section \ref{subhalo based stochastic seed model}. Rather than requiring information about detailed properties of the descendant galaxies of these gas based seeding sites, we show in Section \ref{Building the galaxy mass criterion} that most galaxy properties are well reproduced by simply matching the galaxy mass distribution. We then show in Section \ref{ssec:subhalo_env_criterion} that by additionally imposing a criterion on galaxy environment, we can robustly capture the evolved descendants of seeding sites from our high-resolution simulations.

\subsection{DGB formation and subsequent growth}
\label{Seed formation and subsequent growth}
We have thus far talked about the DGB forming halos and their evolution. In this subsection, we will focus on the formation of the DGBs themselves, and their subsequent growth to assemble higher mass BHs.
\subsubsection{Drivers of DGB formation: Halo growth, star formation and metal enrichment}

\begin{figure*}
\includegraphics[width=\textwidth]{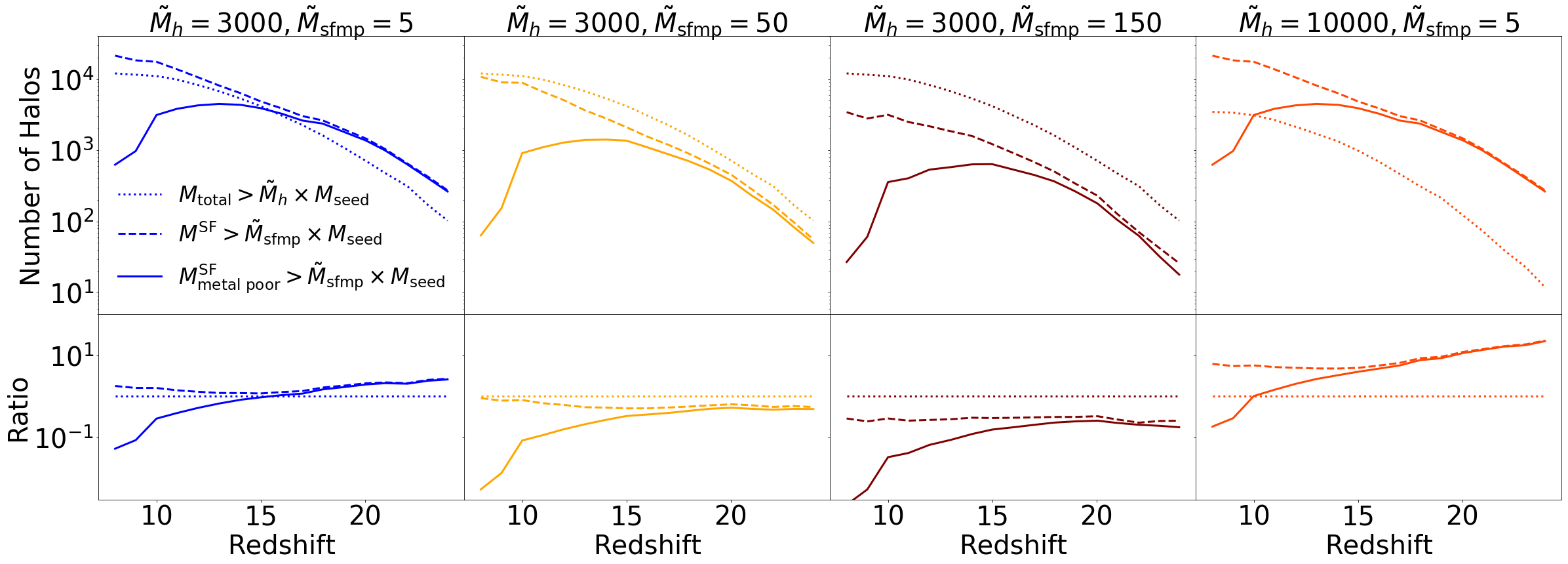}
\caption{The upper panels show the number of halos satisfying different cuts that were used in our gas based seed models: dotted lines correspond to a total mass cut of $\tilde{M}_h\times \seedmass$, dashed lines correspond to a star forming gas mass cut of $\msfmp\times \seedmass$, and solid lines show a star forming \& metal poor gas mass cut of $\msfmp\times \seedmass$. The lower panels show ratio of the normalizations w.r.t. the dotted lines from the top panel. The line with the smallest normalization determines which of the processes between halo growth versus star formation versus metal enrichment is the key driver for DGB formation at a given epoch. %For example, if the dotted lines are significantly below the solid and dashed lines, then halo growth primarily drives the seed formation. If the solid and dashed lines are similar, and they are significantly below the dotted lines, the seed formation is primarily driven by onset of star formation in halos. Lastly, if the solid lines is substantially lower than both dashed and dotted lines, then metal enrichment is the key driver in suppressing seed formation. \laura{\bf[i think it would be clearer for the reader if instead of spelling all this out in theoretical terms, you pointed to a specific panel(s) when comparing the various lines]} 
For $\tilde{M}_h=3000$, we find that metal enrichment becomes the key driver for~(suppressing) DGB formation around $z\sim13$ for all $\msfmp$ values between $5-150$. However, when $\tilde{M}_h=10000$, halo growth continues to be the primary regulator for DGB formation until $z\sim10$, after which metal enrichment takes over.}
\label{no_of_halos}
\end{figure*}

Our gas based seeding criteria %identifies 
identify three main physical processes that govern DGB formation in our simulations, i.e. halo growth, star formation and metal enrichment. Halo growth and star formation %tends 
tend to promote DGB formation with time, whereas metal enrichment suppresses DGB formation with time. The overall rate  
of DGB formation at various redshifts is determined by the complex interplay between these three processes. We study this interplay in Figure \ref{no_of_halos}, wherein we show the number of halos satisfying three different criteria: $M_{\mathrm{total}}>\mh \times \seedmass$~(dotted line), $M^{\mathrm{SF}}>\msfmp \times \seedmass$~(dashed line) and $M^{\mathrm{SF}}_{\mathrm{metal \ poor}}>\msfmp \times \seedmass$~(solid line). $M_{\mathrm{total}}$, $M^{\mathrm{SF}}$ and $M^{\mathrm{SF}}_{\mathrm{metal \ poor}}$ correspond to the total halo mass, star forming gas mass, and star forming \& metal poor gas mass of halos respectively. Amongst the above three criteria, the one %which 
that is most restrictive essentially determines the driving physical process for DGB formation at a given redshift. For example, in the rightmost panel of Figure \ref{no_of_halos}, the dotted lines have the lowest normalization from $z\sim25-10$; this implies that halo growth is primary driver and leads to the production of more DGBs with time. In the 3rd panel from the left, the solid and dashed lines have similar normalization, and both of them are lower than the dotted lines at the highest redshifts; this indicates that star formation is the key driver, which also enhances DGB formation with time. Lastly, in all of the panels, the solid lines have substantially lower normalization than both dashed and dotted lines at the lowest redshifts. In this case, metal enrichment is the primary driver, which leads to slow down and eventual suppression of DGB formation with time. %\laura{\bf[same comment here as in fig caption; also watch out for too much redundancy b/t caption and text - as much as i'm a fan of giving a 'takeaway' in the caption it should be brief \& expanded upon in the text]}

Comparing the different columns in Figure \ref{no_of_halos}, we note that the gas based seeding parameters~($\mh$ and $\msfmp$) have a strong influence in determining which process dominantly drives DGB formation at various redshifts. For $\mh=3000$ and $\msfmp=5$~(leftmost panel), halo growth is the key driver for DGB formation from $z\sim30-15$; at $z\lesssim15$, metal enrichment becomes the primary driver and slows down DGB formation. When $\mh$ is fixed at 3000 and $\msfmp$ is increased to 50 or 150~(2nd and 3rd panels respectively), star formation replaces halo growth to become the primary driver for DGB formation at $z\sim30-15$; however, metal enrichment continues to be the main driver in slowing down DGB formation at $z\lesssim15$. Finally, when $\msfmp$ is fixed at 5 and $\mh$ is increased to 10000~(rightmost panels), halo growth becomes the key driver for DGB formation from $z\sim30-10$. In this case, metal enrichment takes the driving seat at a lower redshift of $z\sim10$ compared to the cases when $\mh=3000$. 

To further summarize the above findings from Figure \ref{no_of_halos}, we find that when $\mh$ is $3000$, DGB formation is ramped up by either star formation or halo growth %till 
until $z\sim15$. After $z\sim15$, it is slowed down by metal enrichment. But when $\mh=10000$, the halo mass criterion becomes much more restrictive and halo growth continues to ramp up DGB formation until $z\sim10$ before it is slowed down by metal enrichment. In the next subsection, we shall see the implications of the foregoing on the rates of DGB formation at various redshifts.

\subsubsection{Formation rates of $\sim10^3~M_{\odot}$ DGBs}

\begin{figure*}
\includegraphics[width=16 cm]{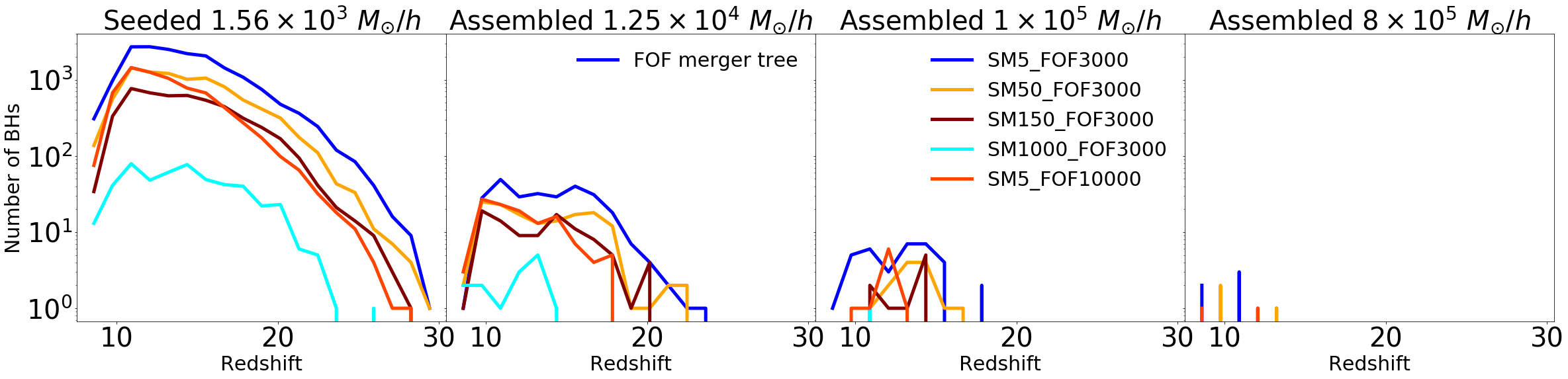}
\caption{We trace the growth of $1.56\times10^3~M_{\odot}/h$ DGBs~(leftmost panels) along merger trees and show the redshifts when they assemble BHs of masses $1.25\times10^4~M_{\odot}/h$, $1\times10^5~M_{\odot}/h$ and $8\times10^5~M_{\odot}/h$~(2nd, 3rd and 4th panels from the left). Different colors correspond to the different gas based seed models with varying $\msfmp=5,50,150~\&~1000, \tilde{M}_{\mathrm{h}}=3000$ and $\msfmp=5,\tilde{M}_{\mathrm{h}}=10000$. We find that the impacts of increasing $\msfmp$ and $\tilde{M}_{\mathrm{h}}$ are qualitatively distinguishible. For $\mh=3000$ and $\msfmp=5-1000$, metal enrichment starts to slow down DGB formation around $z\sim15$. In contrast, when $\tilde{M}_{\mathrm{h}}$ is increased from 3000 to 10000, the slow down of DGB formation due to metal enrichment starts much later~($z\lesssim10$). Similar trends are seen in the assembly rates of higher mass descendants~(particularly $1.25\times10^4~M_{\odot}/h$ BHs).}
\label{seed_formation_bh_assembly}

%\caption{We trace the growth of $1.56\times10^3~M_{\odot}/h$ seeds~(leftmost panels) along \texttt{SUBLINK} merger trees and determine the halo properties when these seeds assemble BHs of masses $1.25\times10^4~M_{\odot}/h$, $1\times10^5~M_{\odot}/h$ and $8\times10^5~M_{\odot}/h$~(2nd, 3rd and 4th panels from the left). In addition to halos which are identified as friends-of-friends~(FOFs) groups of dark matter particles~(solid lines), we also trace their growth within halo substructures identified as ``best friends-of-friends~(bFOFs)'' that have a linking length reduced by $1/3$ compared to the default FOF finder~(dashed lines). The 1st to 4th rows show the total mass, stellar mass, SFR and gas metallicities respectively for the FOFs and bFOFs. The lowermost row shows the redshift at which the BHs of different masses formed or assembled. Different colors correspond to the different gas based seed models with varying $\msfmp=5,50,150~\&~1000$ and $\tilde{M}_{\mathrm{h}}=3000$. The lowermost panels additionally also have $\msfmp=5,\tilde{M}_{\mathrm{h}}=10000$, which we don't show in the first four rows simply to avoid overcrowding of lines. We generally find that BHs of a fixed mass assemble in environments corresponding to a broad range of FOF and bFOF properties. We further find that as the seed models are made more restrictive by increasing $\msfmp$ or $\tilde{M}_{\mathrm{h}}$, BHs of a fixed mass assemble in FOFs and bFOFs of increasing total mass, stellar mass, SFRs as well as metallicities.  }
\end{figure*}
The leftmost panel of Figure \ref{seed_formation_bh_assembly} shows the formation rates of  $1.56\times10^3~M_{\odot}/h$ DGBs for the different gas based seed models. The interplay between halo growth, star formation and metal enrichment discussed in the previous subsection is readily seen in the DGB formation rates. For $\mh=3000$ and $\msfmp=5,50,150~\&~1000$, % ~(blue, orange, maroon and cyan lines), %laura: colors can go in the caption; ppl wont retain all those by the time theyve scrolled to the fig
we find that DGB formation ramps up as the redshift decreases from $z\sim30-15$, driven predominantly either by halo growth~(for $\msfmp=5$) or star formation~(for $\msfmp=50,150~\&~1000$). As the redshift decreases below $z\sim15$, metal enrichment significantly slows down DGB formation. However, when $\mh$ is increased to $10000$~(red line), halo growth continues to ramp up DGB formation till $z\sim10$, after which the suppression of DGB formation due to metal enrichment takes place. Note also that at $z\lesssim10$, DGB formation is finally strongly suppressed due to metal pollution for all the seed models. This is because most of the newly star forming regions are already metal enriched by then, likely due to stellar feedback dispersing the metals throughout the simulation volume.

\subsubsection{Assembly rates of $\sim10^4-10^6~M_{\odot}$ BHs from $\sim10^3~M_{\odot}$ seeds}
The assembly rates of $1.25\times10^4,1\times10^5~\&~8\times10^5~M_{\odot}/h$ BHs are shown in 2nd, 3rd and 4th panels of Figure \ref{seed_formation_bh_assembly} respectively. %Recall from 
As in \cite{2021MNRAS.507.2012B}, we find that nearly $100\%$ of the growth of these DGBs is happening via mergers. This is partly due to the $M_{\rm BH}^2$ scaling of Bondi Hoyle accretion rates, which leads to much slower accretion %driven growth of 
onto low mass DGBs, and it is consistent with the findings of \cite{2014MNRAS.442.2751T}~(see Figure 2 in their paper).  

Let us first focus on the impact of this merger dominated growth on the assembly of $1.25\times10^4~M_{\odot}/h$ BHs~(2nd panel of Figure \ref{seed_formation_bh_assembly}). They generally assemble at rates $\sim50-80$ times lower than the rates at which $1.56\times10^3~M_{\odot}/h$ DGBs form. Notably, the trends seen in the DGB formation rates directly reflect upon the rates at which $1.25\times10^4~M_{\odot}/h$ BHs assemble. In particular, for $\mh=3000$ and $\msfmp=5,50~\&~150$, we see an increase in the assembly rates as the redshift decreases from $z\sim25-15$ wherein DGB formation is driven by halo growth or star formation. The assembly rates slow down at $z\lesssim15$ as metal enrichment slows down DGB formation. For a higher value of $\mh=10000$, halo growth continues to increase the assembly rates until $z\sim10$, before metal enrichment slows it down. Overall, these results suggest that the interplay of halo growth, star formation and metal enrichment processes that we witnessed on the formation rates of $1.56\times10^3~M_{\odot}/h$ DGBs, are also retained in the assembly rates of their higher mass $1.25\times10^4~M_{\odot}/h$ descendants. %Now we focus on the impact of $\mh$ and $\msfmp$ on the assembly rates of $1.25\times10^4~M_{\odot}/h$ BHs. Increasing $\msfmp$ at fixed $\mh=3000$ leads to a broadly uniform suppression in the assembly rates from $z\sim25-10$. The suppression is by factors of $\sim5$ for $\msfmp$ increasing from 5 to 150. In contrast, increasing $\mh$ at fixed $\msfmp=5$ leads to a somewhat stronger suppression in the assembly rates at $z\gtrsim15$~(by factors of $\sim7$), compared to $z\sim10$~(by factors of $\lesssim2$).  Notably, these trends are similar to that of the seed formation rates discussed in the previous subsection. \laura{\bf[same comment as on prev subsection about these trends]}

We also see the assembly of a handful of  $1\times10^5$ and $8\times10^5~M_{\odot}/h$ BHs~(3rd and 4th panels of Figure \ref{seed_formation_bh_assembly}). $1\times10^5~M_{\odot}/h$ BHs generally start assembling at $z\lesssim15$ and $8\times10^5~M_{\odot}/h$ BHs assemble at $z\lesssim12$. However, any potential trends similar to that identified in the previous paragraph for $1.25\times10^4~M_{\odot}/h$ descendants, are difficult to discern for the $1\times10^5$ and $8\times10^5~M_{\odot}/h$ descendants due to very limited statistical power.

\subsection{In which host halos do the $\sim10^4-10^6~M_{\odot}$ descendant BHs assemble?}
\label{where descendants assemble}
\begin{figure*}
\includegraphics[width=5.8 cm]{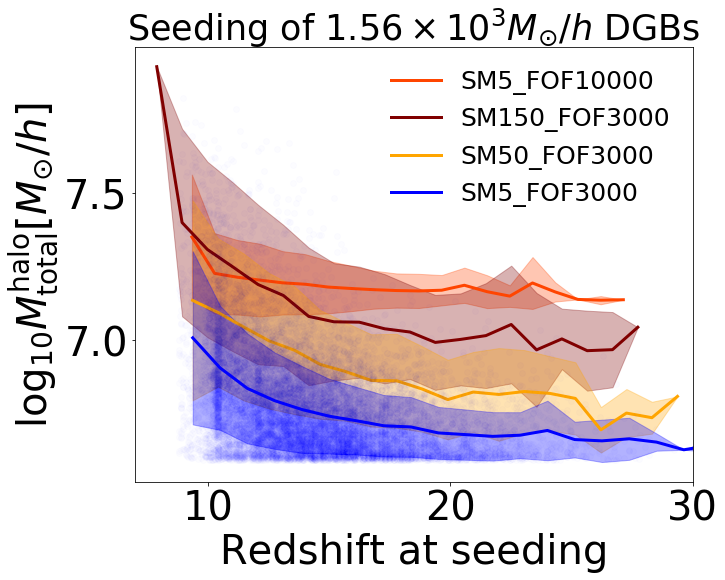}
\includegraphics[width=5.8 cm]{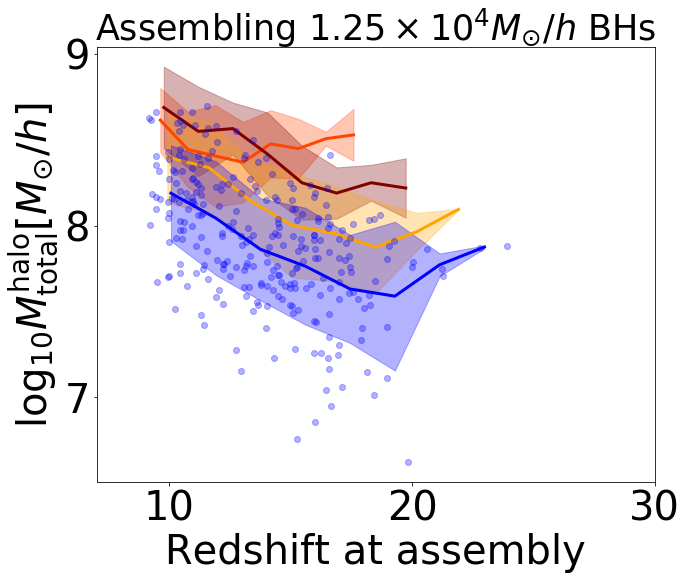}
\includegraphics[width=5.8
 cm]{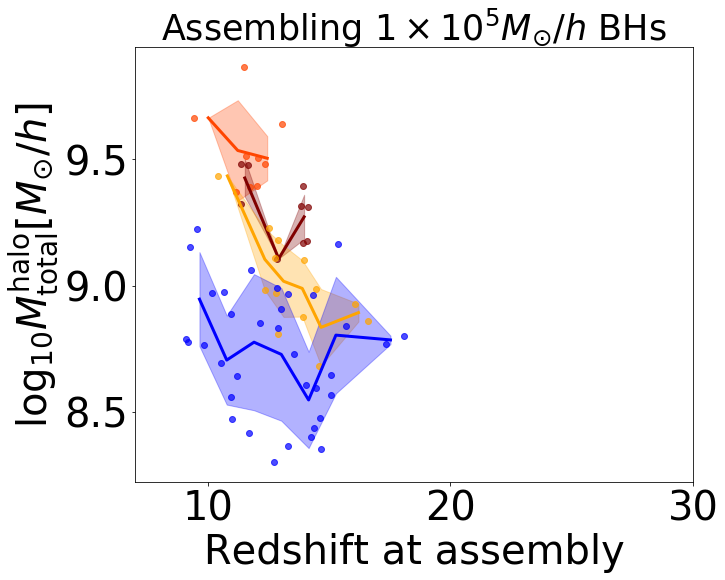}
\caption{The left panel shows the redshifts and the FOF total masses at which $1.56\times10^3~M_{\odot}/h$ DGBs form. Middle and right panels show the redshifts and the FOF total masses at which $1.25\times10^4~M_{\odot}/h$ and $1\times10^5~M_{\odot}/h$ descendant BHs respectively assemble on the FOF merger tree. The different colors correspond to different gas based seed models. Each data point corresponds to a single instance of assembly or seeding. We only show data points for a limited set of models to avoid overcrowding. Solid lines show the mean trend and the shaded regions show $\pm1\sigma$ standard deviations. We find that as metal enrichment takes over as the driving force and suppresses DGB formation at lower redshifts, DGBs form in increasingly massive halos. This also drives a similar redshift dependence for the assembly of $1.25\times10^4~M_{\odot}/h$ BHs.} %Amongst all the models, $\tilde{M}_h=10000,\msfmp=5$ model has a distinct trend with redshift. This is due to the more stringent halo mass criterion, which delays the onset of the influence of metal enrichment on seeding to later times~($z\sim10$) compared to the other models ($z\sim15$).}
\label{host_halos_at_assembly}
\end{figure*}

Figure \ref{host_halos_at_assembly} shows the host halo masses~(denoted by $\massemblyFOF$) and redshifts at which $1.56\times10^{3}~M_{\odot}/h$ DGBs form~(leftmost panel), followed by the assembly of $1.25\times10^{4}~M_{\odot}/h$ and $1\times10^{5}~M_{\odot}/h$ BHs~(middle and right panels respectively). Broadly speaking, $1.56\times10^3~M_{\odot}/h$ DGBs form in $\sim10^{6.5}-10^{7.5}~M_{\odot}/h$ halos, $1.25\times10^4~M_{\odot}/h$ BHs assemble in $\sim10^{7.5}-10^{8.5}~M_{\odot}/h$ haloes, and $1\times10^5~M_{\odot}/h$ BHs assemble in $\sim10^{8.5}-10^{9.5}~M_{\odot}/h$ haloes. Therefore, rates of BH growth versus halo growth are broadly similar. This is a natural expectation from merger-dominated BH growth, since the BH mergers crucially depend on the merging of their host halos. Note however that in the absence of our currently imposed BH repositioning scheme that promptly merges close enough BH pairs, we could expect larger differences between the merger rates of BHs and their host halos. 

The interplay between halo growth, star formation and metal enrichment at different redshifts~(as noted in Section \ref{Seed formation and subsequent growth}) profoundly influences the redshift evolution of the halo masses in which the seeding of $1.56\times10^3~M_{\odot}/h$ DGBs and assembly of higher-mass %$1.25\times10^{4}~\&~1\times10^5~M_{\odot}/h$ 
BHs take place. Let us first focus on the seeding of $1.56\times10^3~M_{\odot}/h$ DGBs~(Figure \ref{host_halos_at_assembly}: left panel). 

We find for $\mh=3000~\&~\msfmp=50,150$ that the halo masses steadily increase with time as star formation drives the formation of DGBs. As described in more detail in Appendix \ref{Evolution of star forming metal poor gas in halos}, this is a simple consequence of cosmological expansion, which makes it more difficult for the gas to cool and form stars at later times within halos of a fixed mass. Notably, as metal enrichment gradually takes over at $z\lesssim15$, the redshift evolution becomes substantially steeper, pushing DGB formation towards even more massive halos at later times. This may seem counterintuitive since we expect more massive halos to have stronger metal enrichment, which should suppress DGB formation within them. However, more massive halos also generally have higher overall star forming gas mass, a portion of which may remain metal poor since star-forming halos are not fully metal enriched instantaneously. As it turns out in our simulations, when metal enrichment increases, it favors DGB formation in more massive halos because they are more likely to have sufficient amount of star forming \& metal poor gas mass. For further details on this, the reader can refer to Appendix \ref{Evolution of star forming metal poor gas in halos}. When $\mh$ is increased to $10000$, the redshift evolution of DGB forming halo mass is flat until $z\sim10$ since the seed formation is primarily driven by the \textit{halo mass criterion}. It is only after $z\sim10$ that the DGB forming halo mass starts to steeply increase due to the full influence of metal enrichment. 

The above trends directly impact the redshift evolution of the host halo masses in which $1.25\times10^4~M_{\odot}/h$ assemble~(middle panel of Figure \ref{host_halos_at_assembly}). For the model with a stricter halo mass criterion (i.e., $\mh=10000~\&~\msfmp=5$), the transition in the slope of the $\massemblyFOF$ versus redshift relation occurs much later~(transition occurs between $z\sim12-10$) compared to models with more lenient halo mass criterion $\mh=3000~\&~\msfmp=5-150$~($z\gtrsim15$). This, again, is because metal enrichment starts to suppress DGB formation much later in the model with stricter halo mass criterion. %Notably, we also find that at $z\gtrsim12$, the halo masses tend to mildly increase with increasing redshift for $\mh=10000~\&~\msfmp=5$. We also see hints of a similar effect for $\mh=3000~\&~\msfmp=50~\&~150$ at the highest redshifts when $1.25\times10^4~M_{\odot}/h$ BHs first start assembling. This is likely a consequence of having very few BHs overall at these epochs; this is because halo growth may be slightly higher than BH growth due to relatively higher contribution of DM accretion to the halo growth between consecutive BH merger events~(recall again the BHs are predominantly growing via mergers in these models). \laura{\bf[i think the detailed discussion of trends in these plots could be trimmed down a bit, given the error bars and amount of overlap b/t models]}
Finally, for the assembly of $1\times10^5~M_{\odot}/h$ BHs, the redshift evolution of the host halo masses cannot be robustly deciphered due to statistical uncertainties. But here too, we see hints of higher host halo masses at lower redshifts in regimes where metal enrichment is the primary driver for~(the suppression of) DGB formation.

Overall, %we see that 
the impact of halo growth, star formation and metal enrichment on DGB formation is well imprinted in the redshift evolution of the host halo masses within which their %higher mass 
descendant BHs assemble. We shall see in later sections how this fact is going to be crucial in building the new seed model to represent (descendants of) $1.56\times10^3~M_{\odot}/h$ DGBs in lower-resolution simulations. % which cannot directly represent them.  

\section{RESULTS II: A new stochastic seed model for larger simulations}
\label{subhalo based stochastic seed model}

We have thus far traced %far studied 
the growth of low mass~($1.56 \times10^3~M_{\odot}/h$) DGBs born in regions with dense \& metal poor gas, in order to determine the host properties of their %. We particularly traced the regions within which these seeds assembled 
higher-mass ($1.25\times10^4~\&~1\times10^5~M_{\odot}/h$) descendant BHs. We will now %focus on using 
use these results to build a new stochastic seed model that can represent these $1.56 \times10^3~M_{\odot}/h$ DGBs within simulations that cannot directly resolve them. In section \ref{Subhalo-based stochastic seeding}, we gave a brief introduction of this seed model and mentioned that this model would rely on a %they'd comprise of a 
\textit{galaxy mass criterion} and a \textit{galaxy environment criterion}. %In this section, we will describe further details underlying 
Here we detail the motivation, construction, and calibration of both of these seeding criteria and demonstrate that the resulting model %. We shall demonstrate that with a calibrated combination of these criteria, the subhalo-based stochastic seed models 
can reproduce reasonably well the high-resolution, gas based seed model predictions in lower-resolution simulations. 

Note that some of our gas based seed parameter  
combinations do not produce enough descendant BHs in our zoom region to perform a robust calibration. These include $\mh=3000; \msfmp=1000$ for the $1.25\times10^4~M_{\odot}/h$ descendants and  $\mh=3000~\&~10000; \msfmp=150~\&~1000$ for the $1\times10^5~M_{\odot}/h$ descendants. Therefore, we shall not consider these parameter values hereafter.

%\laura{\bf[I think we need a shorthand for the two different kinds of seed models too. maybe direct gas based (DGB) and extrapolated seed descendant (ESD) or something like that? i also dont know about calling the latter 'subhalo-based', since bFOFs are not the same as subhalos and regardless this is arguably not the most descriptive term. maybe refer to bFOF galaxies instead of bFOF subhalos throughout]}
In the stochastic seed model, we will directly seed the descendants with initial masses set by the gas mass resolution~($1.25\times10^4~\&~1\times10^5~M_{\odot}/h$ in $L_{\mathrm{max}}=11~\&~10$ respectively). As already mentioned in Section \ref{Subhalo-based stochastic seeding}, because these massive seeds are meant to represent descendants of $1.56 \times10^3~M_{\odot}/h$ DGBs that cannot be resolved directly, we refer to the former as ``extrapolated seed descendants" or ESDs with initial mass denoted by $\descendantseedmass$. %However, we need to keep in mind that these models are still representing $1.56 \times10^3~M_{\odot}/h$ seeds that have presumably grown up to the gas mass resolution limit before being seeded in the lower-resolution simulations. To ensure a very clear distinction between the actual seed mass versus the mass at which the BHs are initialized in these lower resolution simulations, we shall refer to the latter as \textit{ESDs} with mass denoted by $\descendantseedmass$. 
In other words, our new stochastic seeding prescription will place ESDs with $\descendantseedmass$ set by the gas mass resolution of $1.25\times10^4$ or $1\times10^5~M_{\odot}/h$, but they are intended to represent our gas based seed models with unresolvable 
 $1.56\times10^3~M_{\odot}/h$ DGBs. To that end, the next few subsections address the following question: \textit{How do we build a new seed model that can capture the unresolved growth phase from $\seedmass=1.56\times10^3~M_{\odot}/h$ to $\descendantseedmass=1.25\times10^4$ or $1\times10^5~M_{\odot}/h$?}

\subsection{Seeding sites for ESDs: ``Best Friends of Friends~(bFOF)" galaxies}
\label{bFOF_details_sec}
\begin{figure*}
\includegraphics[width=5.5 cm]{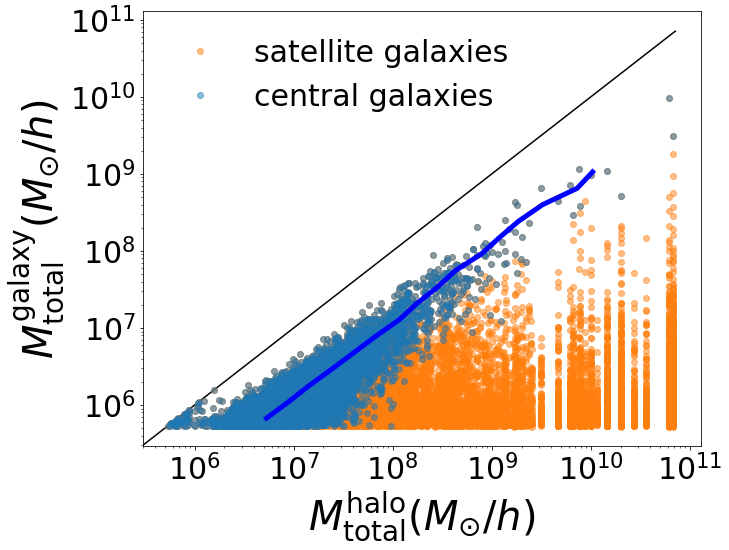}
\includegraphics[width=5.5 cm]{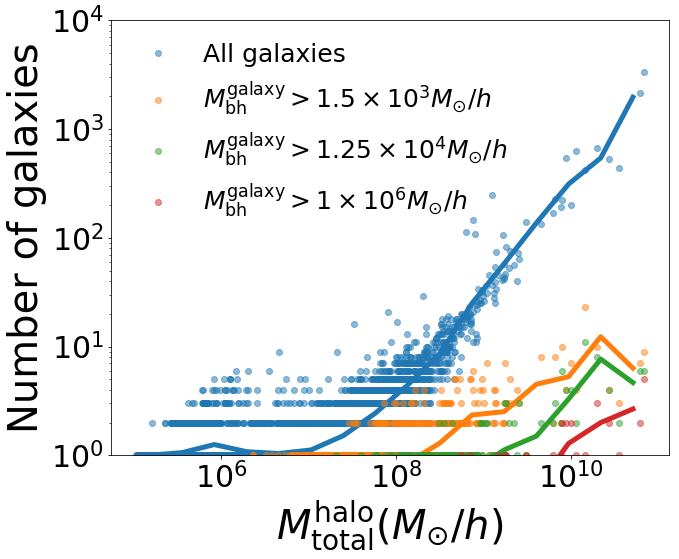}
\includegraphics[width=5.5 cm]{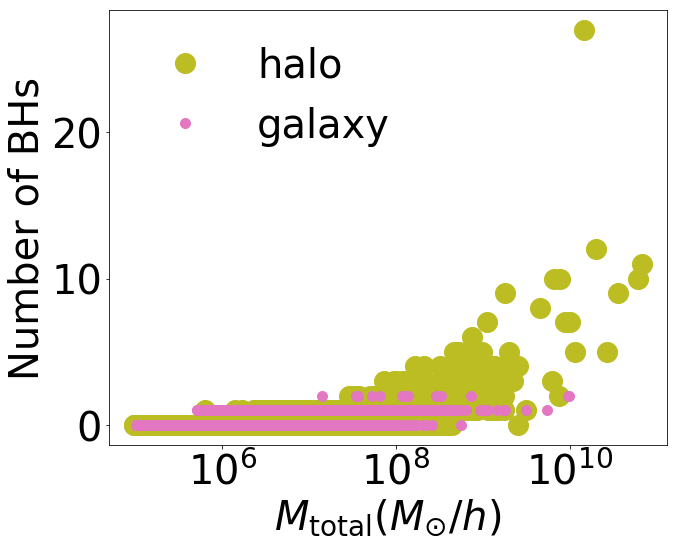}
\caption{Introduction to best friends of friends (bFOF) galaxies, which are identified using the FOF algorithm but with one-third of the linking length used for identifying halos: Left panel shows the relation between halo mass and the mass~($M^{\mathrm{galaxy}}_{\mathrm{total}}$) of the central or most massive bFOF in blue, and satellite bFOF in orange. On an average, the central bFOFs are $\sim7$ times less massive than their host FOFs, but with substantial scatter~($\gtrsim1$ dex) for fixed FOF mass~($M^{\mathrm{halo}}_{\mathrm{total}}$). The middle panel shows the number of bFOFs for FOFs of different total masses. The plots are shown at $z=8$ and for the gas based seed model 
$[\mh,\msfmp=3000,5]$. Blue color shows all bFOFs~(with or without BHs); orange, green and maroon lines show bFOFs with a total BH mass of $1.5\times10^3~M_{\odot}/h$, $1.25\times10^4~M_{\odot}/h$ and $1\times10^5~M_{\odot}/h$ respectively. Right panel shows the number of BHs occupied by FOFs and bFOFs. While $\gtrsim12\%$ of FOFs %occupy 
contain multiple BHs~(up to $\sim30$), only $\sim1\%$ of bFOFs %occupy 
contain multiple BHs. All this motivates us to use bFOFs as seeding sites (instead of FOFs) in our new stochastic seed models that would be able to represent the lowest mass~($\sim10^3~M_{\odot}/h$) DGBs in lower resolution simulations that cannot directly resolve them. These bFOFs are essentially sites of (proto)galaxies residing within the high-z halos. We hereafter refer to these bFOFs as ``galaxies".}
\label{bFOF_properties}
\end{figure*}

\begin{figure*}
\includegraphics[width=16 cm]{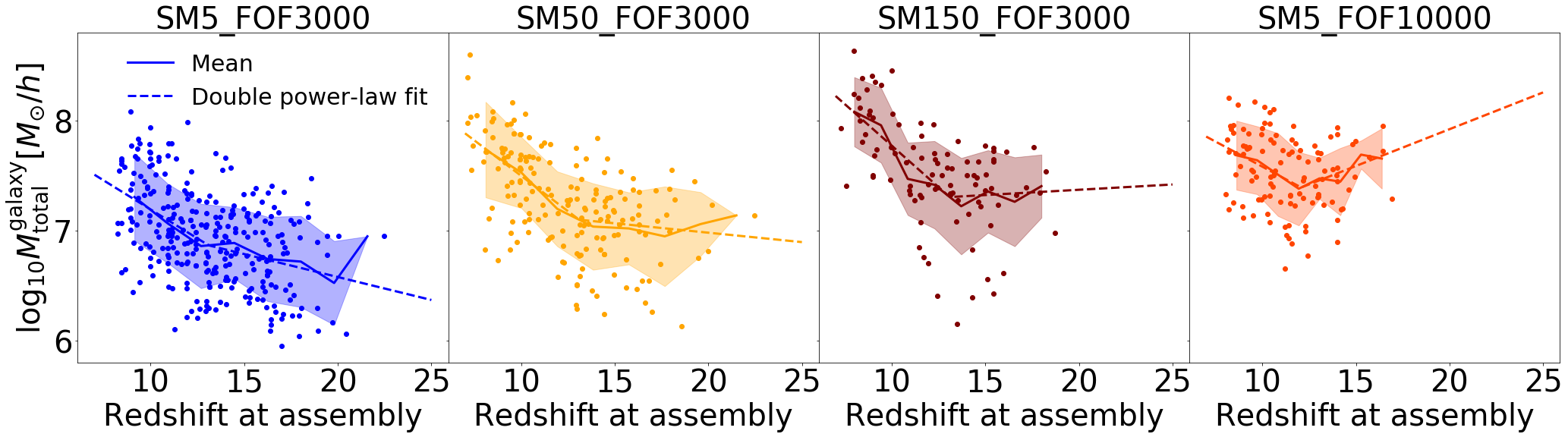}
\includegraphics[width=16 cm]{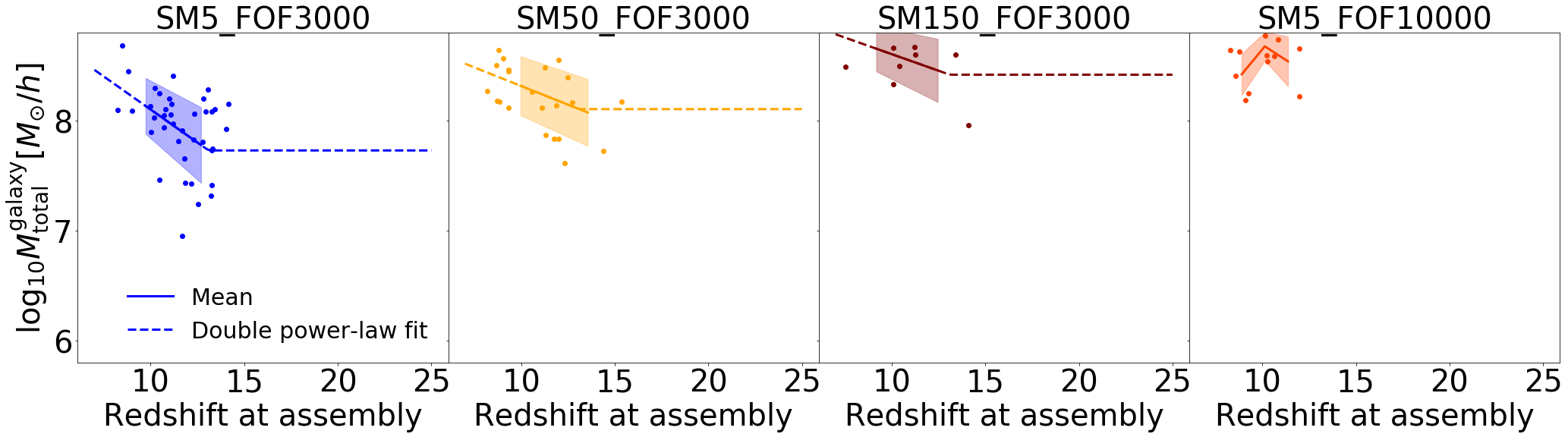}
\caption{Top and bottom rows show the redshifts and the galaxy total masses~($\massembly$ that includes DM, gas and stars) at which $1.25\times10^4~M_{\odot}/h$ and $1\times10^5~M_{\odot}/h$ BHs respectively assemble from $1.56\times10^3~M_{\odot}/h$ DGBs when the BH growth is traced along the galaxy merger tree. The 1st, 2nd and 3rd columns show different gas based seeding models with $\tilde{M}_h=3000$ and $\msfmp=5,50~\&~150$. The 4th column shows $\tilde{M}_h=10000$ and $\msfmp=5$. Solid lines show the mean trend and the shaded regions show $\pm1\sigma$ standard deviations. We find that for all the models, there is a transition in the slope of the mean trend at redshift $z \equiv z_{\mathrm{trans}}\sim12-13$, which is driven by the suppression of seed formation by metal enrichment.  
The trends are reasonably well fit by a double power law~(dashed lines). These fits are used in our stochastic seed models that directly seed the descendants~(referred to as ``extrapolated seed descendants or ESDs) at $1.25\times10^4~M_{\odot}/h$ or $1\times10^5~M_{\odot}/h$ %masses (referred to as ESDs) 
within the lower resolution $L_{\mathrm{max}}=11~\&~10$ zooms, respectively. To obtain fits in the top row, we first assumed $z_{\mathrm{trans}}=13.1$ for $\tilde{M}_h=3000,\msfmp=5,50~\&~150$, and $z_{\mathrm{trans}}=12.1$ for $\tilde{M}_h=10000,\msfmp=5$ via a visual inspection. The fits were then performed to obtain the slopes at $z<z_{\mathrm{trans}}$ and $z>z_{\mathrm{trans}}$ using \texttt{scipy.optimize.curve_fit}. The final fitted parameters are shown in Table \ref{double_power_law_table}.} 
\label{bFOF_at_assembly}

\end{figure*}

\begin{table*}
     \centering
    \begin{tabular}{c|c|c|c|c|c|c|c|c|c|}
         $\msfmp$ & $\tilde{M}_{h}$ & $z_{\mathrm{trans}}$ & $\log_{10}M_{\mathrm{trans}}[M_{\odot}/h]$ & $\alpha$  & $\beta$ & $\sigma$ & $p_0$ & $p_1$ & $\gamma$\\
         \hline
          & & & $\descendantseedmass=1.25\times10^4~M_{\odot}/h$ &   &  & & & &\\
         \hline
         5 & 3000 & 13.1 & 6.86  & -0.105 & -0.041 & 0.330 & NA & NA & NA\\
         50 & 3000 & 13.1 & 7.09 & -0.128  & -0.017 & 0.319 & 0.1 & 0.3 & 1.6 \\
         150 & 3000 & 13.1 & 7.30 & -0.151  & 0.009 & 0.360 & 0.1 & 0.3 & 1.6\\
         5 & 10000 & 12.1 & 7.39 & -0.091  & 0.067 & 0.278 & 0.2 & 0.4 & 1.2\\
         \hline
          & & & $\descendantseedmass=1\times10^5~M_{\odot}/h$ & & &\\
         \hline
         5 & 3000 & 13.1 & 7.72  & -0.120 & 0 & 0.246 & 0.2 & 0.4 & 1.2\\
         50 & 3000 & 13.1 & 8.10 & -0.067  & 0 & 0.286 & 0.2 & 0.4 & 1.2 \\
         150 & 3000 & 13.1 & 8.41 & -0.060 & 0 & 0.298 & 0.2 & 0.4 & 1.2 \\      
         \hline
    \end{tabular}
    \caption{Fiducial model parameters for the stochastic seed model, calibrated for each of the gas based seeding parameters. Columns 1 and 2 show the gas based seeding parameters $\mh$ and $\msfmp$. For each set of $\mh$ and $\msfmp$ values, the remaining columns list the parameters of the stochastic seed model. Columns 3 to 7 show the parameter values used for the \textit{galaxy mass criterion}, which are derived from gas based seed model predictions of the $\massembly$ versus redshift relations~(Figure \ref{bFOF_at_assembly}). $z_{\mathrm{trans}}$, $M_{\mathrm{trans}}$, $\alpha$, \& $\beta$ are obtained by fitting the mean trends using the double power-law function shown in Equation \ref{double_powerlaw_eqn}. $\sigma$ is the standard deviation. Columns 8 to 10 show the parameter values for the \textit{galaxy environment criterion} (i.e., $p_0$, $p_1$ and $\gamma$). These are obtained by exploring a range of possible values to %obtain 
    find the best match with the small-scale BH clustering and overall BH counts predicted by the gas based seed model.}
    \label{double_power_law_table}
\end{table*}

It is common practice in many~(but not all) cosmological simulations to place one seed per halo at a given time step. The advantage to this is that the halo properties~(particularly the total halo mass) show much better resolution convergence compared to the local gas properties. However, this is not quite realistic, as halos typically have a significant amount of substructure and can therefore have multiple seeding sites at a given time. Despite this, subhalos are not typically used to seed BHs, likely because on-the-fly subhalo finders like \texttt{SUBFIND} are much more computationally expensive compared to on-the-fly halo finders like the FOF finder. 

Recall that in our gas based seed model, $1.56\times10^3~M_{\odot}/h$ DGBs were also seeded as ``one seed per halo". But even in this case, as these smaller seed-forming halos and their BHs undergo mergers, %merge to assemble higher-mass BHs and halos respectively, 
configurations with multiple $1.25\times10^4$ or $1\times10^5~M_{\odot}/h$ BHs per halo tend to naturally emerge. We %need 
emulate this in our new seed model by %placing 
seeding ESDs within bFOFs introduced in Section \ref{Subhalo-based stochastic seeding}. The linking length for the bFOFs was chosen to be $1/3$rd of the value adopted for standard FOF halos~(which is $0.2$ times the mean particle separation). This value was chosen after exploring a number of possibilities. On one hand, a much larger linking length does not resolve the substructure adequately. On the other hand, if the linking length is much smaller, a significant number of FOFs end up not containing any bFOFs. 

Figure \ref{bFOF_properties} summarizes the bFOF properties in relation to the familiar FOF halos at $z=8$. The leftmost panel shows the relationship between the masses of FOFs and bFOFs. Within a FOF, the most massive bFOF is assigned as the ``central bFOF"~(blue circles) and the remaining bFOFs are assigned as the ``satellite bFOFs"~(orange circles). The central bFOFs are about $\sim7$ times less massive than the host FOF. Not surprisingly, the satellite bFOFs span a much wider range of masses all the way down to the lowest possible masses at the bFOF/FOF identification limit~($\geq32$ DM particles). The middle panel of Figure \ref{bFOF_properties} shows the bFOF occupation statistics for FOFs of different masses. More massive FOFs tend to host a higher number of bFOFs; the most massive $\sim3\times10^{10}~M_{\odot}/h$ FOF has about $\sim4\times10^3$ bFOFs. We can see that in addition to the central bFOF, the satellite bFOFs can also contain BHs~(orange, green and maroon points in the middle panel). To that end, the right panel of  Figure \ref{bFOF_properties} shows the total BH occupations inside FOFs and bFOFs as a function of their respective masses. We can clearly see that while individual FOFs can contain multiple BHs~(up to a few tens), the vast majority of individual bFOFs contain 0 or 1 BHs. In fact, amongst the $\sim30000$ bFOFs at $z=8$, only 12 of them have more than 1 BH. These results generally hold true at all redshifts.    

%redundant w prev paragraph:
%Overall, we find that almost all bFOFs in our simulations contain at most 1 BH. 
By building our seed model based on bFOFs instead of FOFs~(i.e. one ESD per bFOF), we expect to naturally place multiple $1.25\times10^4~M_{\odot}/h$ or $1\times10^5~M_{\odot}/h$ ESDs in individual halos. As a result, we will successfully capture situations where multiple $1.25\times10^4~M_{\odot}/h$ or $1\times10^5~M_{\odot}/h$ descendant BHs assemble from $1.56\times10^3~M_{\odot}/h$ DGBs in a single halo within close succession. As mentioned in Section \ref{Subhalo-based stochastic seeding}, these bFOFs are essentially the sites where high-z (proto)galaxies reside; we therefore use the phrase ``galaxies" to refer to these bFOFs. 

\subsection{Building the \textit{galaxy mass criterion}}
\label{Building the galaxy mass criterion}
\begin{figure*}
\includegraphics[width=16 cm]{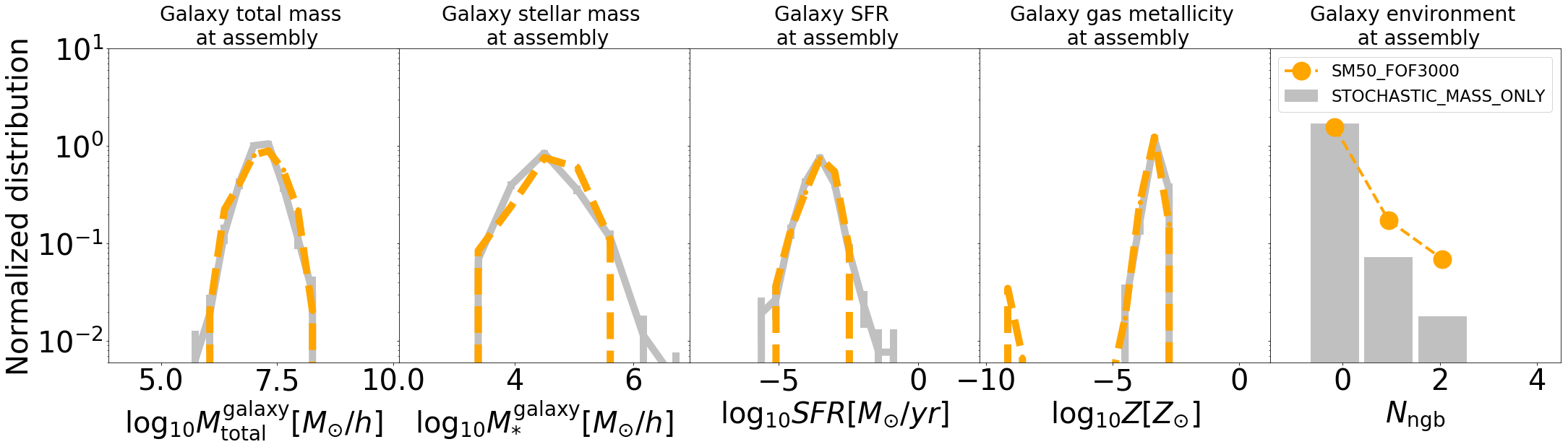}\\
\includegraphics[width=16 cm]{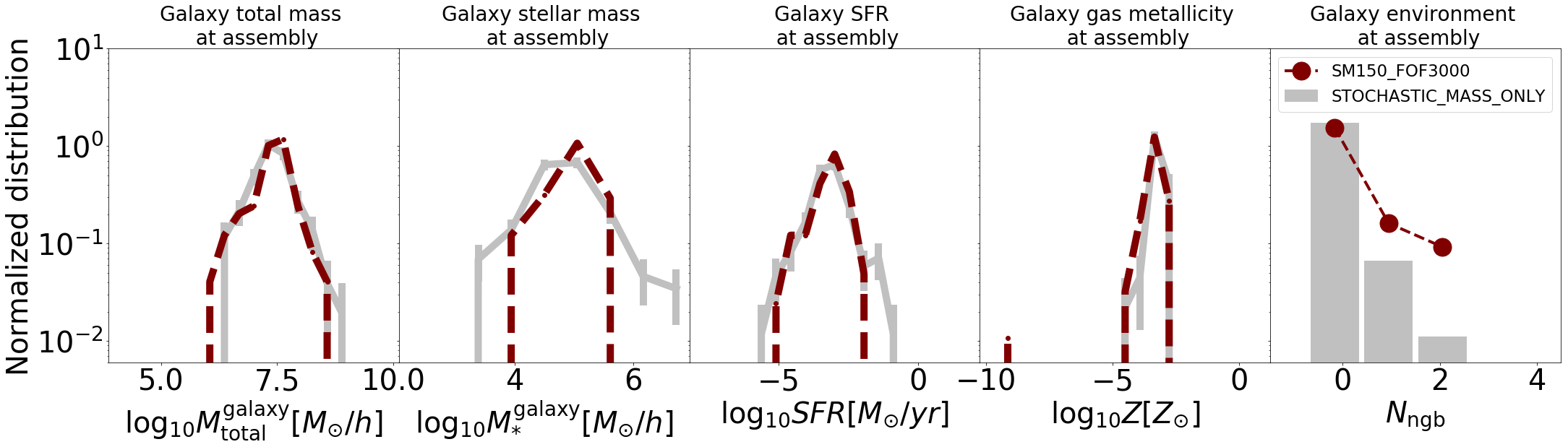}\\
\includegraphics[width=16 cm]{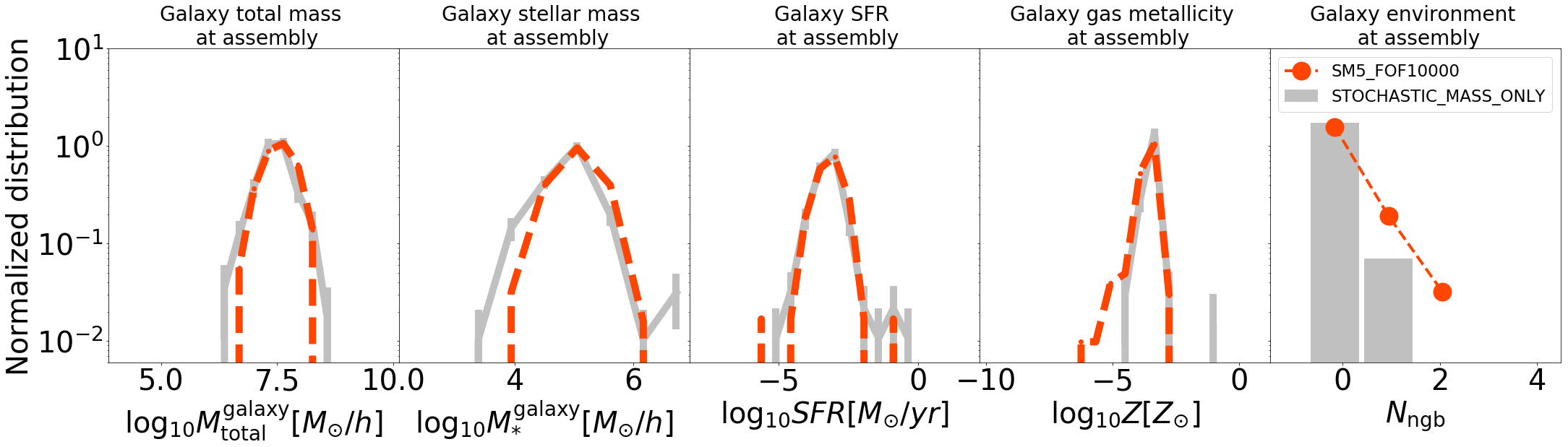}\\

\caption{Colored dashed lines show 1D distributions of galaxy properties in which $1.25\times10^4~M_{\odot}/h$ BHs assemble from $1.56\times10^3~M_{\odot}/h$ DGBs within %simulations using the gas based seed model~(
\texttt{GAS_BASED} simulations. From left to right, the panels in each row %The 1st to 5th panels from the left 
show the total galaxy masses ($M^{\mathrm{galaxy}}_{\rm total}$), stellar masses ($M^{\mathrm{galaxy}}_*$), SFRs, gas metallicities ($Z$), and environments~($N_{\mathrm{ngb}}$ i.e. the number of neighboring halos around the galaxy as defined in Section \ref{Subhalo-based stochastic seeding}). Top, middle and bottom rows correspond to different sets of gas based seed parameters: $[\mh, \msfmp=3000, 50$], $[\mh, \msfmp=3000, 150]$ and $[\mh, \msfmp=10000, 5]$ respectively. In each panel, the light grey lines show host properties for the $1.25\times10^4~M_{\odot}/h$ ESDs in the corresponding %We then compare the \texttt{GAS_BASED} simulations to 
\texttt{STOCHASTIC_MASS_ONLY} simulation. %s that solely use the \textit{subhalo mass criterion} to place ESDs of mass $1.25\times10^4~M_{\odot}/h$~(light grey lines). 
Note that unlike the rest of the paper, here the \texttt{STOCHASTIC_MASS_ONLY} simulations are %also 
run at the highest resolution of $L_{\mathrm{max}}=12$ for a fair comparison of their predicted galaxy baryonic properties with the \texttt{GAS_BASED} simulations run at the same resolution. The total galaxy masses of BH hosts in the %~(leftmost panels) must agree between \texttt{GAS_BASED} and 
\texttt{STOCHASTIC_MASS_ONLY} simulations %because the latter was 
are calibrated match the \texttt{GAS_BASED} simulations, but no other calibration is performed. The agreement of the distributions of baryonic properties (($M_*$, SFR, \& $Z$) between the two types of simulations results naturally from matching the $M^{\mathrm{galaxy}}_{\rm total}$ distribution. 
However, the \texttt{STOCHASTIC_MASS_ONLY} simulations do end up placing the ESDs in significantly less rich environments (smaller $N_{\mathrm{ngb}}$) compared to what is required by the \texttt{GAS_BASED} simulations.}
\label{STOCHASTIC_MASS_ONLY_models}
\end{figure*}

\begin{figure*}
\includegraphics[width=16 cm]{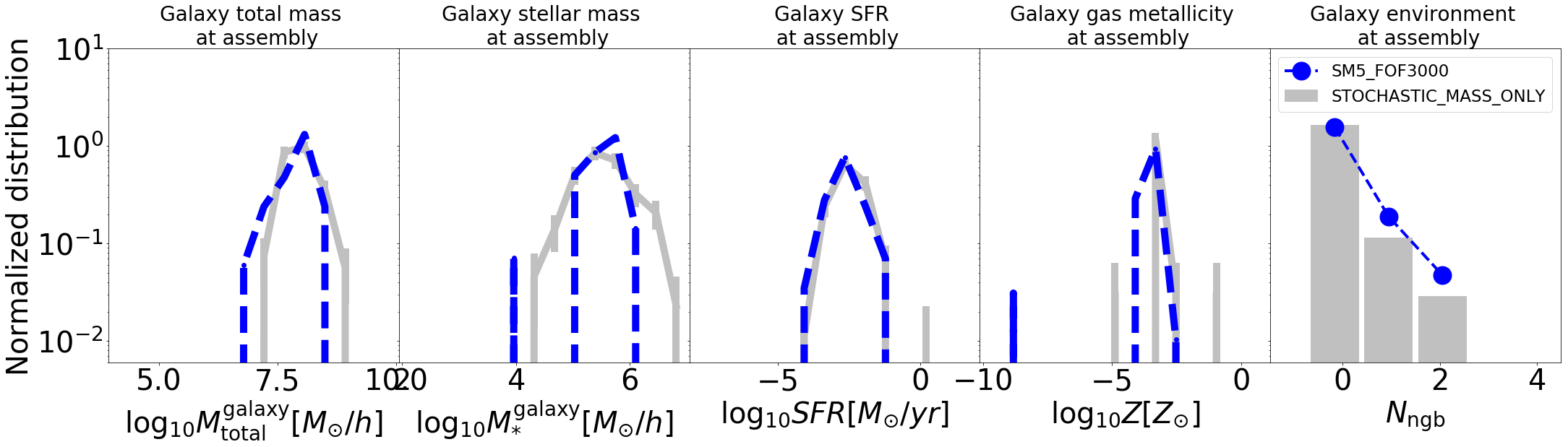}\\

\includegraphics[width=16 cm]{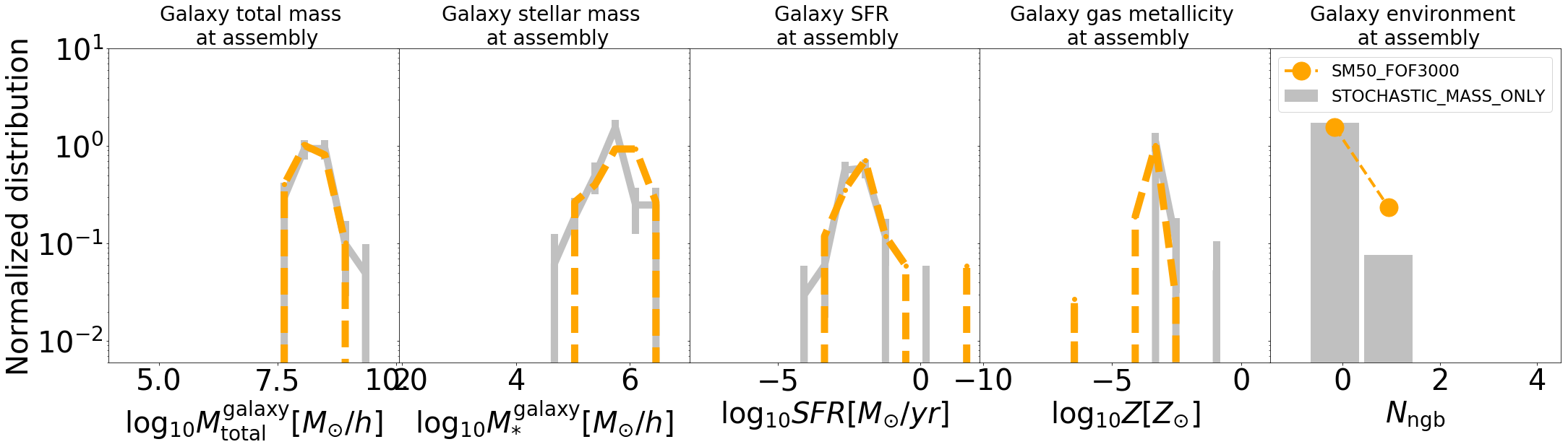}\\

\caption{Similar to Figure \ref{STOCHASTIC_MASS_ONLY_models}, but for the assembly of $1\times10^5~M_{\odot}/h$ BHs from $1.56\times10^3~M_{\odot}/h$ DGBs. Here, the top and bottom rows correspond to $[\mh,\msfmp=3000,5]$ and $[\mh,\msfmp=3000,50]$.}

\label{STOCHASTIC_MASS_ONLY_models_seed5.00}
\end{figure*}

\begin{figure*}
\includegraphics[width=8 cm]{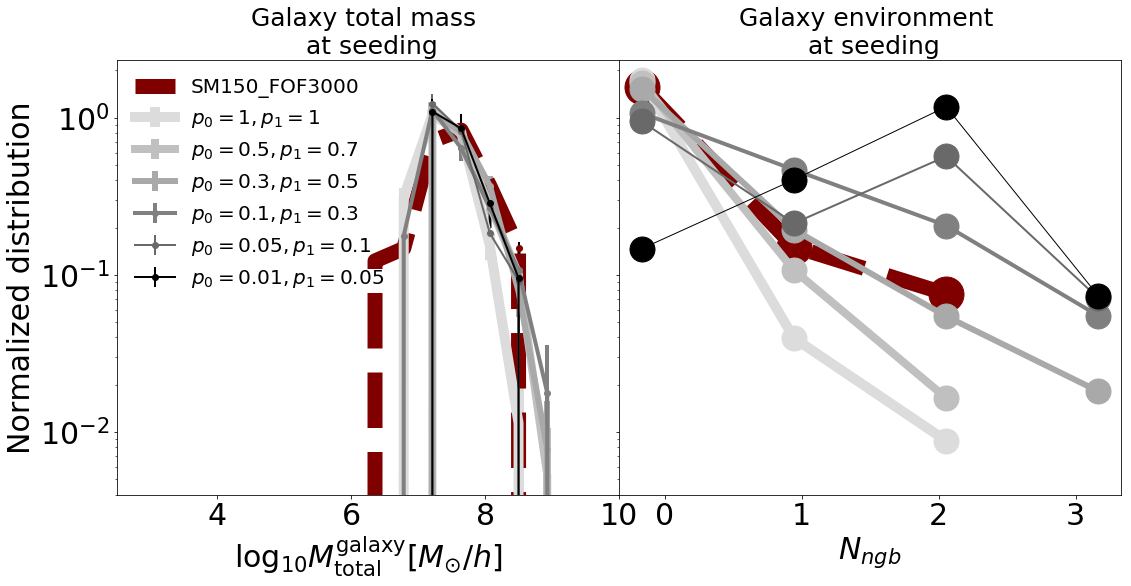} \\

\includegraphics[width=11.8 cm]{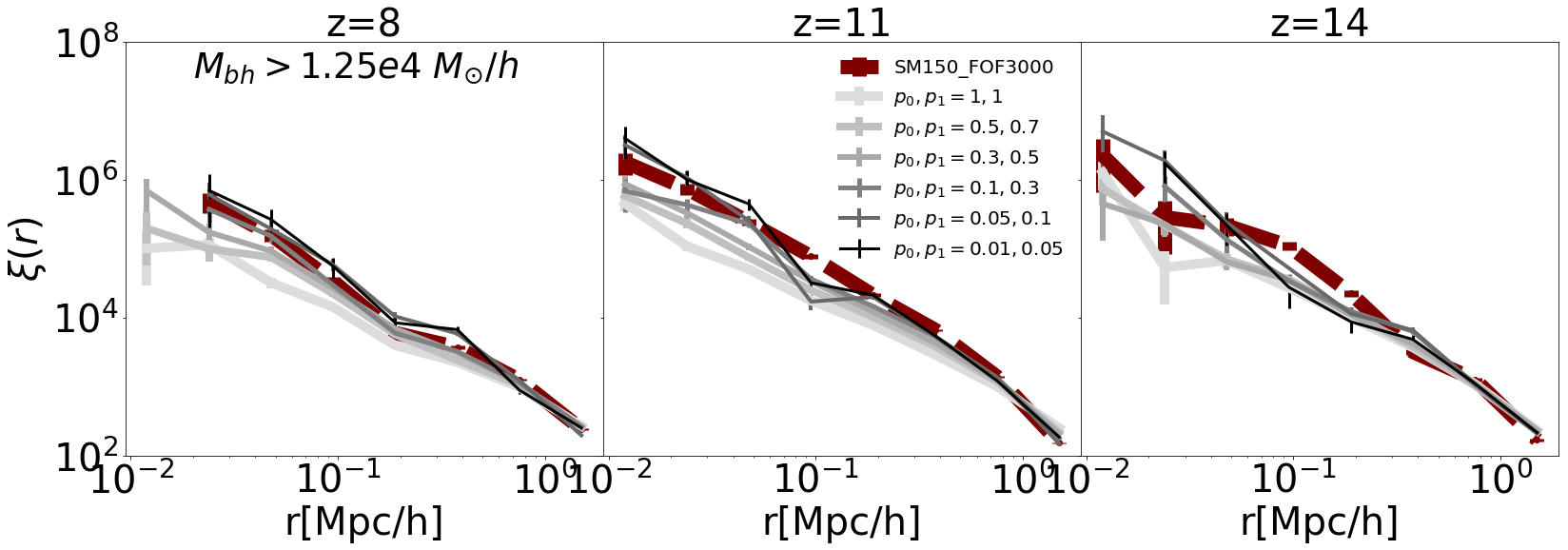} \includegraphics[width= 5 cm]{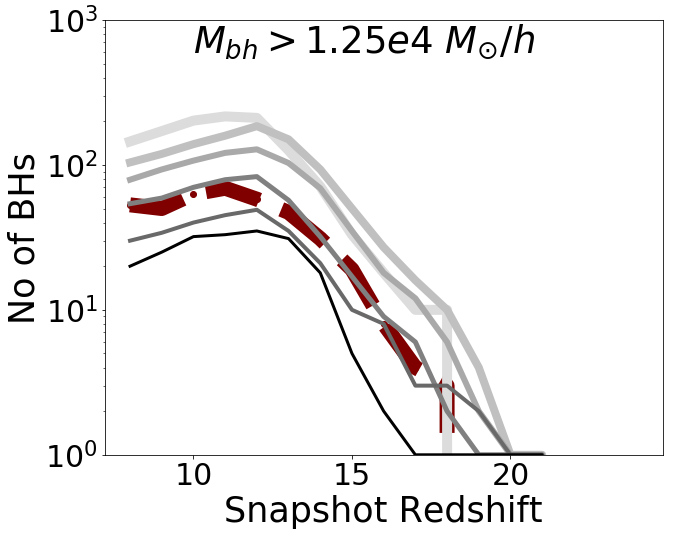} 

\caption{Impact of \textit{galaxy environment criterion} on the two-point clustering and the overall counts of $>1.25\times10^4~M_{\odot}/h$ BHs. The dashed maroon lines show a simulation that uses the gas based seed model $[\mh,\msfmp=3000,150]$ %,  and assembles $1.25\times10^4~M_{\odot}/h$ BHs from 
with $\seedmass = 1.56\times10^3~M_{\odot}/h$. The grey solid lines correspond to simulations that use the stochastic seed model, and directly place ESDs of mass $1.25\times10^4~M_{\odot}/h$ based on both the \textit{galaxy mass criterion} and \textit{galaxy environment criterion}. For the \textit{galaxy environment criterion}, we systematically %increase 
decrease $p_0$ and $p_1$ as the shade gets darker (see legend).  \textit{Upper panels}: The total galaxy mass~(left panel) and galaxy environment~(right panel) during the initial assembly of $1.25\times10^4~M_{\odot}/h$ BHs. \textit{Lower panels}: The left three panels show the two point clustering of $>1.25\times10^4~M_{\odot}/h$ BHs at $z=8,11~\&~14$ respectively, and the rightmost panel shows the overall number of $>1.25\times10^4~M_{\odot}/h$ BHs in each snapshot. We find that the \texttt{STOCHASTIC_MASS_ONLY} simulation ($p_0=1$ and $p_1=1$) significantly underestimates the small-scale clustering and overestimates the BH counts compared to the \texttt{GAS_BASED} simulations. As we introduce the \textit{galaxy environment criterion}~(\texttt{STOCHASTIC_MASS_ENV}) and decrease $p_0$ and $p_1$ to favor seeding in richer environments, we find that the small-scale clustering is enhanced and the BH counts decrease. The model with $p_0,p_1=0.1,0.3$ produces the best match for the small-scale clustering as well as the BH counts.
\label{STOCHASTIC_MASS_ENV_models}
}
\end{figure*}

\begin{figure*}
\includegraphics[width=5.9 cm]{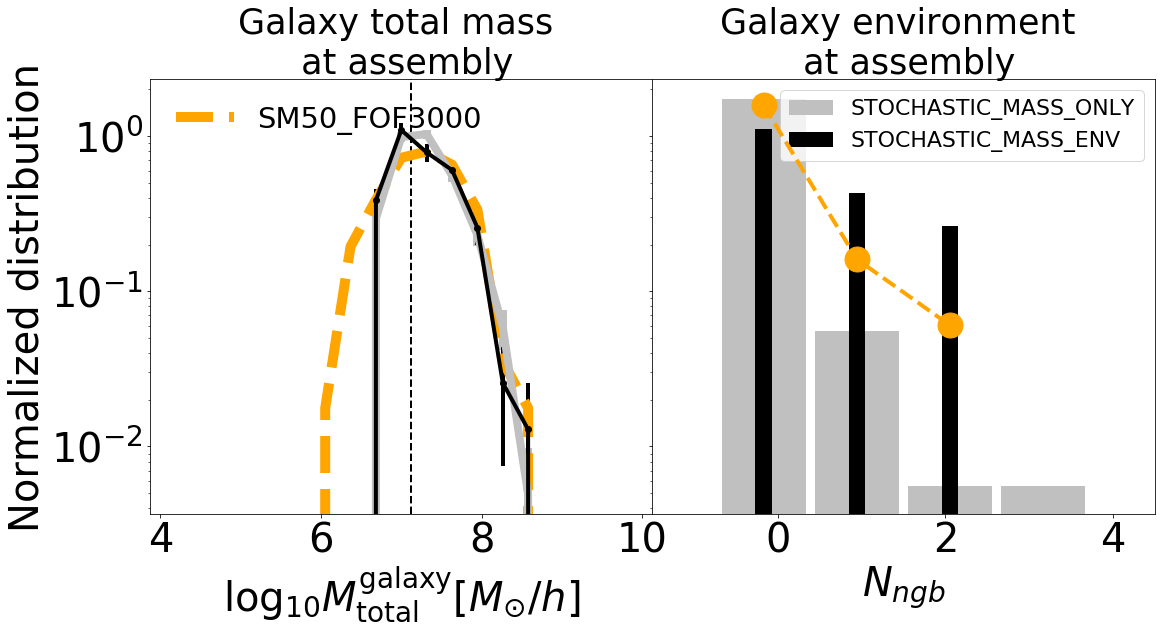}
\includegraphics[width=3.7 cm]{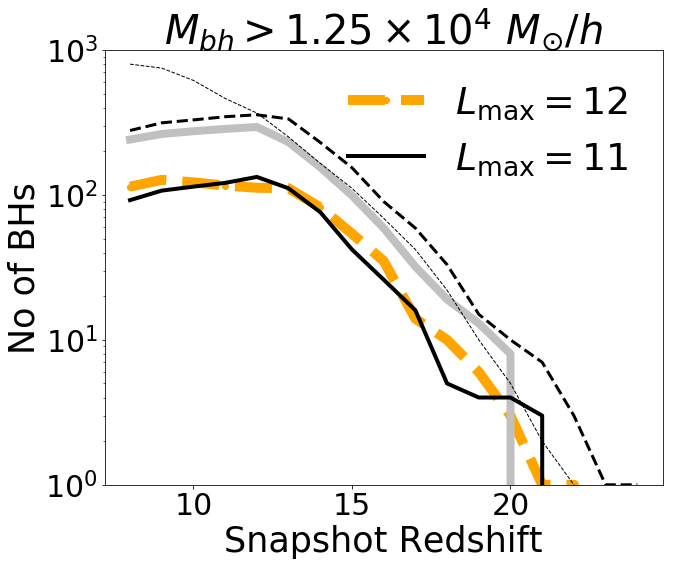}
\includegraphics[width=3.7 cm]{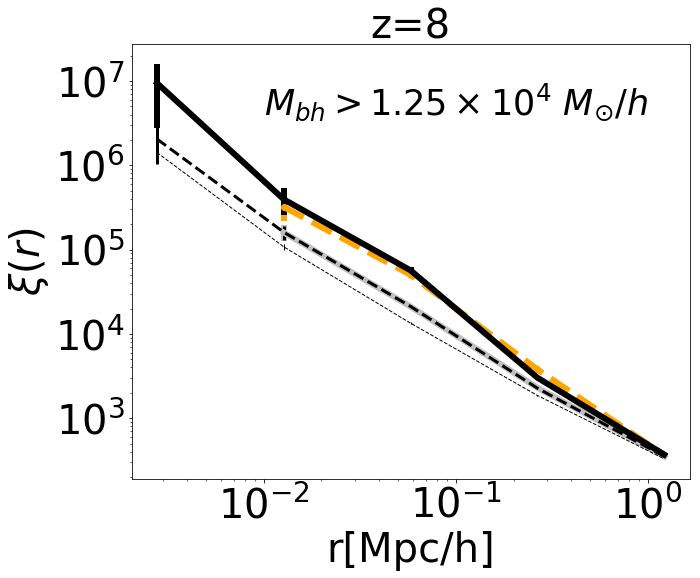}
\includegraphics[width=3.9 cm]{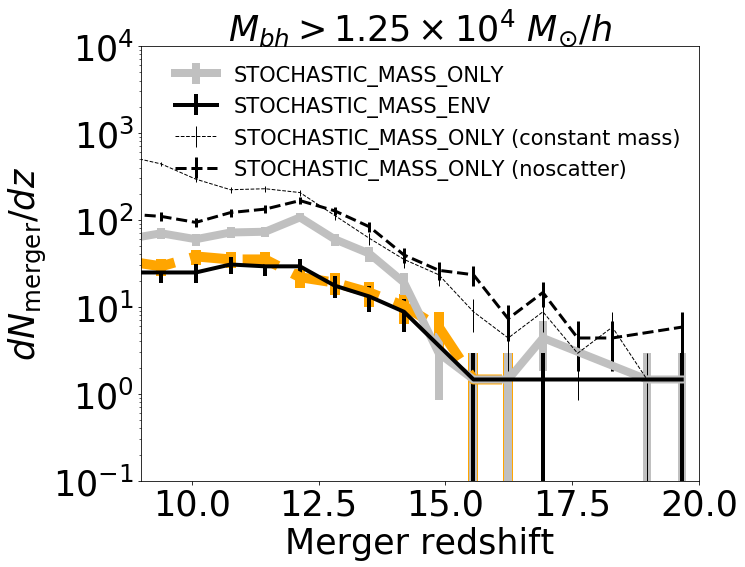}\\

\includegraphics[width=5.9 cm]{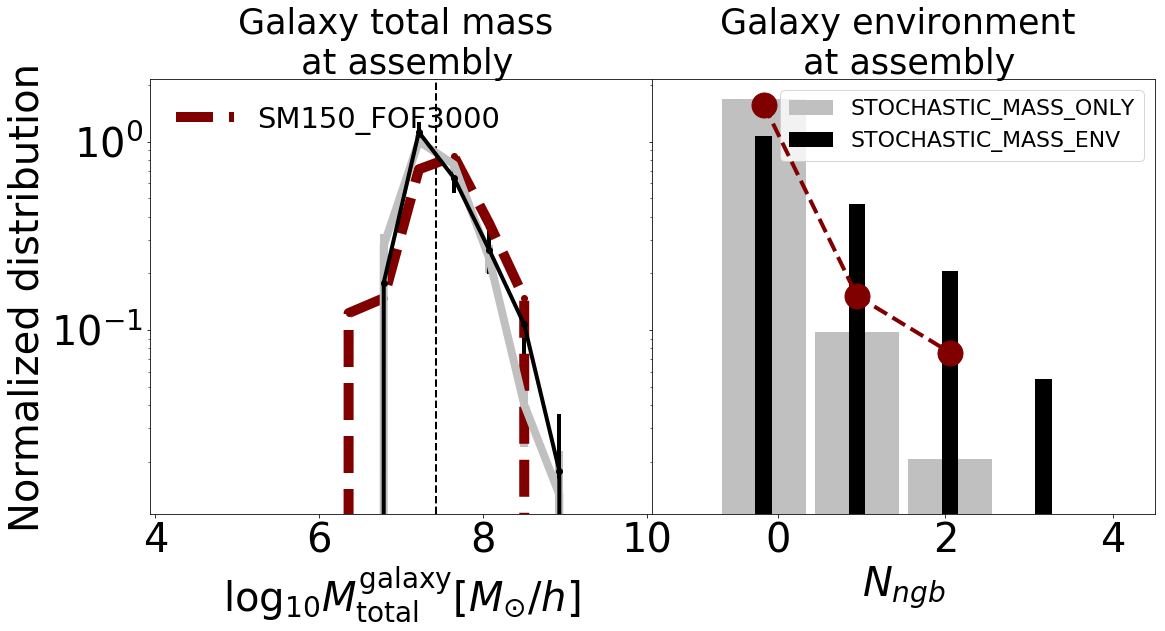}
\includegraphics[width=3.7 cm]{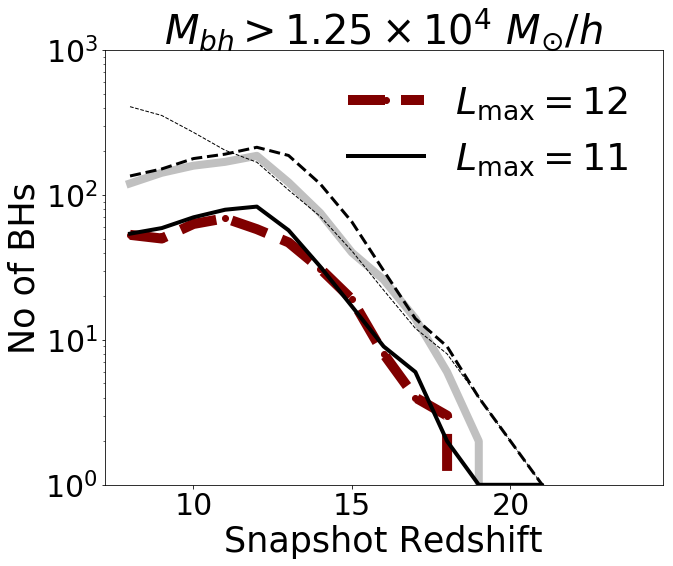}
\includegraphics[width=3.7 cm]{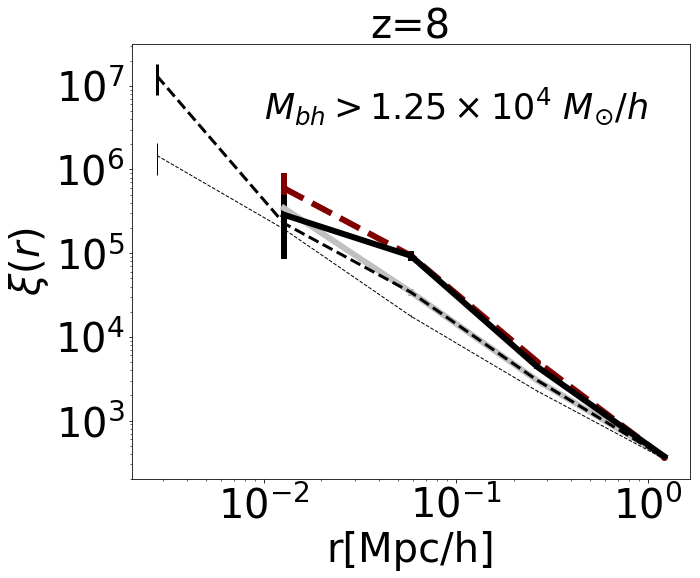}
\includegraphics[width=3.9 cm]{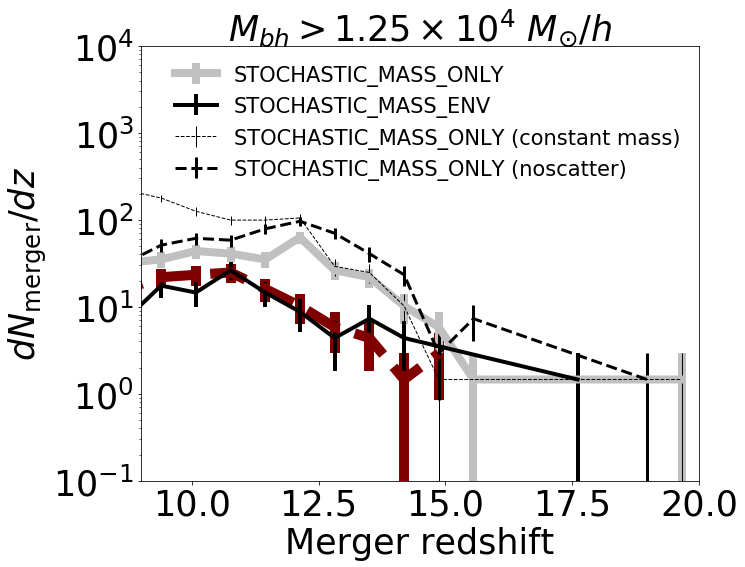}\\

\includegraphics[width=5.9 cm]{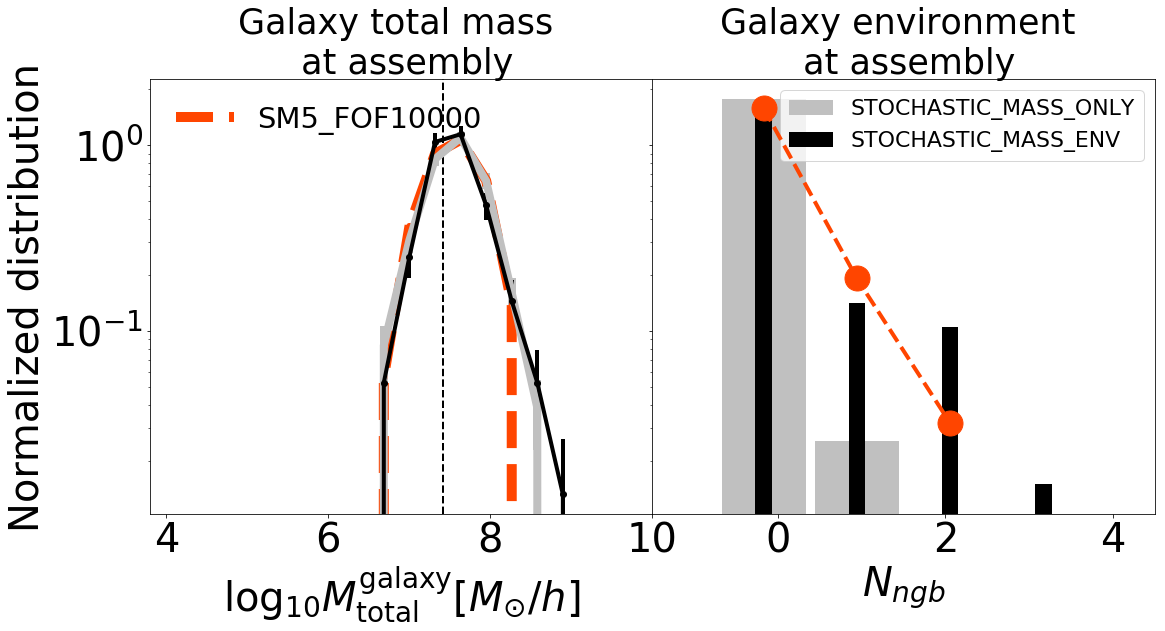}
\includegraphics[width=3.7 cm]{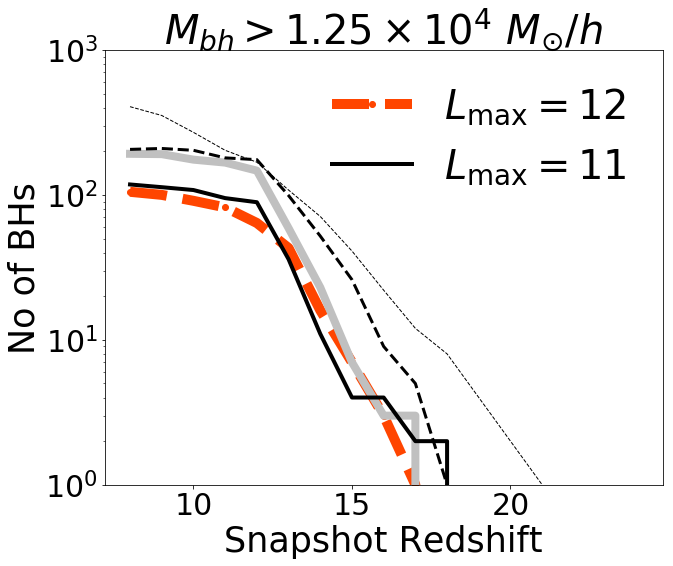}
\includegraphics[width=3.7 cm]{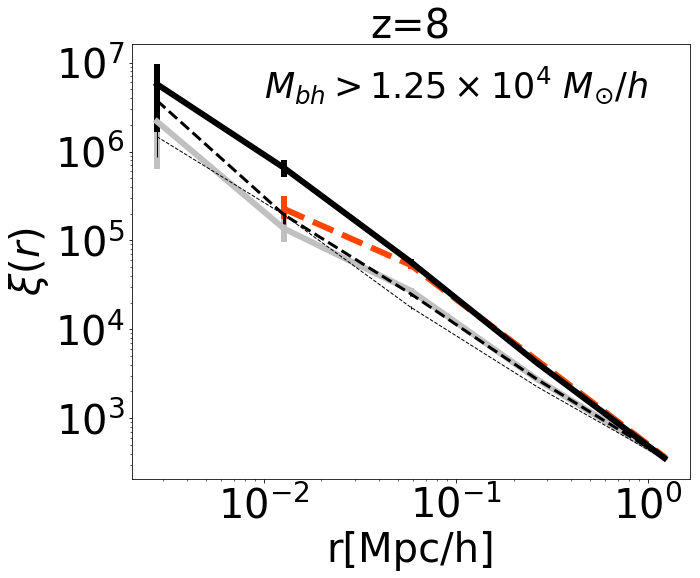}
\includegraphics[width=3.9 cm]{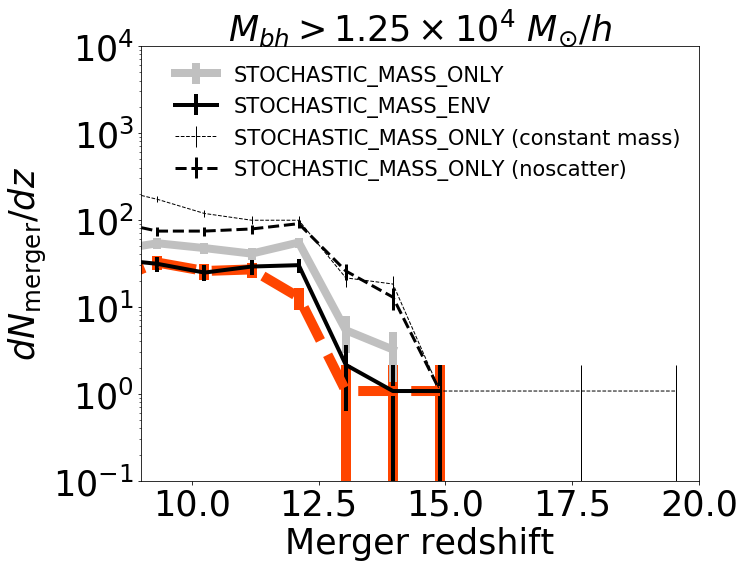}\\
\caption{Here we demonstrate the ability of different $L_{\rm max} = 11$ stochastic seed models to represent the $1.25\times10^4~M_{\odot}/h$ descendants of $1.56\times10^3~M_{\odot}/h$ DGBs formed in $L_{\rm max} = 12$ gas based seed models. % at lower resolutions, for a range of gas based seeding parameters. 
The leftmost two panels show the total galaxy mass and galaxy environment %during the initial 
at the time of assembly of $1.25\times10^4~M_{\odot}/h$ BHs.  The remaining three panels on the right show the statistics of $>1.25\times10^4~M_{\odot}/h$ BHs, namely the total BH counts versus redshift, the two-point clustering at $z=8$, and the merger rates. The colored dashed lines show the \texttt{GAS_BASED} simulations wherein $1.56\times10^3~M_{\odot}/h$ DGBs form and eventually grow to assemble $1.25\times10^4~M_{\odot}/h$ BHs. The different rows correspond to different values of $\msfmp$ and $\tilde{M}_{\mathrm{h}}$~(see legend). The remaining lines correspond to simulations using stochastic seed models that place ESDs directly at $1.25\times10^4~M_{\odot}/h$. The thick and solid silver and black lines and histograms show the \texttt{STOCHASTIC_MASS_ONLY} and \texttt{STOCHASTIC_MASS_ENV} simulations respectively; they use the fiducial seeding parameters calibrated for each set of gas based seeding parameters listed in Table \ref{double_power_law_table}. The thin black dashed lines in the right three panels show \texttt{STOCHASTIC_MASS_ONLY} simulations that  assume zero scatter in the \textit{galaxy mass criterion} i.e $\sigma=0$. The thinnest black solid line in the same panels show simulations that assume a constant galaxy mass threshold fixed at the mean of the distributions from the leftmost panels~(see vertical line). Amongst all the simulations that use stochastic seeding, only the \texttt{STOCHASTIC_MASS_ENV} simulations are able to successfully capture the \texttt{GAS_BASED} simulation predictions.}
\label{subhalo_vs_gas_seeding}
\end{figure*}

\begin{figure*}

\includegraphics[width=5.9 cm]{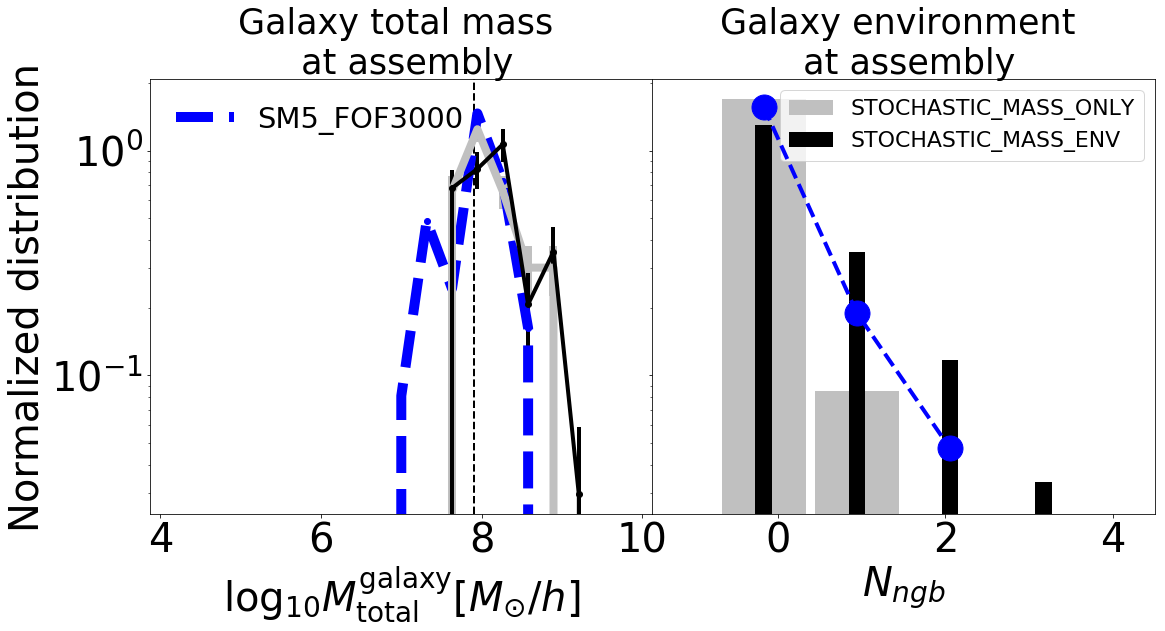}
\includegraphics[width=3.7 cm]{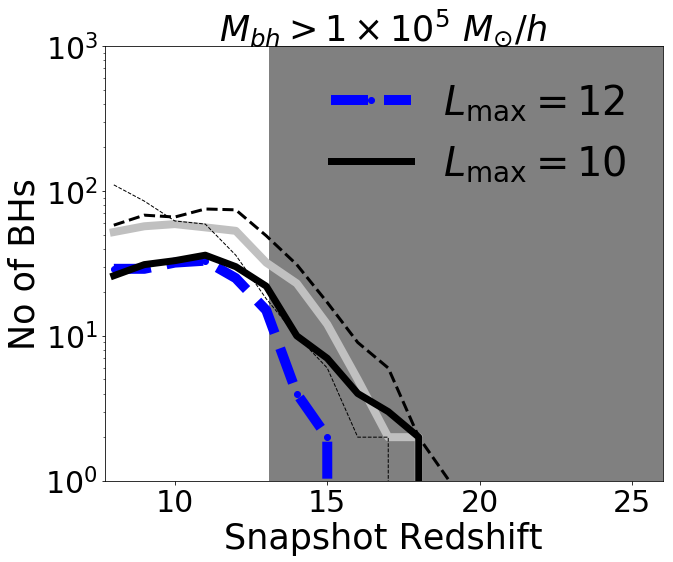}
\includegraphics[width=3.7 cm]{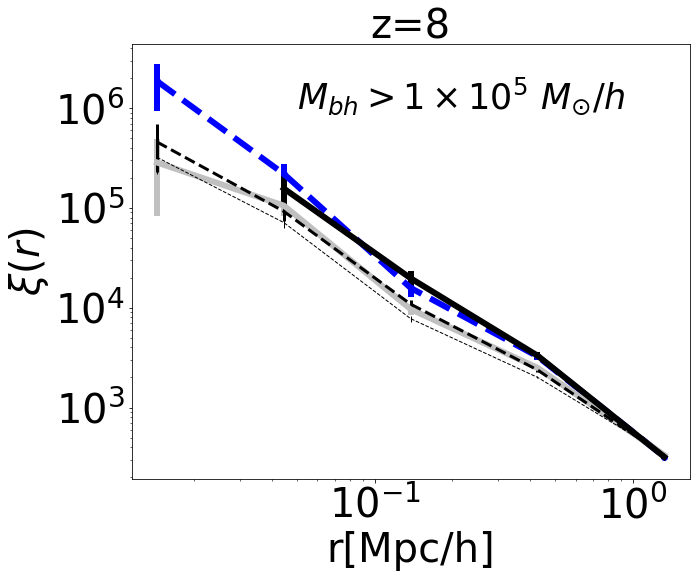}
\includegraphics[width=3.9 cm]{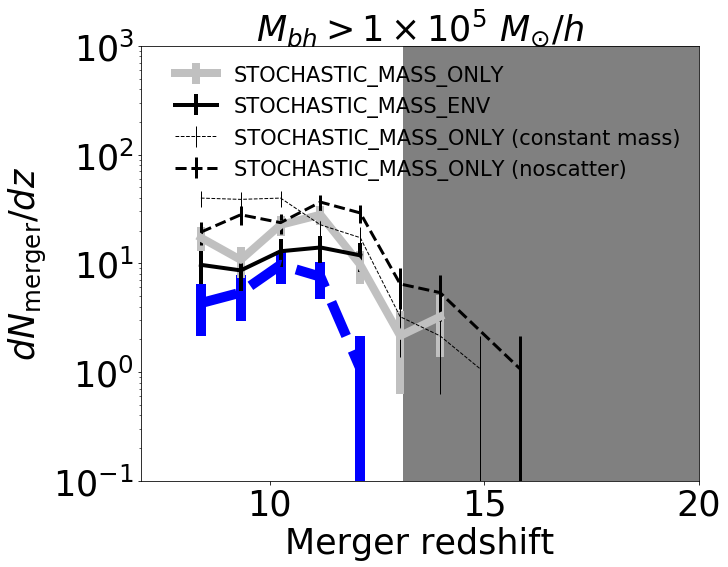}\\

\includegraphics[width=5.9 cm]{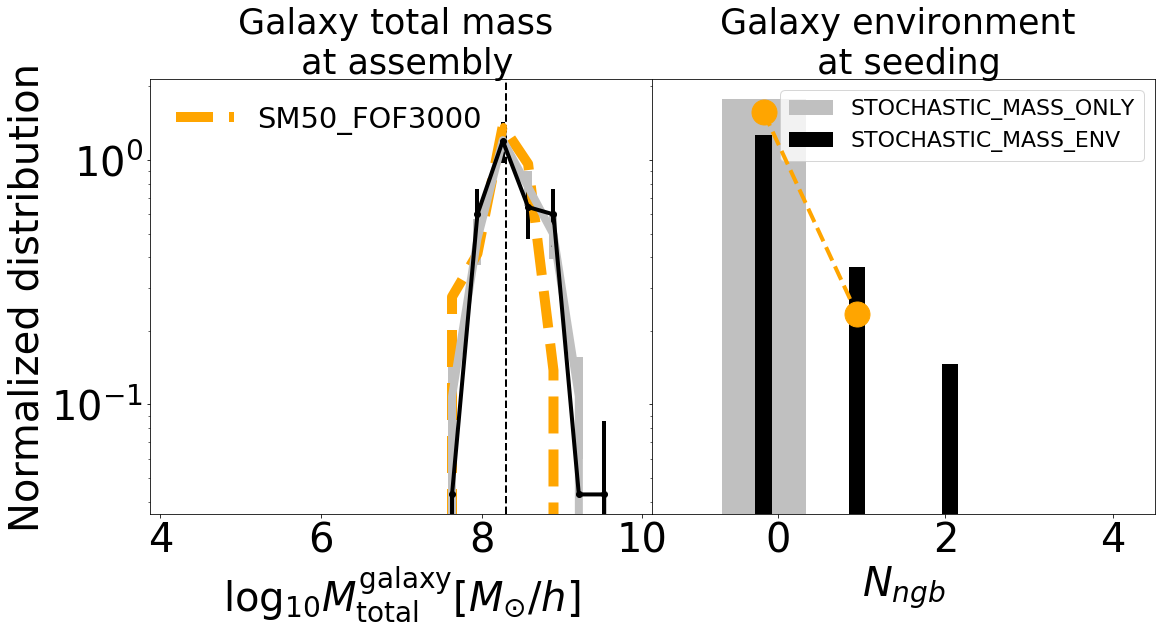}
\includegraphics[width=3.7 cm]{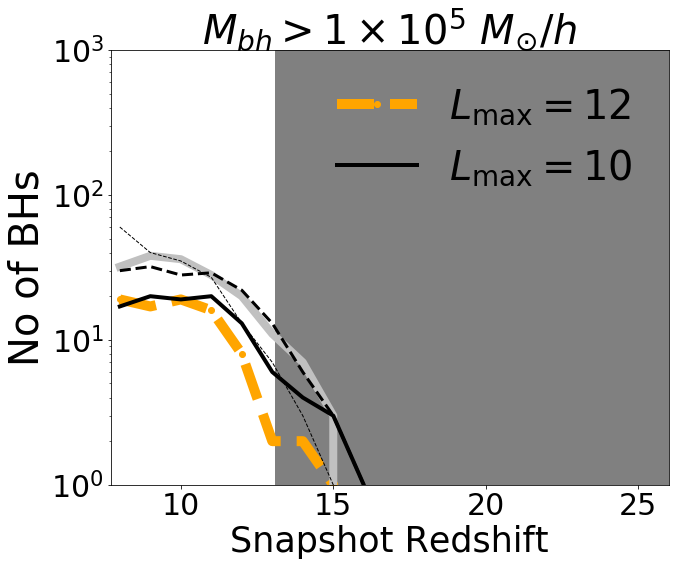}
\includegraphics[width=3.7 cm]{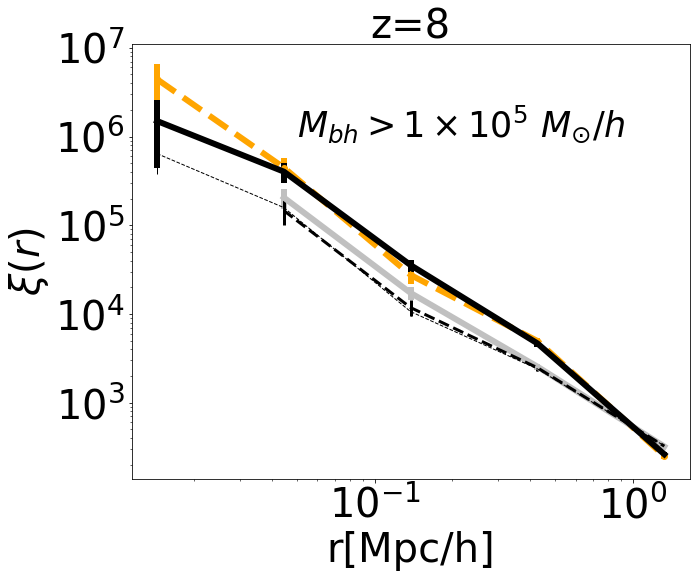}
\includegraphics[width=3.9 cm]{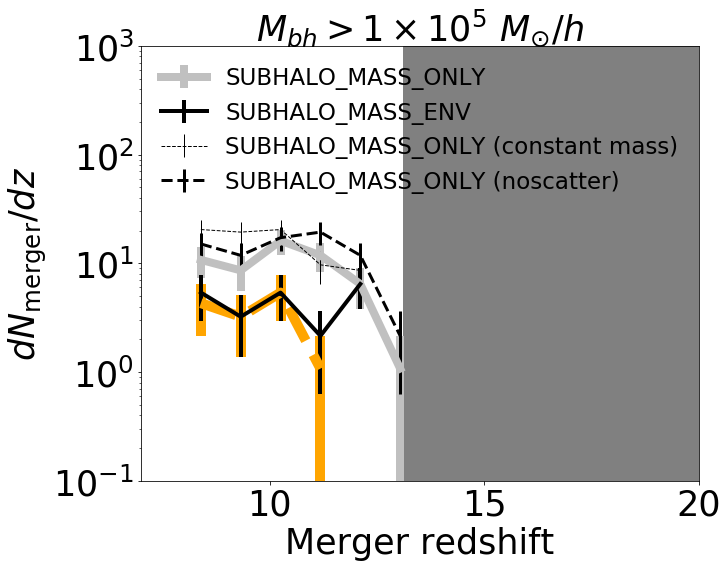}\\

\caption{Same as Figure \ref{subhalo_vs_gas_seeding}, but for the assembly of  $1\times10^{5}~M_{\odot}/h$ BHs. The statistics are more limited compared to the previous figure. The shaded grey regions correspond to $z>13.1$, wherein we could not calibrate the \textit{galaxy mass criterion} due to lack of data points in Figure \ref{bFOF_at_assembly}. But at $z<13.1$ where calibration was possible, we find that the \texttt{STOCHASTIC_MASS_ENV} simulations (at a resolution of $L_{\rm max} = 10$) do reasonably match with the BH counts predicted by the $L_{\rm max} = 12$ \texttt{GAS_BASED} simulations.}
\label{subhalo_vs_gas_seeding_seed5.00}
\end{figure*}

Recall from Section \ref{Rapid metal enrichment after seed formation} that because DGB formation in our gas based seeding model occurs during a transient phase of %due to the 
rapid metal enrichment 
in halos that are otherwise fairly typical, %of seed forming halos, 
their descendents have metallicities~(and SFRs) similar to that of typical halos with similar total masses. This motivates us to first explore low-resolution simulations with seeding criterion that simply matches the galaxy mass distribution of seeding sites in our high-resolution, gas based models. We refer to this seeding criterion as the %based only on 
\textit{galaxy mass criterion}; notably, this differs from typical halo-mass-based seeding models in the use of a distribution of host mass thresholds rather than a single value. The corresponding simulations are referred to as \texttt{STOCHASTIC_MASS_ONLY}. 

\subsubsection{Galaxy masses at assembly of $\sim10^4~\&~10^5~M_{\odot}$ BHs from $\sim10^3~M_{\odot}$ seeds}

To calibrate our seed models, we first determine the galaxy masses ($\massembly$) in which $1.25\times10^4~M_{\odot}/h$ and $1\times10^5~M_{\odot}/h$ BHs assemble from $1.56\times10^3~M_{\odot}/h$ DGBs %~(hereafter denoted by $\massembly$) 
within our \texttt{GAS_BASED} simulations; these are shown in Figure \ref{bFOF_at_assembly}. Let us first focus on the assembly of $1.25\times10^4~M_{\odot}/h$ descendants~(Figure \ref{bFOF_at_assembly}, top panels). Similar to that of $\massemblyFOF$ versus redshift relations~(Figure \ref{host_halos_at_assembly}, middle panel), the $\massembly$ versus redshift relations show features that reflect the interplay between halo growth, star formation and metal enrichment in influencing DGB formation. For $\mh=3000, \msfmp=50~\&~150$, we see that the slope of redshift evolution of the mean~(denoted by $\left<\massembly\right>$ and shown as solid lines) undergoes a gradual transition between $z\sim13-15$. This corresponds to the slow down of DGB formation due to metal enrichment. When $\mh=10000~\&~\msfmp=5$, this transition occurs at comparatively lower redshifts~($z\sim12-10$) as the influence of metal enrichment starts later due to the higher $\mh$. We then fit the mean trend by a double power law~(dashed lines in Figure \ref{bFOF_at_assembly}, upper panels) given by

\begin{eqnarray}
\begin{aligned}
&\log_{10}\left<\massembly\right>= \nonumber \\
&\left\{
    \begin{array}{lr}
        (z-z_{\mathrm{trans}}) \times \alpha + \log_{10}M_{\mathrm{trans}}  , & \text{if } z \geq z_{\mathrm{trans}}\\
        (z-z_{\mathrm{trans}}) \times \beta + \log_{10}M_{\mathrm{trans}}, & \text{if } z < z_{\mathrm{trans}}
    \end{array}
\right\}.
\label{double_powerlaw_eqn}
\end{aligned}
\end{eqnarray}
$z_{\mathrm{trans}}$ roughly marks the transition in the driving physical process for DGB formation. For $z > z_{\mathrm{trans}}$, halo growth or star formation primarily drives DGB formation; for $z < z_{\mathrm{trans}}$, metal enrichment takes over as the primary driver to suppress DGB formation. $M_{\mathrm{trans}}$ is the value of $\left<\massembly\right>$ at the transition redshift. Finally, $\alpha$ and $\beta$ are the slopes of the $\left<\massembly\right>$ versus redshift relation at $z>z_{\mathrm{trans}}$ and $z<z_{\mathrm{trans}}$ respectively.
To simplify our fitting procedure, we first select $z_{\mathrm{trans}}$ for each of the cases via visual inspection and determine $M_{\mathrm{trans}}$ by interpolating the  $\left<\massembly\right>$ versus redshift relation. We then fit for $\alpha$ and $\beta$ using the \texttt{scipy.optimize.curve_fit} python package. Note that the double power-law function assumes a sharp transition in the $\left<\massembly\right>$ versus redshift relation at $z=z_{\mathrm{trans}}$. However, as we can see in Figure \ref{bFOF_at_assembly}, this transition occurs much more gradually as metal enrichment starts to slow down and eventually suppresses DGB formation. Nevertheless, the double power-law model offers a simple~(albeit approximate) framework to capture the intricate convolution of the impact of halo growth, star formation and metal enrichment that leads to the initial rise and eventual suppression of DGB formation. 

The values of $z_{\mathrm{trans}}$, $M_{\mathrm{trans}}$, $\alpha$ and $\beta$ for the different gas based seed models are listed in the top four rows of Table \ref{double_power_law_table}. We choose $z_{\mathrm{trans}}=13.1$ for $\mh=3000, \msfmp=5,50~\&~150$. $z_{\mathrm{trans}}$ is the same for all three $\msfmp$ values to encode that the slow down of seed formation due to metal enrichment starts at similar redshifts for all these models. For $\mh=10000,\msfmp=5$, we choose a lower transition redshift of  
$z_{\mathrm{trans}}=12.1$ as halo growth continues to drive up seed formation up to lower redshifts compared to the models with $\mh=3000$. 

The impact of $\mh$ and $\msfmp$ on $M_{\mathrm{trans}}$, $\alpha$ and $\beta$ is noteworthy. As $\mh$ or $\msfmp$ increases, the value of $M_{\mathrm{trans}}$ also increases to generally reflect the fact that descendant BHs of a fixed mass are assembling in more massive halos. $\alpha$ is significantly more sensitive to $\msfmp$ compared to $\mh$; this is not surprising as $\alpha$ corresponds to the regime where metal enrichment primarily governs seed formation. A higher value of $\msfmp$ produces a steeper $\alpha$, as it leads to stronger suppression of DGB formation by metal enrichment. Lastly, $\beta$ is impacted by both $\msfmp$ and $\mh$. This also makes sense because $\beta$ corresponds to the regime where either star formation or halo growth can drive seed formation. Increasing $\msfmp$ enhances the role of star formation, and increasing $\mh$ enhances the role of halo growth. Generally, we see that as the number of DGBs forming at the highest redshifts is decreased due to increase in $\mh$ or $\msfmp$, $\beta$ tends to go from negative to positive values thereby favoring higher $\massembly$ at higher redshifts. This is likely because when BHs are very few, merger driven growth is slow and galaxies have more time to grow via DM accretion between successive mergers. As a result, galaxy growth is slightly faster than merger dominated BH growth at these highest redshifts where there are very few BHs. 
 %As $\msfmp$ increases from 5 to 50, $\beta$ becomes less negative, and becomes slightly positive for $\msfmp=150$. When $\mh$ is increased to $10000$, $\beta$ becomes even more positive. 
%Recall that we also noted similar trends upon tracking the \textit{halo masses} at assembly of $1.25\times10^4~M_{\odot}/h$ in Figure \ref{host_halos_at_assembly}~(middle panels), but they were not as pronounced. This may be because the subhalos are smaller in mass overall, so the relative impact of DM accretion~(between successive mergers) would be more for subhalos compared to halos. \laura{\bf[i think this paragraph could be tightened up a bit]}

We now turn our attention to the assembly of $10^5~M_{\odot}/h$ descendant BHs~(bottom panels of Figure \ref{bFOF_at_assembly}). In this case, we do not have adequate statistics to robustly determine the $\left<\massembly\right>$ versus redshift relations. We can see that data points only exist at $z\lesssim13$, wherein $\left<\massembly\right>$ tends to increase with decreasing redshift, %for $\mh=3000,\msfmp=5,50~\&~150$~(
except for $\mh=10000,\msfmp=5$, where statistics are too poor to reveal any useful trends). Here, we only fit for $\alpha$ after assuming the same values of $z_{\mathrm{trans}}$ that were used for the assembly of $1.25\times10^4~M_{\odot}/h$ BHs~(dashed lines in Figure \ref{bFOF_at_assembly}, lower panels). The best fit values are shown in the bottom three rows of Table \ref{double_power_law_table}. Overall, we should still keep in mind that there are very few $10^5~M_{\odot}/h$ descendants. Therefore, these fits are not very statistically robust. Nevertheless,  
they will still be useful to test our stochastic seed models in the next subsection.

In addition to the mean trends, the $\massembly$ versus redshift relations show a significant amount of scatter ($\sigma$). This is defined to be the 1 sigma standard deviation shown by the shaded regions in Figure \ref{bFOF_at_assembly}. Generally we see that the scatter does not have a strong redshift evolution. The overall mean scatter~(averaged over the entire redshift range) for the different gas based seed models is shown in the seventh column of Table \ref{double_power_law_table}.  The scatter decreases slightly as we make the gas based seeding criterion more restrictive by increasing $\mh$ or $\msfmp$. This is likely because for more restrictive seed models, assembly of higher-mass BHs occurs in more massive galaxies for which the underlying galaxy mass function is steeper. For the same reason, the scatter is also smaller for the assembly of $1\times10^5~M_{\odot}/h$ BHs compared to that of $1.25\times10^4~M_{\odot}/h$ BHs. 

\subsubsection{Properties of galaxies that form ESDs: Comparison with gas based seed model predictions} 

We finally use the $\massembly$ versus redshift relations to formulate our \textit{galaxy mass criterion}. More specifically, we place ESDs of mass $1.25\times10^4~M_{\odot}/h$ and $1\times10^5~M_{\odot}/h$ based on minimum galaxy mass thresholds. The threshold value~($M_{\mathrm{th}}$) is stochastically drawn from redshift dependent distributions described by a log-normal function,  i.e $\propto \exp{[-\frac{1}{2}(\log_{10}{M_{\mathrm{th}}^2-\mu^2)/\sigma^2]}}$, with mean $\mu\equiv\left<\massembly\right>(z)$ described by the double power-law fits shown in Figure \ref{bFOF_at_assembly} and Table \ref{double_power_law_table}. The standard deviation $\sigma$ is shown in Table \ref{double_power_law_table}~(column 7).   

In Figure \ref{STOCHASTIC_MASS_ONLY_models}, we show the 1D distributions~(marginalized over all redshifts until %till 
$z=7$) of the various galaxy properties wherein $1.25\times10^4~M_{\odot}/h$ descendants assemble (i.e., total mass, stellar mass, SFRs, gas metallicities and environments). We compare the predictions for the \texttt{GAS_BASED} simulations that assemble the $1.25\times10^4~M_{\odot}/h$ descendants from $1.56\times10^3~M_{\odot}/h$ DGBs~(colored lines), and the \texttt{STOCHASTIC_MASS_ONLY} simulations that directly seed the $1.25\times10^4~M_{\odot}/h$ ESDs~(grey lines). We can clearly see that after calibrating the \texttt{STOCHASTIC_MASS_ONLY} simulations to reproduce the total galaxy masses~(1st panels from the left) predicted by the \texttt{GAS_BASED} simulation, it also broadly reproduces the baryonic properties of the galaxies such as stellar masses, SFRs and metallicities~(2nd, 3rd and 4th panels). This further solidifies our findings from Figures \ref{evolution_of_seed_forming_gas} to \ref{unbiased_halo_at_assembly_fig}, that the galaxies wherein the $1.25\times10^4~M_{\odot}/h$ descendants assemble are reasonably well characterized by their total mass alone. Recall that this is attributed to the transience of the rapid metal enrichment phase in which halos form %of gas within halos forming the 
$1.56\times10^3~M_{\odot}/h$ DGBs in the \texttt{GAS_BASED} suite. 

However, we see that the \textit{galaxy mass criterion} places the %$1.25\times10^4~M_{\odot}/h$ 
ESDs %within subhalos living 
in sparser environments~(hosts with fewer %number of 
neighboring halos) compared to the \texttt{GAS_BASED} simulation predictions~(rightmost panels in Figure \ref{STOCHASTIC_MASS_ONLY_models}). This reflects the fact that when the low-mass %$1.56\times10^3~M_{\odot}/h$ 
DGBs assemble higher-mass BHs through merger-dominated BH growth, their descendants naturally grow faster in regions with more frequent major halo and galaxy mergers. Therefore, for a %sample of subhalos with a 
given distribution of total galaxy masses, those living in richer environments are more likely to %occupy 
contain higher-mass descendant BHs.

%The foregoing 
These results for the assembly of $1.25\times 10^4~M_{\odot}/h$ BHs %descendants 
also hold true for the assembly of $1\times 10^5~M_{\odot}/h$ BHs, as shown in Figure \ref{STOCHASTIC_MASS_ONLY_models_seed5.00}. In the next section, we develop an additional seeding criterion to account for this %shall explore the implications of these results on the 
small-scale clustering of the assembly sites of higher mass descendants in our \texttt{GAS_BASED} models. %the BH populations predicted by the different seed models.

\subsection{Building the \textit{galaxy environment criterion}}
\label{ssec:subhalo_env_criterion}

In this section, we describe an %will implement the 
additional \textit{galaxy environment criterion} to favor the placement of ESDs in galaxies in richer environments~(at fixed galaxy mass). We then %, and 
explore its implications on their two-point clustering and the overall BH population. %counts. We first describe the formalism for the \textit{subhalo environment criterion} implementation as follows: 

First, we assume that any potential seeding site with two or more neighbors ($N_{\rm ngb} \geq 1$) will always seed an ESD. Potential seeding sites with zero or one neighbors will seed an ESD with a probability $0 \leq \environmentseedprobability \leq 1$. For these cases, we assign a different linear dependence of
%Let us suppose we plan to seed descendant BHs of mass $\descendantseedmass$. We implement an overall seeding probability for seeding descendants in subhalos with 0 and 1 neighbors denoted by $\environmentseedprobability$ . 
$\environmentseedprobability$ %linearly increases with 
on the galaxy mass $M^{\mathrm{galaxy}}_{\mathrm{total}}$, such that the probability for any potential seeding site to actually form an ESD is given by %respectively as

\begin{equation}
\environmentseedprobability=
\left\{
    \begin{array}{lr}
        \left(M^{\mathrm{galaxy}}_{\mathrm{total}}-\left<\massembly\right>\right)\gamma + p_0 , & \text{if } N_{\mathrm{ngb}}=0\\
        \left(M^{\mathrm{galaxy}}_{\mathrm{total}}-\left<\massembly\right>\right)\gamma + p_1 , & \text{if } N_{\mathrm{ngb}}=1 \\
        1, & \text{if } N_{\mathrm{ngb}}>1
    \end{array}
\right\}.
\label{environment_based_seed_probability}
\end{equation}
Here, $p_0$ and $p_1$ denote the seeding probability in galaxies with 0 and 1 neighbors respectively, at the mean~$\left(\left<\massembly\right>\right)$ of the total mass distributions of galaxies wherein the descendant BHs assemble. 

The parameter $\gamma$ defines the slope for the linear dependence of $\environmentseedprobability$ on the galaxy mass; it varies slightly between the underlying gas based seed models used for calibration, as %. Its values for the various gas based seeding parameters are 
listed in Table \ref{double_power_law_table}. The motivation for this linear dependence and the adopted $\gamma$ values are described in Appendix \ref{Relationship between subhalo environment and subhalo mass}. But to briefly summarize the main physical motivation, we use a $\gamma>0$ to encode the natural expectation that for fixed $N_{\mathrm{ngb}}$, descendants will grow faster within galaxies with higher total mass. This is because $N_{\mathrm{ngb}}$, by definition, counts the number of halos with masses \textit{higher than} the host halo mass of the galaxy that are within $5 R_{\rm vir}$. %~(revisit Section \ref{Subhalo-based stochastic seeding} for the definition of $N_{\mathrm{ngb}}$). 
As a result, a higher-mass galaxy with $N_{\mathrm{ngb}}$ neighbors is in a more overdense %extreme~(higher overdensity) 
region than a lower-mass galaxy with the same $N_{\mathrm{ngb}}$ neighbors. 

We add the \textit{galaxy environment criterion} %in addition 
to the already applied \textit{galaxy mass criterion}. We shall refer to the resulting suite of simulations as \texttt{STOCHASTIC_MASS_ENV}.  In Figure \ref{STOCHASTIC_MASS_ENV_models}, we systematically compare the \texttt{GAS_BASED}~
%($\mh=3000~\&~\msfmp=150$; maroon lines) 
simulations~(maroon lines) to the %that of 
\texttt{STOCHASTIC_MASS_ENV} simulations that trace $1.25\times10^4~M_{\odot}/h$ descendants~(%different shades of 
grey lines) for a range of parameter values for $p_0$ and $p_1$. %Here we focus on the assembly of $1.25\times10^4~M_{\odot}/h$ descendants~(also implying that $\descendantseedmass=1.25\times10^4~M_{\odot}/h$). 
We start with $p_0=1,p_1=1$, which is basically the \texttt{STOCHASTIC_MASS_ONLY} simulation~(lightest grey lines), and find that it significantly underestimates the two point clustering~(by factors up to $\sim5$) of the $\geq1.25\times10^4~M_{\odot}/h$ BHs compared to the \texttt{GAS_BASED} simulations~(lower left three panels). At the same time, the \texttt{STOCHASTIC_MASS_ONLY} simulation also over-estimates the overall counts of the $\geq1.25\times10^4~M_{\odot}/h$ BHs~(lower right most panel). Upon decreasing the probabilities as $p_0<p_1<1$, we can see that the two-point clustering starts to increase while the overall BH counts simultaneously decrease. %Very notably, 
For $p_0=0.1~\&~p_1=0.3$, we produce the best agreement of the two-point clustering as well as the overall BH counts. Further decreasing $p_0$ and $p_1$ mildly enhances the two-point clustering, but leads to too much suppression of the BH counts compared to \texttt{GAS_BASED} simulations. Therefore, we identify $p_0=0.1~\&~p_1=0.3$ as the best set of parameter values for the gas based seeding parameters $[\mh,\msfmp=3000,150]$. 

As a caveat, we must also note in Figure \ref{STOCHASTIC_MASS_ENV_models} that while $p_0=0.1~\&~p_1=0.3$ produces the best agreement with the two point correlation function between \texttt{GAS_BASED} and \texttt{STOCHASTIC_MASS_ENV} simulations, it does %end up placing 
place the ESDs in galaxies with somewhat higher $N_{\mathrm{ngb}}$ compared to the \texttt{GAS_BASED} simulations~(upper right panels). To that end, recall that $N_{\mathrm{ngb}}$ only measures the galaxy environment at a fixed separation scale of $D_{\mathrm{ngb}}=5~R_{\mathrm{vir}}$~(revisit Section \ref{Subhalo-based stochastic seeding}). Therefore, we cannot expect $N_{\mathrm{ngb}}$ to fully determine the two-point correlation profile, which measures the environment over a wide range of separation scales~($\sim0.01-1~\mathrm{Mpc}/h$ in our case). In other words, one could come up with alternative set of \textit{galaxy environment criteria}~(for example, using $N_{\mathrm{ngb}}$ within a  different $D_{\mathrm{ngb}}\neq5~R_{\mathrm{vir}}$ or even multiple set of $N_{\mathrm{ngb}}$ values within different multiple $D_{\mathrm{ngb}}$ values) and still be able simultaneously reproduce the two-point correlation function as well as the BH counts. Finding all these different possibilities of \textit{galaxy environment criteria} is not the focus of this work. Instead, our objective here is simply to demonstrate that to reproduce the \texttt{GAS_BASED} simulation predictions, we need a \textit{galaxy environment criterion} to favor the placing of ESDs in galaxies with richer environments. Furthermore, we showed that by applying a \textit{galaxy environment criterion} that brings %and bringing 
the two point correlation function %to agree 
into agreement with the \texttt{GAS_BASED} simulations, our \texttt{STOCHASTIC_MASS_ENV} simulations achieve the primary goal for our sub-grid seeding model: faithfully representing %reproduce 
the %overall counts of the 
descendants of $1.56\times10^3~M_{\odot}/h$ seeds produced in the \texttt{GAS_BASED} simulations. 

Thus far we have calibrated a \texttt{STOCHASTIC_MASS_ENV} simulation to reproduce the $1.25\times10^4~M_{\odot}/h$ descendant BH population from a
%So far we focused on the 
gas based seed model with  $[\mh,\msfmp=3000,150]$ and $M_{\rm seed} = 1.56 \times 10^3~M_{\odot}/h$. %assembling $1.25\times10^4~M_{\odot}/h$ descendant BHs. 
We can perform the same calibration
%But we also demonstrate this 
for the %wider range of 
remaining gas based seed models in our suite, and for the assembly of $1\times10^5~M_{\odot}/h$ descendant BHs in addition to $1.25\times10^4~M_{\odot}/h$ descendants. % assembling $1.25\times10^4~M_{\odot}/h$ as well as  $1\times10^5~M_{\odot}/h$ descendants. 
The resulting $p_0$ and $p_1$ values for all the gas based seeding parameters are listed in Table \ref{double_power_law_table}. Broadly speaking, we require $p_0\sim0.1-0.2$ and $p_1\sim0.3-0.4$ to simultaneously reproduce the gas based seed model predictions for the small-scale clustering and BH counts of the descendant BHs. Slightly higher $p_0$ and $p_1$ values are favored for more restrictive gas based criteria and for %Notably, as we make 1) the gas based seeding criterion more restrictive, or 2) consider the assembly of 
higher-mass descendant BHs, %the fiducial $p_0$ and $p_1$ values have a tendency to slightly increase. This may be 
possibly because in both cases %both 1) and 2) causes 
the descendant BHs assemble in higher-mass galaxies. Note that higher-mass galaxies tend to be more strongly clustered than lower mass galaxies. As a result, during the calibration of the \texttt{STOCHASTIC_MASS_ENV} simulations, the \textit{galaxy mass criterion} alone will already produce a slightly stronger clustering for the ESDs. This lessens the burden on the  \textit{galaxy environment criterion} to achieve the desired clustering predicted by the gas based seed models.        

In Figures \ref{subhalo_vs_gas_seeding} and \ref{subhalo_vs_gas_seeding_seed5.00}, we show the \texttt{STOCHASTIC_MASS_ENV}~(solid black lines) versus \texttt{GAS_BASED}~(colored dashed lines) seed model predictions. % for $\msfmp=5,50~\&~150$ and $\mh=3000~\&~10000$. Recall that the \texttt{GAS_BASED} simulations are run at $L_{\mathrm{max}}=12$ with $\seedmass=1.56\times10^3~M_\odot/h$, and the  \texttt{STOCHASTIC_MASS_ENV} simulations are run at $L_{\mathrm{max}}=11~\&~10$ and $\descendantseedmass=1.25\times10^4~\&~1\times10^5~M_{\odot}/h$, respectively. 
%laura: STOCHASTIC_MASS_ENV sims already defined
%For the \texttt{STOCHASTIC_MASS_ENV} simulations, we use our fiducial seed model parameters for the \textit{subhalo mass criterion} and \textit{subhalo environment criterion}, listed in Table \ref{double_power_law_table}. 
For $\descendantseedmass=1.25\times10^4~M
_{\odot}/h$~(Figure \ref{subhalo_vs_gas_seeding}), we calibrate models corresponding to %do this excercise for 
$[\mh,\msfmp=3000,50~\&~3000,150]$ and $[\mh,\msfmp=10000,5]$. We exclude the most lenient gas based seed parameters of $[\mh,\msfmp=3000,5]$, since it leads to a significant portion of $1.25\times10^4~M
_{\odot}/h$ descendants to assemble in galaxies that cannot be resolved in the $L_{\mathrm{max}}=11$ runs. For the remaining gas based seed parameters, the \texttt{STOCHASTIC_MASS_ENV} simulations well reproduce the \texttt{GAS_BASED} simulation predictions for the BH counts, two-point correlation functions and merger rates of $>1.25\times10^4~M_{\odot}/h$ BHs. 

For $\descendantseedmass=1\times10^5~M
_{\odot}/h$~(Figure \ref{subhalo_vs_gas_seeding_seed5.00}), we only do this exercise for the most lenient gas based seed models i.e. $[\mh,\msfmp=3000,5~\&~3000,50]$. This is because for the stricter gas based seed models, there are too few BHs produced overall. Here, the \texttt{STOCHASTIC_MASS_ENV} simulations well reproduce the counts of $>1\times10^5~M_{\odot}/h$ BHs at $z<13.1$ (wherein there is enough data to calibrate the slope $\alpha$; revisit Figure \ref{bFOF_at_assembly}, bottom row). For $z>13.1$, %recall that we assumed 
$\beta=0$ is assumed due to the absence of enough data points to perform any fitting; here, 
the \texttt{STOCHASTIC_MASS_ENV} seed model overestimates the number of $>1\times10^5~M_{\odot}/h$ BHs and their high-$z$ merger rates. % compared to the \texttt{GAS_BASED} seed model. We can see a similar overestimation in the merger rates at the highest redshifts~(rightmost panels). %A higher value of $\beta\sim0.1-0.3$ can bring the \texttt{STOCHASTIC_MASS_ENV} and \texttt{GAS_BASED} seed model predictions for the BH counts to agree at $z\geq13.1$.
Regardless, where enough data exist for robust calibration, these results imply that with a calibrated combination of \textit{galaxy mass criterion} and \textit{galaxy environment criterion}, the \texttt{STOCHASTIC_MASS_ENV}
simulations can well reproduce the \texttt{GAS_BASED} simulation predictions for a wide range of gas based seeding parameters.

Figures \ref{subhalo_vs_gas_seeding} and \ref{subhalo_vs_gas_seeding_seed5.00} also disentangle the impact of the various components of our final stochastic seed model, and they highlight the importance of each component in the successful representation of the gas based seed models. As seen previously, %in the absence of the \textit{subhalo environment criterion}~(grey solid lines), 
the \texttt{STOCHASTIC_MASS_ONLY} seed model overestimates the BH counts and merger rates %of $>1.25\times10^4~M_{\odot}/h$ and $>1\times10^5~M_{\odot}/h$ BHs 
by factors between $\sim2-5$. Next, when we assume zero scatter in the \textit{galaxy mass criterion} ($\Sigma=0$, black dashed lines), it further overestimates the BH counts and merger rates up to factors of $\sim1.5$~(grey solid versus black dashed lines). Finally, if we remove the redshift dependence in the \textit{galaxy mass criterion} and instead assume a constant threshold value~(thin dotted lines), the BH counts and merger rates monotonically increase with time. Not surprisingly, this is because such a model cannot capture the suppression of seed formation due to metal enrichment. 

Overall, we can clearly see that in order to represent our $L_{\rm max} = 12$ gas based seed models forming %lowest mass~(
$1.56\times10^3~M_{\odot}/h$ BH seeds in lower-resolution, larger-volume simulations, we need a stochastic seed model that %seeds 
places their resolvable descendant BHs~(ESDs) using the following two criteria

\begin{itemize}
\item A \textit{galaxy mass criterion} with a galaxy mass seeding threshold that is drawn from a distribution that evolves with redshift. The redshift evolution encodes the impact of star formation, halo growth and metal encrichment on seed formation. 

\item A \textit{galaxy environment criterion} that favors seeding within galaxies living in rich environments. %~(higher number of neighboring halos). 
This encodes the impact of the unresolved, hierarchical-merger-dominated growth of these seeds from $\seedmass$ to $\descendantseedmass$.

\end{itemize}
    
\subsection{Accounting for unresolved minor mergers}
\label{Modeling the BH growth of ESDs}

\begin{figure*}
\includegraphics[width=16 cm]{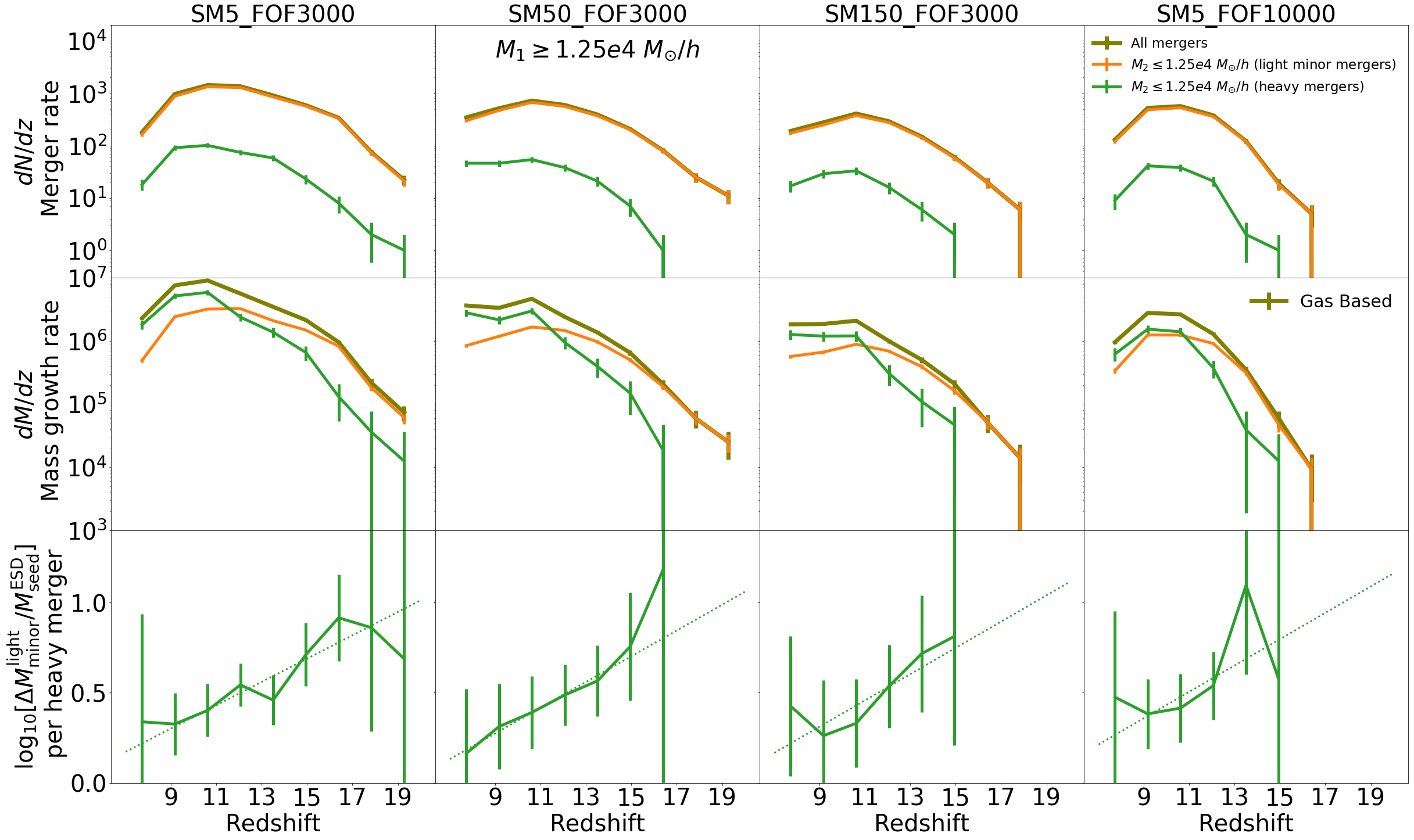}
\caption{Comparing the contributions of heavy mergers versus light minor mergers to the merger driven BH growth within the \texttt{GAS_BASED} suite. The green lines show heavy mergers where the masses of both primary and secondary BHs are $\geq1.25\times10^4~M_{\odot}/h$. The orange lines show the light minor mergers where the secondary BH mass is $<1.25\times10^4~M_{\odot}/h$ but the primary BH mass is $\geq1.25\times10^4~M_{\odot}/h$. The olive lines show the total contribution from both types of mergers i.e. all mergers with primary BHs $\geq1.25\times10^4~M_{\odot}/h$. The different columns show different gas based seed models. Middle panels show the mass growth rate due to mergers as a function of redshift, which is defined as the total mass of all merging secondary BHs per unit redshift. The light minor mergers show a dominant contribution at $z\gtrsim11$, whereas heavy mergers tend to be more prevalent at $z\lesssim11$. The bottom panels show the mass growth~($\Delta M^{\mathrm{light}}_{\mathrm{minor}}$) due to the light minor mergers between successive heavy mergers. This contribution needs to be explicitly included in simulations that use the stochastic seed models, to produce BH growth consistent with the \texttt{GAS_BASED} simulations.}
%\includegraphics[width=16 cm]{Minor_vs_major_mergers_seed5.00.png}\\
%\caption{Same as the previous figure, but for $1\times10^5~M_{\odot}/h$ BHs. Statistics are limited, but here too, the \textit{light mergers}~(secondary BH mass $<1\times10^5~M_{\odot}/h$) generally tend to dominate the BH growth at $z\gtrsim9$ and heavy mergers~(secondary BH mass $>1\times10^5~M_{\odot}/h$) tend to have a greater contribution at $z\lesssim9$.}
\label{major_vs_minor}
\end{figure*} 

We have thus far successfully built a new stochastic BH seed model that places ESDs which represent the %places extrapolated seeds of masses 
$\sim10^4-10^5~M_{\odot}/h$ %that represent 
descendants of %the lowest mass 
$\sim10^3~M_{\odot}/h$ DGBs in simulations that cannot directly resolve these lowest-mass BHs. %born within dense and metal poor gas. 
In this section, we %will focus on modeling 
model the subsequent growth of these ESDs. To do so, we must first account for one additional contribution to their growth: unresolved minor mergers. 

Recall %again 
from \cite{2021MNRAS.507.2012B} that the earliest growth of these $\sim10^3~M_{\odot}/h$ DGBs is completely driven by BH mergers, with negligible contribution from gas accretion. For our present purposes, these BH mergers can be classified into three types:
\begin{itemize}
\item \textit{Heavy mergers:} In these mergers, both the primary and secondary black holes~(with masses $M_1$ and $M_2$ respectively) are greater than the mass of the ESDs~($M_1>M_2>\descendantseedmass$). Therefore, these mergers will be fully resolvable within \texttt{STOCHASTIC_MASS_ENV} simulations.   
\item \textit{Light major mergers:} In these mergers, both the primary and secondary black holes are less massive than the ESDs~($\seedmass<M_2<M_1<\descendantseedmass$). %Therefore, these 
These mergers cannot be resolved in \texttt{STOCHASTIC_MASS_ENV} simulations. % that use the subhalo-based stochastic seed models. Note however 
However, these are the mergers that lead to the initial assembly of the descendants represented by the ESDs, such that their contribution to BH assembly %. Threfore, the contribution from these mergers 
is already implicitly captured within the %different components of 
stochastic seed model. %, particularly the \textit{subhalo mass criterion} and the \textit{subhalo environment criterion}. 
\item \textit{Light minor mergers:} In these mergers, the primary black hole is more massive than the ESD mass, but %not 
the secondary black hole is not~($M_1 > \descendantseedmass$ \& $\seedmass<M_2<\descendantseedmass$). These mergers %also 
cannot be resolved in \texttt{STOCHASTIC_MASS_ENV} simulations, % using subhalo-based seed models. Additionally, their contributions are neither 
and their contributions to BH mass assembly cannot be captured by the \textit{galaxy mass criterion} or the \textit{galaxy environment criterion}. Therefore, we must modify our prescription to explicitly add their contribution to the growth of the ESDs.
\end{itemize}

We first determine the contribution of light minor mergers within the \texttt{GAS_BASED} simulations. Here we only show the results for $\descendantseedmass=1.25\times10^4~M_{\odot}/h$, since there are too few $1\times10^5~M_{\odot}$ BHs formed in the \texttt{GAS_BASED} simulations to robustly perform this analysis for the latter. %$\descendantseedmass=1\times10^5~M_{\odot}/h$. To that end, t
The light minor mergers are thus defined to have %be having 
$M_1>1.25\times10^4~M_{\odot}/h$ and $1.56\times10^3<M_2<1.25\times10^4~M_{\odot}/h$, and heavy mergers are defined to be those with $M_1>M_2>1.25\times10^4~M_{\odot}/h$. In Figure \ref{major_vs_minor}, we compare the contributions of the light minor mergers and heavy mergers to the %merger-driven BH 
growth of $>1.25\times10^4~M_{\odot}/h$ BHs in the \texttt{GAS_BASED} simulations. The light minor mergers are $\sim30$ times more frequent than the heavy mergers~(top row); this is simply due to higher overall number of %$1.56\times10^3<
$M_{\rm BH}<1.25\times10^4~M_{\odot}/h$ BHs compared to $M_{bh}>1.25\times10^4~M_{\odot}/h$ BHs. When we compare the mass growth contributed by light minor mergers versus heavy mergers~(middle row), we find that the light minor mergers dominate at the highest redshifts~($z\sim15-19$). % where the earliest merger events are happening. With time, 
As BH growth proceeds over time, the mass growth contributed by heavy mergers increases and eventually exceeds that of the light minor mergers at $z\lesssim12$, even though the overall merger rates are still dominated by light minor mergers. This is because the masses of the BHs involved in the heavy mergers continue to increase with time. 
Eventually, when new DGB formation is strongly suppressed by metal enrichment, the %We can expect the 
mass growth due to the light minor mergers becomes small. %to become negligibly small at low enough redshifts, as the formation of new seeds is suppressed by metal enrichment; 
We clearly see these trends in the third row of Figure \ref{major_vs_minor} which shows %, wherein we 
$\Delta M^{\mathrm{light}}_{\mathrm{minor}}$ defined as the %quantify 
amount of mass growth due to light minor mergers between successive \textit{heavy merger} events. %~(denoted by $\Delta M^{\mathrm{light}}_{\mathrm{minor}}$). 
$\Delta M^{\mathrm{light}}_{\mathrm{minor}}$ monotonically decreases with redshift and its evolution is reasonably well fit by power laws. %~(dotted green lines in the bottom row of Figure \ref{major_vs_minor}). In the following paragraph, we describe how these power-law fits are used to include the missing contribution from light minor mergers in simulations that use the subhalo-based stochastic seed models. %We do note that all these   $\descendantseedmass=1\times10^5~M_{\odot}/h$, albeit the results are not as statistically robust. 

\begin{figure*}
\includegraphics[width=16 cm]{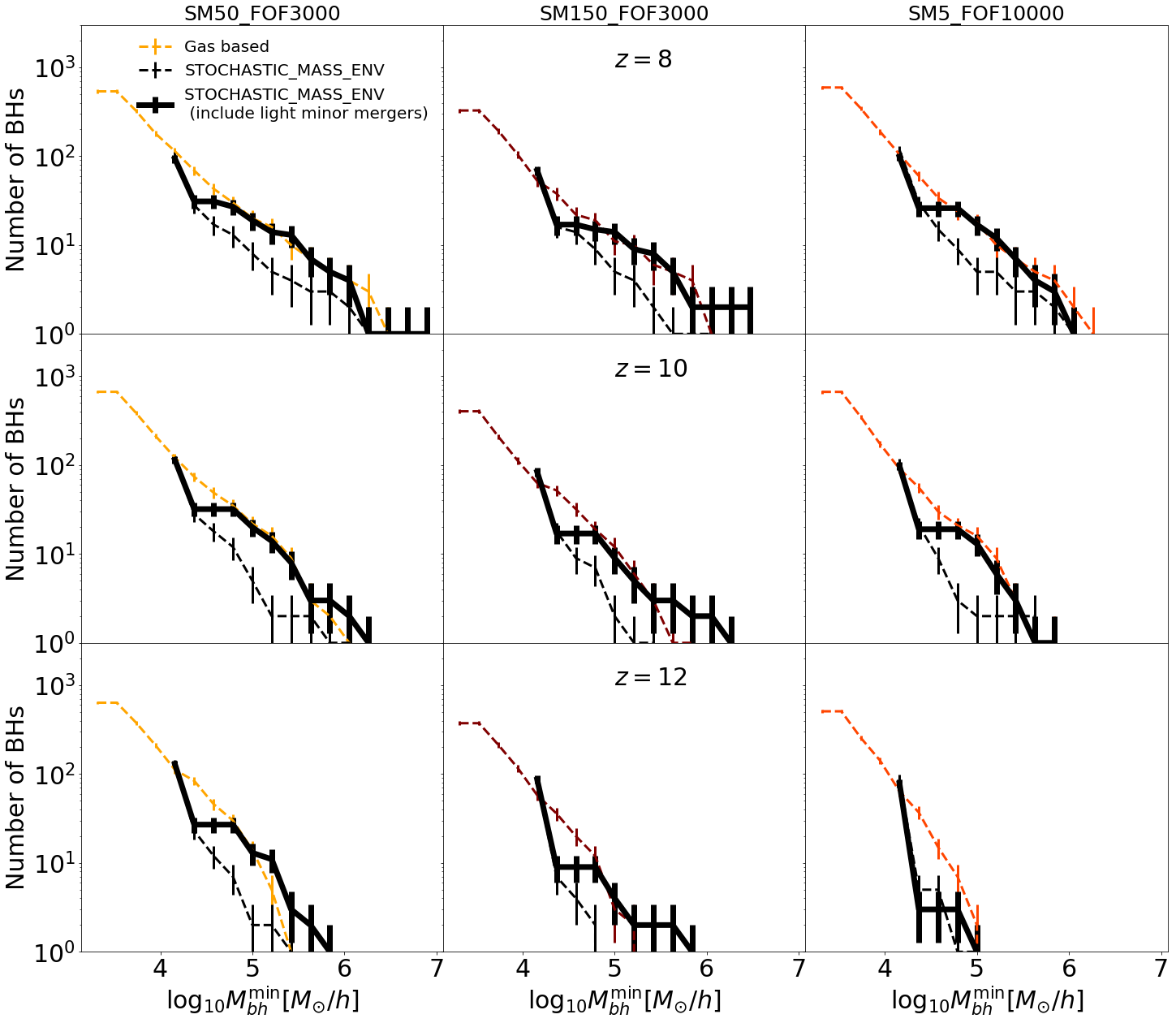}
\caption{Comparison of the cumulative mass functions~(i.e. the number of BHs above a minimum BH mass threshold $M^{\mathrm{min}}_{bh}$) between the \texttt{GAS_BASED}~(colored lines) and \texttt{STOCHASTIC_MASS_ENV}~(black lines) simulations. The top, middle and bottom rows show $z=8$, 10 and 12, respectively. The black dashed and solid lines show the \texttt{STOCHASTIC_MASS_ENV} predictions with and without the explicit inclusion of the contribution from the unresolved light minor mergers. %We can see that 
Without the light minor mergers, the \texttt{STOCHASTIC_MASS_ENV} BH mass functions are significantly steeper than in the \texttt{GAS_BASED} simulations. After including the contribution from the \textit{unresolved light mergers}, % using the power-law fits to $M^{\mathrm{light}}_{\mathrm{minor}}$ shown in the lowermost panels of Figure \ref{major_vs_minor}, 
the \texttt{STOCHASTIC_MASS_ENV} simulations are able to achieve reasonable agreement with the BH mass functions predicted by the \texttt{GAS_BASED} simulations.}
\label{mass_functions}
\end{figure*}

%In Figure \ref{mass_functions}, we compare the mass function predictions between the \texttt{GAS_BASED} simulations versus their corresponding calibrated \texttt{STOCHASTIC_MASS_ENV} simulations with $\descendantseedmass=1.25\times10^4~M_{\odot}/h$. We can clearly see that the \texttt{STOCHASTIC_MASS_ENV} simulations predict significantly steeper mass functions compared to the \texttt{GAS_BASED} simulations~(colored versus solid dashed lines). This is a consequence of slower BH growth in the \texttt{STOCHASTIC_MASS_ENV} simulations due to the absence of unresolved light minor mergers. We then rerun the \texttt{STOCHASTIC_MASS_ENV} simulations with a simple prescription to include the contribution of the light minor mergers~(solid black lines). We do this using
We use the power law fits of $\Delta M^{\mathrm{light}}_{\mathrm{minor}}$ (shown in the last row of Figure \ref{major_vs_minor}) to determine the missing BH growth contribution from light minor mergers. More specifically, for each %occurrence of 
heavy merger event in a %during the 
\texttt{STOCHASTIC_MASS_ENV} simulation, we added extra mass growth of $\Delta M^{\mathrm{light}}_{\mathrm{minor}}$ due to light minor mergers, calculated based on these power law fits. %We find that  
Figure \ref{mass_functions} shows that it is only after the inclusion of these unresolved light minor mergers, we achieve reasonable agreement between the BH mass functions predicted by the \texttt{GAS_BASED} and the \texttt{STOCHASTIC_MASS_ENV} simulations~(colored dashed lines versus solid black lines). Note that at masses between $\descendantseedmass$ and $2\descendantseedmass$, the \texttt{STOCHASTIC_MASS_ENV} simulations will inevitably continue to slightly underpredict the mass functions. This is because within our prescription, the contribution from light minor mergers does not occur until %starts to get included only after 
the first heavy merger event %occurs among 
between the ESDs.

\section{Summary and Conclusions}
\label{Summary and Conclusions}
In this work, we tackle one of the longstanding challenges in modeling BH seeds in cosmological hydrodynamic simulations: how do we simulate low mass~($\lesssim10^3~M_{\odot}$) seeds in simulations that cannot directly resolve them? We address this challenge by building a new sub-grid seed model that can stochastically seed the smallest resolvable descendants of low mass seeds in lower-resolution simulations~(hereafter referred to as ``stochastic seed model"). Our new seed model is motivated and calibrated based on highest resolution simulations that directly resolve the low mass seeds. With this new tool, we have bridged a critical gap between high-resolution simulations that directly resolves low mass seeds, and larger-volume simulations that can generate sufficient numbers of BHs to compare against observational measurements. This paves the way for making statistically robust predictions for signatures of low-mass seeds using cosmological hydrodynamic simulations, which is a crucial step in preparation for the wealth of observations with ongoing JWST, as well as upcoming facilities such as LISA.

The core objective of this work has been to determine the key ingredients needed to construct such a seed model. To do this, we study the growth of the lowest mass $1.56\times10^3~M_{\odot}/h$ seeds that were fully resolved using highest resolution zoom simulations. These seeds are placed in halos containing gas that is simultaneously star forming as well as metal poor~($<10^{-4} Z_{\odot}$), consistent with proposed low mass seeding candidates such as Pop III stellar remnants. We trace the growth of these $1.56\times10^3~M_{\odot}/h$ seeds until they assemble descendants with masses that are close to different possible gas mass resolutions~($\sim10^4-10^6~M_{\odot}$) expected in larger cosmological volumes. We characterize the environments in which these descendants assemble; for e.g. they assemble in halos with masses ranging from $\sim10^7-10^9~M_{\odot}$. The results are used to build our stochastic seed model that directly seeds these descendants in lower resolution simulations. To distinguish against the \textit{actual} $1.56\times10^3~M_{\odot}/h$ seeds, we refer to the ``seeds" formed by the stochastic seed model as ``extrapolated seed descendants" or ESDs~(with mass $\descendantseedmass$). We consider $1.25\times10^4~\&~1\times10^5~M_{\odot}/h$ ESDs that are aimed at faithfully representing the descendants of  $1.56\times10^3~M_{\odot}/h$ seeds born out of star forming and metal poor gas. Specifically, we explore a wide range of stochastic seed models on lower resolution versions of our zoom region, and determine the crucial ingredients required to reproduce the results of the highest resolution zoom simulations that explicitly resolve the $1.56\times10^3~M_{\odot}/h$ seeds. Following are the key features of our new seed model: 

\begin{itemize}
\item We seed the ESDs in high-z (proto)galaxies which are bound substructures within high-z halos. Since halos can contain multiple galaxies, this naturally allows the placement of multiple ESDs per halo. This is important because even if $1.56\times10^3~M_{\odot}/h$ seeds are placed as one seed per halo, their subsequent hierarchical growth inevitably assembles multiple higher mass descendants within individual halos.
\item We introduce a \textit{galaxy mass criterion} which places the ESDs based on galaxy mass thresholds. These thresholds are stochastically drawn from galaxy mass (including DM, stars and gas) distributions wherein $1.25\times10^4~\&~1\times10^5~M_{\odot}/h$ BHs assemble from $1.56\times10^3~M_{\odot}/h$ seeds. We find that the \textit{galaxy mass criterion} effortlessly also replicates the baryonic properties of the galaxies at the time of assembly of the seed descendants, including stellar mass, SFRs, and gas metallicities. This is because, although $1.56\times10^3~M_{\odot}/h$ seeds form within halos exhibiting a bias towards lower metallicities in comparison to typical halos of similar masses, they undergo a transient phase characterized by rapid metal enrichment. As a result, the higher mass $1.25\times10^4~\&~1\times10^5~M_{\odot}/h$ descendants end up in unbiased halos with metallicities similar to halos with similar masses. The redshift dependence of the distributions underlying the galaxy mass thresholds capture the complex influence of processes such as halo growth, star formation and metal enrichment, on the formation of $1.56\times10^3~M_{\odot}/h$ seeds.

\item However, if our stochastic seed model only contains the \textit{galaxy mass criterion}, it underestimates the two-point clustering~(at scales of $0.01-0.1~\mathrm{Mpc}/h$) of $\geq1.25\times10^4~\&~1\times10^5~M_{\odot}/h$ BHs by factors of $\sim5$. At the same time, it overestimates the BH abundances and merger rates of $\geq1.25\times10^4~\&~1\times10^5~M_{\odot}/h$ BHs by factors up to $\sim5$. This is a direct consequence of the fact that in our highest resolution zooms, the $1.56\times10^3~M_{\odot}/h$ seeds grow primarily via BH-BH mergers. As a result, the assembly of the higher mass descendants is more efficient in galaxies with richer environments~(higher number of neighboring halos) with a more extensive merger history. This cannot be captured solely by the \textit{galaxy mass criterion}.  

\item To successfully capture the two-point clustering of the  $\geq1.25\times10^4~\&~1\times10^5~M_{\odot}/h$ descendant BHs, we introduce a \textit{galaxy environment criterion}, where we assign seeding probabilities less than unity for galaxies with $\leq1$ neighbors. By doing this, we preferentially place ESDs in richer environments, which enhances the two-point clustering. We demonstrate that by adding a \textit{galaxy-environment criterion} that is calibrated to produce the correct two-point clustering, our stochastic seed models can simultaneously also reproduce the BH abundances and merger rates of the $\geq1.25\times10^4~\&~1\times10^5~M_{\odot}/h$ BHs.

\item Lastly, the BH growth in our stochastic seed models is underestimated due to the absence of light minor mergers, defined as those involving a resolved primary~($M_1>\descendantseedmass$) but an unresolved secondary~($M_2 <\descendantseedmass$). We compute the contribution of these mergers from the highest resolution zooms that resolve the $1.56\times10^3~M_{\odot}/h$ seeds, and explicitly add them to the simulations that use the stochastic seed models. It is only after adding the contribution from light minor mergers, do our stochastic seed models achieve success in accurately reproducing the BH mass functions predicted by the highest resolution zooms.

\end{itemize}

Overall, our stochastic seed model requires three main seeding components to successfully represent low mass seeds in lower resolution-larger volume simulations: 1) a \textit{galaxy mass criterion}, 2) \textit{galaxy environment criterion}, and 3) inclusion of unresolved light minor mergers. In our upcoming companion paper~(Bhowmick et al. in prep), we apply these stochastic seed models to uniform volume cosmological simulations, and thereby make predictions that would be directly comparable to facilities such as JWST and LISA for different seeding scenarios.

The construction of our stochastic seed model essentially rests only on two important aspects of the formation of low mass seeds. First, these seeds are forming in regions which are already in the process of rapid metal enrichment, which is a natural consequence of seeding within star forming \& metal poor gas. Second, the BH growth is dominantly driven by BH-BH mergers. Therefore, our stochastic seed model could be tuned to represent \textit{any} low mass seeding scenario for which the foregoing assumptions hold true. These include scenarios beyond the ones we consider in this work. Furthermore, we can calibrate our stochastic seed model against any high resolution simulation run with different galaxy formation models or using different state-of-the-art numerical solvers such as \texttt{GADGET-4}~\citep{2021MNRAS.506.2871S}, \texttt{GIZMO}~\citep{2015MNRAS.450...53H} etc. %Our stochastic seed model could also be readily implemented within these different solvers. 
Lastly, a key advantage of our seed model is that it depends solely only on galaxy total mass~(which is dark matter dominated) and galaxy environment. Therefore, it can also be readily applied to DM only simulations as well as semi-analytic models that are typically much less expensive compared to full hydrodynamic simulations.  

In the near future, we shall test our stochastic seed model for their ability to represent low mass seeds when coupled with alternate accretion and dynamics models. For example, having a smaller scaling exponent between BH accretion rate and BH mass~(such as $\alpha=1/6$ for gravitational torque driven accretion model) may significantly enhance the role of gas accretion in the growth of low mass seeds at high redshifts. Similarly, having a more physically motivated BH dynamics prescription will likely impact the merger rates and change the relative importance of accretion versus mergers in driving BH growth. In such a case, we can envision requiring additional ingredient(s) in our stochastic seed model to capture the impact of unresolved accretion driven growth of low mass seeds, similar to how the galaxy environment criterion was needed to account for the impact of unresolved merger dominated BH growth.

Nevertheless, our new stochastic seed model offers a substantial improvement from existing cosmological simulations that have either relied on a threshold halo / stellar mass, or on poorly resolved gas properties for seeding. Unlike most of these currently used seed models, our models will allow us to represent low-mass seeds in cosmological simulations without the need to either explicitly resolve the seeds, or seed below the gas mass resolution of the simulation. Overall, this work is an important step towards the next generation of cosmological hydrodynamic simulations in terms of improved modeling of high redshift SMBHs, to finally understand their role in shaping high redshift galaxy evolution in the ongoing JWST and upcoming LISA era.

\section*{Acknowledgements}
LB acknowledges support from NSF award AST-1909933 and Cottrell Scholar Award \#27553 from the Research Corporation for Science Advancement.
PT acknowledges support from NSF-AST 2008490.
RW acknowledges funding of a Leibniz Junior Research Group (project number J131/2022).
\section*{Data availablity}
The underlying data used in this work shall be made available upon reasonable request to the corresponding author.

%\begin{figure*}

%\includegraphics[width=6 cm]{variance2.png}
%\includegraphics[width=6.3 cm]{variance2_Dndz.png}

%\end{figure*}
\bibliography{references}

\begin{thebibliography}{}
\makeatletter
\relax
\def\mn@urlcharsother{\let\do\@makeother \do\$\do\&\do\#\do\^\do\_\do\%\do\~}
\def\mn@doi{\begingroup\mn@urlcharsother \@ifnextchar [ {\mn@doi@}
  {\mn@doi@[]}}
\def\mn@doi@[#1]#2{\def\@tempa{#1}\ifx\@tempa\@empty \href
  {http://dx.doi.org/#2} {doi:#2}\else \href {http://dx.doi.org/#2} {#1}\fi
  \endgroup}
\def\mn@eprint#1#2{\mn@eprint@#1:#2::\@nil}
\def\mn@eprint@arXiv#1{\href {http://arxiv.org/abs/#1} {{\tt arXiv:#1}}}
\def\mn@eprint@dblp#1{\href {http://dblp.uni-trier.de/rec/bibtex/#1.xml}
  {dblp:#1}}
\def\mn@eprint@#1:#2:#3:#4\@nil{\def\@tempa {#1}\def\@tempb {#2}\def\@tempc
  {#3}\ifx \@tempc \@empty \let \@tempc \@tempb \let \@tempb \@tempa \fi \ifx
  \@tempb \@empty \def\@tempb {arXiv}\fi \@ifundefined
  {mn@eprint@\@tempb}{\@tempb:\@tempc}{\expandafter \expandafter \csname
  mn@eprint@\@tempb\endcsname \expandafter{\@tempc}}}

\bibitem[\protect\citeauthoryear{{Abbott} et~al.,}{{Abbott}
  et~al.}{2009}]{2009RPPh...72g6901A}
{Abbott} B.~P.,  et~al., 2009, \mn@doi [Reports on Progress in Physics]
  {10.1088/0034-4885/72/7/076901}, \href
  {https://ui.adsabs.harvard.edu/abs/2009RPPh...72g6901A} {72, 076901}

\bibitem[\protect\citeauthoryear{{Abbott} et~al.,}{{Abbott}
  et~al.}{2020}]{2020ApJ...900L..13A}
{Abbott} R.,  et~al., 2020, \mn@doi [\apjl] {10.3847/2041-8213/aba493}, \href
  {https://ui.adsabs.harvard.edu/abs/2020ApJ...900L..13A} {900, L13}

\bibitem[\protect\citeauthoryear{{Agazie} et~al.,}{{Agazie}
  et~al.}{2023}]{2023ApJ...951L...8A}
{Agazie} G.,  et~al., 2023, \mn@doi [\apjl] {10.3847/2041-8213/acdac6}, \href
  {https://ui.adsabs.harvard.edu/abs/2023ApJ...951L...8A} {951, L8}

\bibitem[\protect\citeauthoryear{{Amaro-Seoane} et~al.,}{{Amaro-Seoane}
  et~al.}{2017}]{2017arXiv170200786A}
{Amaro-Seoane} P.,  et~al., 2017, arXiv e-prints, \href
  {https://ui.adsabs.harvard.edu/abs/2017arXiv170200786A} {p. arXiv:1702.00786}

\bibitem[\protect\citeauthoryear{{Ba{\~n}ados} et~al.,}{{Ba{\~n}ados}
  et~al.}{2018}]{2018Natur.553..473B}
{Ba{\~n}ados} E.,  et~al., 2018, \mn@doi [\nat] {10.1038/nature25180}, \href
  {https://ui.adsabs.harvard.edu/abs/2018Natur.553..473B} {553, 473}

\bibitem[\protect\citeauthoryear{{Baker} et~al.,}{{Baker}
  et~al.}{2019}]{2019arXiv190706482B}
{Baker} J.,  et~al., 2019, arXiv e-prints, \href
  {https://ui.adsabs.harvard.edu/abs/2019arXiv190706482B} {p. arXiv:1907.06482}

\bibitem[\protect\citeauthoryear{{Barcons} et~al.,}{{Barcons}
  et~al.}{2017}]{2017AN....338..153B}
{Barcons} X.,  et~al., 2017, \mn@doi [Astronomische Nachrichten]
  {10.1002/asna.201713323}, \href
  {https://ui.adsabs.harvard.edu/abs/2017AN....338..153B} {338, 153}

\bibitem[\protect\citeauthoryear{{Barnes} \& {Hut}}{{Barnes} \&
  {Hut}}{1986}]{1986Natur.324..446B}
{Barnes} J.,  {Hut} P.,  1986, \mn@doi [\nat] {10.1038/324446a0}, \href
  {https://ui.adsabs.harvard.edu/abs/1986Natur.324..446B} {324, 446}

\bibitem[\protect\citeauthoryear{Bañados et~al.,}{Bañados
  et~al.}{2016}]{2016Banados}
Bañados E.,  et~al., 2016, \mn@doi [The Astrophysical Journal Supplement
  Series] {10.3847/0067-0049/227/1/11}, 227, 11

\bibitem[\protect\citeauthoryear{{Begelman} \& {Silk}}{{Begelman} \&
  {Silk}}{2023}]{2023arXiv230519081B}
{Begelman} M.~C.,  {Silk} J.,  2023, \mn@doi [arXiv e-prints]
  {10.48550/arXiv.2305.19081}, \href
  {https://ui.adsabs.harvard.edu/abs/2023arXiv230519081B} {p. arXiv:2305.19081}

\bibitem[\protect\citeauthoryear{{Begelman}, {Volonteri}  \& {Rees}}{{Begelman}
  et~al.}{2006}]{2006MNRAS.370..289B}
{Begelman} M.~C.,  {Volonteri} M.,   {Rees} M.~J.,  2006, \mn@doi [\mnras]
  {10.1111/j.1365-2966.2006.10467.x}, \href
  {https://ui.adsabs.harvard.edu/abs/2006MNRAS.370..289B} {370, 289}

\bibitem[\protect\citeauthoryear{{Bhowmick} et~al.,}{{Bhowmick}
  et~al.}{2021}]{2021MNRAS.507.2012B}
{Bhowmick} A.~K.,  et~al., 2021, \mn@doi [\mnras] {10.1093/mnras/stab2204},
  \href {https://ui.adsabs.harvard.edu/abs/2021MNRAS.507.2012B} {507, 2012}

\bibitem[\protect\citeauthoryear{{Bhowmick}, {Blecha}, {Torrey}, {Kelley},
  {Vogelsberger}, {Nelson}, {Weinberger}  \& {Hernquist}}{{Bhowmick}
  et~al.}{2022a}]{2022MNRAS.510..177B}
{Bhowmick} A.~K.,  {Blecha} L.,  {Torrey} P.,  {Kelley} L.~Z.,  {Vogelsberger}
  M.,  {Nelson} D.,  {Weinberger} R.,   {Hernquist} L.,  2022a, \mn@doi
  [\mnras] {10.1093/mnras/stab3439}, \href
  {https://ui.adsabs.harvard.edu/abs/2022MNRAS.510..177B} {510, 177}

\bibitem[\protect\citeauthoryear{{Bhowmick} et~al.,}{{Bhowmick}
  et~al.}{2022b}]{2022MNRAS.516..138B}
{Bhowmick} A.~K.,  et~al., 2022b, \mn@doi [\mnras] {10.1093/mnras/stac2238},
  \href {https://ui.adsabs.harvard.edu/abs/2022MNRAS.516..138B} {516, 138}

\bibitem[\protect\citeauthoryear{{Bromm} \& {Loeb}}{{Bromm} \&
  {Loeb}}{2003}]{2003ApJ...596...34B}
{Bromm} V.,  {Loeb} A.,  2003, \mn@doi [\apj] {10.1086/377529}, \href
  {https://ui.adsabs.harvard.edu/abs/2003ApJ...596...34B} {596, 34}

\bibitem[\protect\citeauthoryear{{Cann}, {Satyapal}, {Abel}, {Ricci},
  {Secrest}, {Blecha}  \& {Gliozzi}}{{Cann} et~al.}{2018}]{2018ApJ...861..142C}
{Cann} J.~M.,  {Satyapal} S.,  {Abel} N.~P.,  {Ricci} C.,  {Secrest} N.~J.,
  {Blecha} L.,   {Gliozzi} M.,  2018, \mn@doi [\apj]
  {10.3847/1538-4357/aac64a}, \href
  {https://ui.adsabs.harvard.edu/abs/2018ApJ...861..142C} {861, 142}

\bibitem[\protect\citeauthoryear{{Das}, {Schleicher}, {Basu}  \&
  {Boekholt}}{{Das} et~al.}{2021a}]{2021MNRAS.tmp.1381D}
{Das} A.,  {Schleicher} D. R.~G.,  {Basu} S.,   {Boekholt} T. C.~N.,  2021a,
  \mn@doi [\mnras] {10.1093/mnras/stab1428}, \href
  {https://ui.adsabs.harvard.edu/abs/2021MNRAS.tmp.1381D} {}

\bibitem[\protect\citeauthoryear{{Das}, {Schleicher}, {Leigh}  \&
  {Boekholt}}{{Das} et~al.}{2021b}]{2021MNRAS.503.1051D}
{Das} A.,  {Schleicher} D. R.~G.,  {Leigh} N. W.~C.,   {Boekholt} T. C.~N.,
  2021b, \mn@doi [\mnras] {10.1093/mnras/stab402}, \href
  {https://ui.adsabs.harvard.edu/abs/2021MNRAS.503.1051D} {503, 1051}

\bibitem[\protect\citeauthoryear{{Davies}, {Miller}  \& {Bellovary}}{{Davies}
  et~al.}{2011}]{2011ApJ...740L..42D}
{Davies} M.~B.,  {Miller} M.~C.,   {Bellovary} J.~M.,  2011, \mn@doi [\apjl]
  {10.1088/2041-8205/740/2/L42}, \href
  {https://ui.adsabs.harvard.edu/abs/2011ApJ...740L..42D} {740, L42}

\bibitem[\protect\citeauthoryear{{Davis}, {Efstathiou}, {Frenk}  \&
  {White}}{{Davis} et~al.}{1985}]{1985ApJ...292..371D}
{Davis} M.,  {Efstathiou} G.,  {Frenk} C.~S.,   {White} S.~D.~M.,  1985,
  \mn@doi [\apj] {10.1086/163168}, \href
  {https://ui.adsabs.harvard.edu/abs/1985ApJ...292..371D} {292, 371}

\bibitem[\protect\citeauthoryear{{Di Matteo}, {Khandai}, {DeGraf}, {Feng},
  {Croft}, {Lopez}  \& {Springel}}{{Di Matteo}
  et~al.}{2012}]{2012ApJ...745L..29D}
{Di Matteo} T.,  {Khandai} N.,  {DeGraf} C.,  {Feng} Y.,  {Croft} R.~A.~C.,
  {Lopez} J.,   {Springel} V.,  2012, \mn@doi [\apjl]
  {10.1088/2041-8205/745/2/L29}, \href
  {https://ui.adsabs.harvard.edu/abs/2012ApJ...745L..29D} {745, L29}

\bibitem[\protect\citeauthoryear{{Donnari} et~al.,}{{Donnari}
  et~al.}{2019}]{2019MNRAS.485.4817D}
{Donnari} M.,  et~al., 2019, \mn@doi [\mnras] {10.1093/mnras/stz712}, \href
  {https://ui.adsabs.harvard.edu/abs/2019MNRAS.485.4817D} {485, 4817}

\bibitem[\protect\citeauthoryear{{Dubois}, {Peirani}, {Pichon}, {Devriendt},
  {Gavazzi}, {Welker}  \& {Volonteri}}{{Dubois}
  et~al.}{2016}]{2016MNRAS.463.3948D}
{Dubois} Y.,  {Peirani} S.,  {Pichon} C.,  {Devriendt} J.,  {Gavazzi} R.,
  {Welker} C.,   {Volonteri} M.,  2016, \mn@doi [\mnras]
  {10.1093/mnras/stw2265}, \href
  {https://ui.adsabs.harvard.edu/abs/2016MNRAS.463.3948D} {463, 3948}

\bibitem[\protect\citeauthoryear{{Fan} et~al.,}{{Fan}
  et~al.}{2001}]{2001AJ....122.2833F}
{Fan} X.,  et~al., 2001, \mn@doi [\aj] {10.1086/324111}, \href
  {https://ui.adsabs.harvard.edu/abs/2001AJ....122.2833F} {122, 2833}

\bibitem[\protect\citeauthoryear{{Feng}, {Di-Matteo}, {Croft}, {Bird},
  {Battaglia}  \& {Wilkins}}{{Feng} et~al.}{2016}]{2016MNRAS.455.2778F}
{Feng} Y.,  {Di-Matteo} T.,  {Croft} R.~A.,  {Bird} S.,  {Battaglia} N.,
  {Wilkins} S.,  2016, \mn@doi [\mnras] {10.1093/mnras/stv2484}, \href
  {https://ui.adsabs.harvard.edu/abs/2016MNRAS.455.2778F} {455, 2778}

\bibitem[\protect\citeauthoryear{{Fryer}, {Woosley}  \& {Heger}}{{Fryer}
  et~al.}{2001}]{2001ApJ...550..372F}
{Fryer} C.~L.,  {Woosley} S.~E.,   {Heger} A.,  2001, \mn@doi [\apj]
  {10.1086/319719}, \href
  {https://ui.adsabs.harvard.edu/abs/2001ApJ...550..372F} {550, 372}

\bibitem[\protect\citeauthoryear{{Gardner} et~al.,}{{Gardner}
  et~al.}{2006}]{2006SSRv..123..485G}
{Gardner} J.~P.,  et~al., 2006, \mn@doi [\ssr] {10.1007/s11214-006-8315-7},
  \href {https://ui.adsabs.harvard.edu/abs/2006SSRv..123..485G} {123, 485}

\bibitem[\protect\citeauthoryear{{Genel} et~al.,}{{Genel}
  et~al.}{2018}]{2018MNRAS.474.3976G}
{Genel} S.,  et~al., 2018, \mn@doi [\mnras] {10.1093/mnras/stx3078}, \href
  {https://ui.adsabs.harvard.edu/abs/2018MNRAS.474.3976G} {474, 3976}

\bibitem[\protect\citeauthoryear{{Habouzit}, {Pisani}, {Goulding}, {Dubois},
  {Somerville}  \& {Greene}}{{Habouzit} et~al.}{2020}]{2020MNRAS.493..899H}
{Habouzit} M.,  {Pisani} A.,  {Goulding} A.,  {Dubois} Y.,  {Somerville} R.~S.,
    {Greene} J.~E.,  2020, \mn@doi [\mnras] {10.1093/mnras/staa219}, \href
  {https://ui.adsabs.harvard.edu/abs/2020MNRAS.493..899H} {493, 899}

\bibitem[\protect\citeauthoryear{{Habouzit} et~al.,}{{Habouzit}
  et~al.}{2021}]{2021MNRAS.503.1940H}
{Habouzit} M.,  et~al., 2021, \mn@doi [\mnras] {10.1093/mnras/stab496}, \href
  {https://ui.adsabs.harvard.edu/abs/2021MNRAS.503.1940H} {503, 1940}

\bibitem[\protect\citeauthoryear{{Hahn} \& {Abel}}{{Hahn} \&
  {Abel}}{2011}]{2011MNRAS.415.2101H}
{Hahn} O.,  {Abel} T.,  2011, \mn@doi [\mnras]
  {10.1111/j.1365-2966.2011.18820.x}, \href
  {https://ui.adsabs.harvard.edu/abs/2011MNRAS.415.2101H} {415, 2101}

\bibitem[\protect\citeauthoryear{{Harikane} et~al.,}{{Harikane}
  et~al.}{2023}]{2023arXiv230311946H}
{Harikane} Y.,  et~al., 2023, \mn@doi [arXiv e-prints]
  {10.48550/arXiv.2303.11946}, \href
  {https://ui.adsabs.harvard.edu/abs/2023arXiv230311946H} {p. arXiv:2303.11946}

\bibitem[\protect\citeauthoryear{{Hopkins}}{{Hopkins}}{2015}]{2015MNRAS.450...53H}
{Hopkins} P.~F.,  2015, \mn@doi [\mnras] {10.1093/mnras/stv195}, \href
  {https://ui.adsabs.harvard.edu/abs/2015MNRAS.450...53H} {450, 53}

\bibitem[\protect\citeauthoryear{{Inayoshi}, {Onoue}, {Sugahara}, {Inoue}  \&
  {Ho}}{{Inayoshi} et~al.}{2022}]{2022ApJ...931L..25I}
{Inayoshi} K.,  {Onoue} M.,  {Sugahara} Y.,  {Inoue} A.~K.,   {Ho} L.~C.,
  2022, \mn@doi [\apjl] {10.3847/2041-8213/ac6f01}, \href
  {https://ui.adsabs.harvard.edu/abs/2022ApJ...931L..25I} {931, L25}

\bibitem[\protect\citeauthoryear{{Jiang} et~al.,}{{Jiang}
  et~al.}{2016}]{2016ApJ...833..222J}
{Jiang} L.,  et~al., 2016, \mn@doi [\apj] {10.3847/1538-4357/833/2/222}, \href
  {https://ui.adsabs.harvard.edu/abs/2016ApJ...833..222J} {833, 222}

\bibitem[\protect\citeauthoryear{{Kaviraj} et~al.,}{{Kaviraj}
  et~al.}{2017}]{2017MNRAS.467.4739K}
{Kaviraj} S.,  et~al., 2017, \mn@doi [\mnras] {10.1093/mnras/stx126}, \href
  {https://ui.adsabs.harvard.edu/abs/2017MNRAS.467.4739K} {467, 4739}

\bibitem[\protect\citeauthoryear{{Khandai}, {Di Matteo}, {Croft}, {Wilkins},
  {Feng}, {Tucker}, {DeGraf}  \& {Liu}}{{Khandai}
  et~al.}{2015}]{2015MNRAS.450.1349K}
{Khandai} N.,  {Di Matteo} T.,  {Croft} R.,  {Wilkins} S.,  {Feng} Y.,
  {Tucker} E.,  {DeGraf} C.,   {Liu} M.-S.,  2015, \mn@doi [\mnras]
  {10.1093/mnras/stv627}, \href
  {https://ui.adsabs.harvard.edu/abs/2015MNRAS.450.1349K} {450, 1349}

\bibitem[\protect\citeauthoryear{{Kroupa}, {Subr}, {Jerabkova}  \&
  {Wang}}{{Kroupa} et~al.}{2020}]{2020MNRAS.498.5652K}
{Kroupa} P.,  {Subr} L.,  {Jerabkova} T.,   {Wang} L.,  2020, \mn@doi [\mnras]
  {10.1093/mnras/staa2276}, \href
  {https://ui.adsabs.harvard.edu/abs/2020MNRAS.498.5652K} {498, 5652}

\bibitem[\protect\citeauthoryear{{Larson} et~al.,}{{Larson}
  et~al.}{2023}]{2023arXiv230308918L}
{Larson} R.~L.,  et~al., 2023, \mn@doi [arXiv e-prints]
  {10.48550/arXiv.2303.08918}, \href
  {https://ui.adsabs.harvard.edu/abs/2023arXiv230308918L} {p. arXiv:2303.08918}

\bibitem[\protect\citeauthoryear{{Latif}, {Schleicher}  \& {Hartwig}}{{Latif}
  et~al.}{2016}]{2016MNRAS.458..233L}
{Latif} M.~A.,  {Schleicher} D.~R.~G.,   {Hartwig} T.,  2016, \mn@doi [\mnras]
  {10.1093/mnras/stw297}, \href
  {https://ui.adsabs.harvard.edu/abs/2016MNRAS.458..233L} {458, 233}

\bibitem[\protect\citeauthoryear{{Luo}, {Ardaneh}, {Shlosman}, {Nagamine},
  {Wise}  \& {Begelman}}{{Luo} et~al.}{2018}]{2018MNRAS.476.3523L}
{Luo} Y.,  {Ardaneh} K.,  {Shlosman} I.,  {Nagamine} K.,  {Wise} J.~H.,
  {Begelman} M.~C.,  2018, \mn@doi [\mnras] {10.1093/mnras/sty362}, \href
  {https://ui.adsabs.harvard.edu/abs/2018MNRAS.476.3523L} {476, 3523}

\bibitem[\protect\citeauthoryear{{Luo}, {Shlosman}, {Nagamine}  \&
  {Fang}}{{Luo} et~al.}{2020}]{2020MNRAS.492.4917L}
{Luo} Y.,  {Shlosman} I.,  {Nagamine} K.,   {Fang} T.,  2020, \mn@doi [\mnras]
  {10.1093/mnras/staa153}, \href
  {https://ui.adsabs.harvard.edu/abs/2020MNRAS.492.4917L} {492, 4917}

\bibitem[\protect\citeauthoryear{{Lupi}, {Colpi}, {Devecchi}, {Galanti}  \&
  {Volonteri}}{{Lupi} et~al.}{2014}]{2014MNRAS.442.3616L}
{Lupi} A.,  {Colpi} M.,  {Devecchi} B.,  {Galanti} G.,   {Volonteri} M.,  2014,
  \mn@doi [\mnras] {10.1093/mnras/stu1120}, \href
  {https://ui.adsabs.harvard.edu/abs/2014MNRAS.442.3616L} {442, 3616}

\bibitem[\protect\citeauthoryear{{Ma}, {Hopkins}, {Ma},
  {Angl{\'e}s-Alc{\'a}zar}, {Faucher-Gigu{\`e}re}  \& {Kelley}}{{Ma}
  et~al.}{2021}]{2021MNRAS.508.1973M}
{Ma} L.,  {Hopkins} P.~F.,  {Ma} X.,  {Angl{\'e}s-Alc{\'a}zar} D.,
  {Faucher-Gigu{\`e}re} C.-A.,   {Kelley} L.~Z.,  2021, \mn@doi [\mnras]
  {10.1093/mnras/stab2713}, \href
  {https://ui.adsabs.harvard.edu/abs/2021MNRAS.508.1973M} {508, 1973}

\bibitem[\protect\citeauthoryear{{Madau} \& {Rees}}{{Madau} \&
  {Rees}}{2001}]{2001ApJ...551L..27M}
{Madau} P.,  {Rees} M.~J.,  2001, \mn@doi [\apjl] {10.1086/319848}, \href
  {https://ui.adsabs.harvard.edu/abs/2001ApJ...551L..27M} {551, L27}

\bibitem[\protect\citeauthoryear{{Maiolino} et~al.,}{{Maiolino}
  et~al.}{2023}]{2023arXiv230512492M}
{Maiolino} R.,  et~al., 2023, \mn@doi [arXiv e-prints]
  {10.48550/arXiv.2305.12492}, \href
  {https://ui.adsabs.harvard.edu/abs/2023arXiv230512492M} {p. arXiv:2305.12492}

\bibitem[\protect\citeauthoryear{{Marinacci} et~al.,}{{Marinacci}
  et~al.}{2018}]{2018MNRAS.480.5113M}
{Marinacci} F.,  et~al., 2018, \mn@doi [\mnras] {10.1093/mnras/sty2206}, \href
  {https://ui.adsabs.harvard.edu/abs/2018MNRAS.480.5113M} {480, 5113}

\bibitem[\protect\citeauthoryear{{Matsuoka} et~al.,}{{Matsuoka}
  et~al.}{2018}]{2018ApJS..237....5M}
{Matsuoka} Y.,  et~al., 2018, \mn@doi [\apjs] {10.3847/1538-4365/aac724}, \href
  {https://ui.adsabs.harvard.edu/abs/2018ApJS..237....5M} {237, 5}

\bibitem[\protect\citeauthoryear{{Matsuoka} et~al.,}{{Matsuoka}
  et~al.}{2019}]{2019ApJ...872L...2M}
{Matsuoka} Y.,  et~al., 2019, \mn@doi [\apjl] {10.3847/2041-8213/ab0216}, \href
  {https://ui.adsabs.harvard.edu/abs/2019ApJ...872L...2M} {872, L2}

\bibitem[\protect\citeauthoryear{{Mayer}, {Capelo}, {Zwick}  \& {Di
  Matteo}}{{Mayer} et~al.}{2023}]{2023arXiv230402066M}
{Mayer} L.,  {Capelo} P.~R.,  {Zwick} L.,   {Di Matteo} T.,  2023, \mn@doi
  [arXiv e-prints] {10.48550/arXiv.2304.02066}, \href
  {https://ui.adsabs.harvard.edu/abs/2023arXiv230402066M} {p. arXiv:2304.02066}

\bibitem[\protect\citeauthoryear{{Mortlock} et~al.,}{{Mortlock}
  et~al.}{2011}]{2011Natur.474..616M}
{Mortlock} D.~J.,  et~al., 2011, \mn@doi [\nat] {10.1038/nature10159}, \href
  {https://ui.adsabs.harvard.edu/abs/2011Natur.474..616M} {474, 616}

\bibitem[\protect\citeauthoryear{{Mushotzky} et~al.,}{{Mushotzky}
  et~al.}{2019}]{2019BAAS...51g.107M}
{Mushotzky} R.,  et~al., 2019, in Bulletin of the American Astronomical
  Society. p.~107 (\mn@eprint {arXiv} {1903.04083}),
  \mn@doi{10.48550/arXiv.1903.04083}

\bibitem[\protect\citeauthoryear{{Naiman} et~al.,}{{Naiman}
  et~al.}{2018}]{2018MNRAS.477.1206N}
{Naiman} J.~P.,  et~al., 2018, \mn@doi [\mnras] {10.1093/mnras/sty618}, \href
  {https://ui.adsabs.harvard.edu/abs/2018MNRAS.477.1206N} {477, 1206}

\bibitem[\protect\citeauthoryear{{Natarajan}, {Pacucci}, {Ferrara}, {Agarwal},
  {Ricarte}, {Zackrisson}  \& {Cappelluti}}{{Natarajan}
  et~al.}{2017}]{2017ApJ...838..117N}
{Natarajan} P.,  {Pacucci} F.,  {Ferrara} A.,  {Agarwal} B.,  {Ricarte} A.,
  {Zackrisson} E.,   {Cappelluti} N.,  2017, \mn@doi [\apj]
  {10.3847/1538-4357/aa6330}, \href
  {https://ui.adsabs.harvard.edu/abs/2017ApJ...838..117N} {838, 117}

\bibitem[\protect\citeauthoryear{{Nelson} et~al.,}{{Nelson}
  et~al.}{2018}]{2018MNRAS.475..624N}
{Nelson} D.,  et~al., 2018, \mn@doi [\mnras] {10.1093/mnras/stx3040}, \href
  {https://ui.adsabs.harvard.edu/abs/2018MNRAS.475..624N} {475, 624}

\bibitem[\protect\citeauthoryear{{Nelson} et~al.,}{{Nelson}
  et~al.}{2019a}]{2019ComAC...6....2N}
{Nelson} D.,  et~al., 2019a, \mn@doi [Computational Astrophysics and Cosmology]
  {10.1186/s40668-019-0028-x}, \href
  {https://ui.adsabs.harvard.edu/abs/2019ComAC...6....2N} {6, 2}

\bibitem[\protect\citeauthoryear{{Nelson} et~al.,}{{Nelson}
  et~al.}{2019b}]{2019MNRAS.490.3234N}
{Nelson} D.,  et~al., 2019b, \mn@doi [\mnras] {10.1093/mnras/stz2306}, \href
  {https://ui.adsabs.harvard.edu/abs/2019MNRAS.490.3234N} {490, 3234}

\bibitem[\protect\citeauthoryear{{Ni} et~al.,}{{Ni}
  et~al.}{2022}]{2022MNRAS.513..670N}
{Ni} Y.,  et~al., 2022, \mn@doi [\mnras] {10.1093/mnras/stac351}, \href
  {https://ui.adsabs.harvard.edu/abs/2022MNRAS.513..670N} {513, 670}

\bibitem[\protect\citeauthoryear{{Pakmor}, {Bauer}  \& {Springel}}{{Pakmor}
  et~al.}{2011}]{2011MNRAS.418.1392P}
{Pakmor} R.,  {Bauer} A.,   {Springel} V.,  2011, \mn@doi [\mnras]
  {10.1111/j.1365-2966.2011.19591.x}, \href
  {https://ui.adsabs.harvard.edu/abs/2011MNRAS.418.1392P} {418, 1392}

\bibitem[\protect\citeauthoryear{{Pakmor}, {Pfrommer}, {Simpson}, {Kannan}  \&
  {Springel}}{{Pakmor} et~al.}{2016}]{2016MNRAS.462.2603P}
{Pakmor} R.,  {Pfrommer} C.,  {Simpson} C.~M.,  {Kannan} R.,   {Springel} V.,
  2016, \mn@doi [\mnras] {10.1093/mnras/stw1761}, \href
  {https://ui.adsabs.harvard.edu/abs/2016MNRAS.462.2603P} {462, 2603}

\bibitem[\protect\citeauthoryear{{Pillepich} et~al.,}{{Pillepich}
  et~al.}{2018a}]{2018MNRAS.473.4077P}
{Pillepich} A.,  et~al., 2018a, \mn@doi [\mnras] {10.1093/mnras/stx2656}, \href
  {https://ui.adsabs.harvard.edu/abs/2018MNRAS.473.4077P} {473, 4077}

\bibitem[\protect\citeauthoryear{{Pillepich} et~al.,}{{Pillepich}
  et~al.}{2018b}]{2018MNRAS.475..648P}
{Pillepich} A.,  et~al., 2018b, \mn@doi [\mnras] {10.1093/mnras/stx3112}, \href
  {https://ui.adsabs.harvard.edu/abs/2018MNRAS.475..648P} {475, 648}

\bibitem[\protect\citeauthoryear{{Pillepich} et~al.,}{{Pillepich}
  et~al.}{2019}]{2019MNRAS.490.3196P}
{Pillepich} A.,  et~al., 2019, \mn@doi [\mnras] {10.1093/mnras/stz2338}, \href
  {https://ui.adsabs.harvard.edu/abs/2019MNRAS.490.3196P} {490, 3196}

\bibitem[\protect\citeauthoryear{{Planck Collaboration} et~al.,}{{Planck
  Collaboration} et~al.}{2016}]{2016A&A...594A..13P}
{Planck Collaboration} et~al., 2016, \mn@doi [\aap]
  {10.1051/0004-6361/201525830}, \href
  {https://ui.adsabs.harvard.edu/abs/2016A&A...594A..13P} {594, A13}

\bibitem[\protect\citeauthoryear{{Reed} et~al.,}{{Reed}
  et~al.}{2017}]{2017MNRAS.468.4702R}
{Reed} S.~L.,  et~al., 2017, \mn@doi [\mnras] {10.1093/mnras/stx728}, \href
  {https://ui.adsabs.harvard.edu/abs/2017MNRAS.468.4702R} {468, 4702}

\bibitem[\protect\citeauthoryear{{Regan}, {Johansson}  \& {Wise}}{{Regan}
  et~al.}{2014}]{2014ApJ...795..137R}
{Regan} J.~A.,  {Johansson} P.~H.,   {Wise} J.~H.,  2014, \mn@doi [\apj]
  {10.1088/0004-637X/795/2/137}, \href
  {https://ui.adsabs.harvard.edu/abs/2014ApJ...795..137R} {795, 137}

\bibitem[\protect\citeauthoryear{{Rodriguez-Gomez} et~al.,}{{Rodriguez-Gomez}
  et~al.}{2015}]{2015MNRAS.449...49R}
{Rodriguez-Gomez} V.,  et~al., 2015, \mn@doi [\mnras] {10.1093/mnras/stv264},
  \href {https://ui.adsabs.harvard.edu/abs/2015MNRAS.449...49R} {449, 49}

\bibitem[\protect\citeauthoryear{{Rodriguez-Gomez} et~al.,}{{Rodriguez-Gomez}
  et~al.}{2019}]{2019MNRAS.483.4140R}
{Rodriguez-Gomez} V.,  et~al., 2019, \mn@doi [\mnras] {10.1093/mnras/sty3345},
  \href {https://ui.adsabs.harvard.edu/abs/2019MNRAS.483.4140R} {483, 4140}

\bibitem[\protect\citeauthoryear{{Schaye} et~al.,}{{Schaye}
  et~al.}{2015}]{2015MNRAS.446..521S}
{Schaye} J.,  et~al., 2015, \mn@doi [\mnras] {10.1093/mnras/stu2058}, \href
  {https://ui.adsabs.harvard.edu/abs/2015MNRAS.446..521S} {446, 521}

\bibitem[\protect\citeauthoryear{{Sijacki}, {Vogelsberger}, {Genel},
  {Springel}, {Torrey}, {Snyder}, {Nelson}  \& {Hernquist}}{{Sijacki}
  et~al.}{2015}]{2015MNRAS.452..575S}
{Sijacki} D.,  {Vogelsberger} M.,  {Genel} S.,  {Springel} V.,  {Torrey} P.,
  {Snyder} G.~F.,  {Nelson} D.,   {Hernquist} L.,  2015, \mn@doi [\mnras]
  {10.1093/mnras/stv1340}, \href
  {https://ui.adsabs.harvard.edu/abs/2015MNRAS.452..575S} {452, 575}

\bibitem[\protect\citeauthoryear{{Smith}, {Regan}, {Downes}, {Norman}, {O'Shea}
   \& {Wise}}{{Smith} et~al.}{2018}]{2018MNRAS.480.3762S}
{Smith} B.~D.,  {Regan} J.~A.,  {Downes} T.~P.,  {Norman} M.~L.,  {O'Shea}
  B.~W.,   {Wise} J.~H.,  2018, \mn@doi [\mnras] {10.1093/mnras/sty2103}, \href
  {https://ui.adsabs.harvard.edu/abs/2018MNRAS.480.3762S} {480, 3762}

\bibitem[\protect\citeauthoryear{{Springel}}{{Springel}}{2010}]{2010MNRAS.401..791S}
{Springel} V.,  2010, \mn@doi [\mnras] {10.1111/j.1365-2966.2009.15715.x},
  \href {https://ui.adsabs.harvard.edu/abs/2010MNRAS.401..791S} {401, 791}

\bibitem[\protect\citeauthoryear{{Springel}, {White}, {Tormen}  \&
  {Kauffmann}}{{Springel} et~al.}{2001}]{2001MNRAS.328..726S}
{Springel} V.,  {White} S. D.~M.,  {Tormen} G.,   {Kauffmann} G.,  2001,
  \mn@doi [\mnras] {10.1046/j.1365-8711.2001.04912.x}, \href
  {https://ui.adsabs.harvard.edu/abs/2001MNRAS.328..726S} {328, 726}

\bibitem[\protect\citeauthoryear{{Springel} et~al.,}{{Springel}
  et~al.}{2018}]{2018MNRAS.475..676S}
{Springel} V.,  et~al., 2018, \mn@doi [\mnras] {10.1093/mnras/stx3304}, \href
  {https://ui.adsabs.harvard.edu/abs/2018MNRAS.475..676S} {475, 676}

\bibitem[\protect\citeauthoryear{{Springel}, {Pakmor}, {Zier}  \&
  {Reinecke}}{{Springel} et~al.}{2021}]{2021MNRAS.506.2871S}
{Springel} V.,  {Pakmor} R.,  {Zier} O.,   {Reinecke} M.,  2021, \mn@doi
  [\mnras] {10.1093/mnras/stab1855}, \href
  {https://ui.adsabs.harvard.edu/abs/2021MNRAS.506.2871S} {506, 2871}

\bibitem[\protect\citeauthoryear{{Taylor} \& {Kobayashi}}{{Taylor} \&
  {Kobayashi}}{2014}]{2014MNRAS.442.2751T}
{Taylor} P.,  {Kobayashi} C.,  2014, \mn@doi [\mnras] {10.1093/mnras/stu983},
  \href {https://ui.adsabs.harvard.edu/abs/2014MNRAS.442.2751T} {442, 2751}

\bibitem[\protect\citeauthoryear{{Torrey} et~al.,}{{Torrey}
  et~al.}{2019}]{2019MNRAS.484.5587T}
{Torrey} P.,  et~al., 2019, \mn@doi [\mnras] {10.1093/mnras/stz243}, \href
  {https://ui.adsabs.harvard.edu/abs/2019MNRAS.484.5587T} {484, 5587}

\bibitem[\protect\citeauthoryear{{Tremmel}, {Karcher}, {Governato},
  {Volonteri}, {Quinn}, {Pontzen}, {Anderson}  \& {Bellovary}}{{Tremmel}
  et~al.}{2017}]{2017MNRAS.470.1121T}
{Tremmel} M.,  {Karcher} M.,  {Governato} F.,  {Volonteri} M.,  {Quinn} T.~R.,
  {Pontzen} A.,  {Anderson} L.,   {Bellovary} J.,  2017, \mn@doi [\mnras]
  {10.1093/mnras/stx1160}, \href
  {https://ui.adsabs.harvard.edu/abs/2017MNRAS.470.1121T} {470, 1121}

\bibitem[\protect\citeauthoryear{{{\"U}bler} et~al.,}{{{\"U}bler}
  et~al.}{2021}]{2021MNRAS.500.4597U}
{{\"U}bler} H.,  et~al., 2021, \mn@doi [\mnras] {10.1093/mnras/staa3464}, \href
  {https://ui.adsabs.harvard.edu/abs/2021MNRAS.500.4597U} {500, 4597}

\bibitem[\protect\citeauthoryear{{Venemans} et~al.,}{{Venemans}
  et~al.}{2015}]{2015MNRAS.453.2259V}
{Venemans} B.~P.,  et~al., 2015, \mn@doi [\mnras] {10.1093/mnras/stv1774},
  \href {https://ui.adsabs.harvard.edu/abs/2015MNRAS.453.2259V} {453, 2259}

\bibitem[\protect\citeauthoryear{{Vogelsberger} et~al.,}{{Vogelsberger}
  et~al.}{2014a}]{2014MNRAS.444.1518V}
{Vogelsberger} M.,  et~al., 2014a, \mn@doi [\mnras] {10.1093/mnras/stu1536},
  \href {https://ui.adsabs.harvard.edu/abs/2014MNRAS.444.1518V} {444, 1518}

\bibitem[\protect\citeauthoryear{{Vogelsberger} et~al.,}{{Vogelsberger}
  et~al.}{2014b}]{2014Natur.509..177V}
{Vogelsberger} M.,  et~al., 2014b, \mn@doi [\nat] {10.1038/nature13316}, \href
  {https://ui.adsabs.harvard.edu/abs/2014Natur.509..177V} {509, 177}

\bibitem[\protect\citeauthoryear{{Vogelsberger}, {Marinacci}, {Torrey}  \&
  {Puchwein}}{{Vogelsberger} et~al.}{2020}]{2020NatRP...2...42V}
{Vogelsberger} M.,  {Marinacci} F.,  {Torrey} P.,   {Puchwein} E.,  2020,
  \mn@doi [Nature Reviews Physics] {10.1038/s42254-019-0127-2}, \href
  {https://ui.adsabs.harvard.edu/abs/2020NatRP...2...42V} {2, 42}

\bibitem[\protect\citeauthoryear{{Volonteri}}{{Volonteri}}{2007}]{2007ApJ...663L...5V}
{Volonteri} M.,  2007, \mn@doi [\apjl] {10.1086/519525}, \href
  {https://ui.adsabs.harvard.edu/abs/2007ApJ...663L...5V} {663, L5}

\bibitem[\protect\citeauthoryear{{Volonteri}, {Dubois}, {Pichon}  \&
  {Devriendt}}{{Volonteri} et~al.}{2016}]{2016MNRAS.460.2979V}
{Volonteri} M.,  {Dubois} Y.,  {Pichon} C.,   {Devriendt} J.,  2016, \mn@doi
  [\mnras] {10.1093/mnras/stw1123}, \href
  {https://ui.adsabs.harvard.edu/abs/2016MNRAS.460.2979V} {460, 2979}

\bibitem[\protect\citeauthoryear{{Volonteri} et~al.,}{{Volonteri}
  et~al.}{2020}]{2020MNRAS.498.2219V}
{Volonteri} M.,  et~al., 2020, \mn@doi [\mnras] {10.1093/mnras/staa2384}, \href
  {https://ui.adsabs.harvard.edu/abs/2020MNRAS.498.2219V} {498, 2219}

\bibitem[\protect\citeauthoryear{{Wang} et~al.,}{{Wang}
  et~al.}{2018}]{2018ApJ...869L...9W}
{Wang} F.,  et~al., 2018, \mn@doi [\apjl] {10.3847/2041-8213/aaf1d2}, \href
  {https://ui.adsabs.harvard.edu/abs/2018ApJ...869L...9W} {869, L9}

\bibitem[\protect\citeauthoryear{{Wang}, {Taylor}, {Federrath}  \&
  {Kobayashi}}{{Wang} et~al.}{2019}]{2019MNRAS.483.4640W}
{Wang} E.~X.,  {Taylor} P.,  {Federrath} C.,   {Kobayashi} C.,  2019, \mn@doi
  [\mnras] {10.1093/mnras/sty3491}, \href
  {https://ui.adsabs.harvard.edu/abs/2019MNRAS.483.4640W} {483, 4640}

\bibitem[\protect\citeauthoryear{{Wang} et~al.,}{{Wang}
  et~al.}{2021}]{2021ApJ...907L...1W}
{Wang} F.,  et~al., 2021, \mn@doi [\apjl] {10.3847/2041-8213/abd8c6}, \href
  {https://ui.adsabs.harvard.edu/abs/2021ApJ...907L...1W} {907, L1}

\bibitem[\protect\citeauthoryear{{Weinberger} et~al.,}{{Weinberger}
  et~al.}{2017}]{2017MNRAS.465.3291W}
{Weinberger} R.,  et~al., 2017, \mn@doi [\mnras] {10.1093/mnras/stw2944}, \href
  {https://ui.adsabs.harvard.edu/abs/2017MNRAS.465.3291W} {465, 3291}

\bibitem[\protect\citeauthoryear{{Weinberger} et~al.,}{{Weinberger}
  et~al.}{2018}]{2018MNRAS.479.4056W}
{Weinberger} R.,  et~al., 2018, \mn@doi [\mnras] {10.1093/mnras/sty1733}, \href
  {https://ui.adsabs.harvard.edu/abs/2018MNRAS.479.4056W} {479, 4056}

\bibitem[\protect\citeauthoryear{{Weinberger}, {Springel}  \&
  {Pakmor}}{{Weinberger} et~al.}{2020}]{2020ApJS..248...32W}
{Weinberger} R.,  {Springel} V.,   {Pakmor} R.,  2020, \mn@doi [\apjs]
  {10.3847/1538-4365/ab908c}, \href
  {https://ui.adsabs.harvard.edu/abs/2020ApJS..248...32W} {248, 32}

\bibitem[\protect\citeauthoryear{{Willott} et~al.,}{{Willott}
  et~al.}{2010}]{2010AJ....139..906W}
{Willott} C.~J.,  et~al., 2010, \mn@doi [\aj] {10.1088/0004-6256/139/3/906},
  \href {https://ui.adsabs.harvard.edu/abs/2010AJ....139..906W} {139, 906}

\bibitem[\protect\citeauthoryear{{Wise}, {Regan}, {O'Shea}, {Norman}, {Downes}
  \& {Xu}}{{Wise} et~al.}{2019}]{2019Natur.566...85W}
{Wise} J.~H.,  {Regan} J.~A.,  {O'Shea} B.~W.,  {Norman} M.~L.,  {Downes}
  T.~P.,   {Xu} H.,  2019, \mn@doi [\nat] {10.1038/s41586-019-0873-4}, \href
  {https://ui.adsabs.harvard.edu/abs/2019Natur.566...85W} {566, 85}

\bibitem[\protect\citeauthoryear{{Xu}, {Wise}  \& {Norman}}{{Xu}
  et~al.}{2013}]{2013ApJ...773...83X}
{Xu} H.,  {Wise} J.~H.,   {Norman} M.~L.,  2013, \mn@doi [\apj]
  {10.1088/0004-637X/773/2/83}, \href
  {https://ui.adsabs.harvard.edu/abs/2013ApJ...773...83X} {773, 83}

\bibitem[\protect\citeauthoryear{{Yang} et~al.,}{{Yang}
  et~al.}{2019}]{2019AJ....157..236Y}
{Yang} J.,  et~al., 2019, \mn@doi [\aj] {10.3847/1538-3881/ab1be1}, \href
  {https://ui.adsabs.harvard.edu/abs/2019AJ....157..236Y} {157, 236}

\makeatother
\end{thebibliography}
\appendix

\section{Relationship between subhalo environment and subhalo mass}
\label{Relationship between subhalo environment and subhalo mass}

\begin{figure}
\includegraphics[width=8 cm]{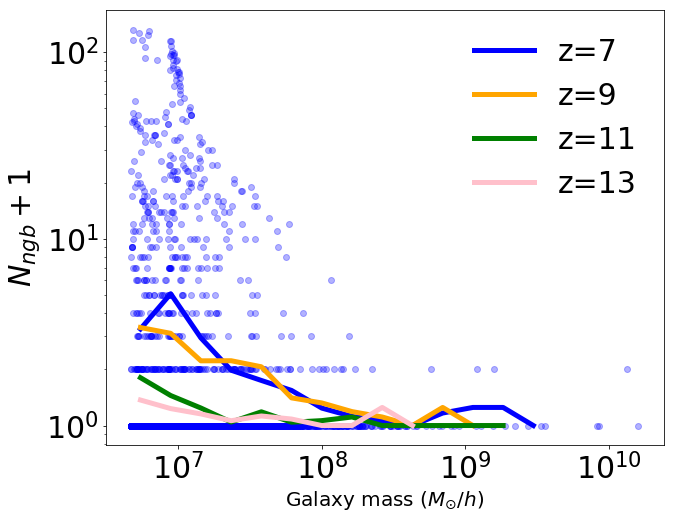}
\caption{Relationship between galaxy mass and galaxy environment~(number of neighboring halos $N_{\mathrm{ngb}}$ as defined in Section \ref{Subhalo-based stochastic seeding}) for the galaxy populations in our \texttt{GAS_BASED} simulations. We plot `$N_{\rm ngb}+1$' on the y-axis in order to also show galaxies with no neighbors on the log scale. The circles show data points at $z=7$, and the solid lines show the mean trends at $z=7,9,11~\&~13$. We can see that smaller mass galaxies generally have higher number of neighbors. This is not unexpected, given the fact that $N_{\mathrm{ngb}}$ counts only those neighbors which exceed the host halo mass of the galaxy. And as expected from hierarchical structure formation, %Not surprisingly, 
galaxies of a given mass have fewer number of neighbors at higher redshifts.}
\label{FOF_environment_mass_scaling}
\end{figure}

While our stochastic seed models apply seeding criteria based on galaxy mass and galaxy environment~(number of neighboring halos $N_{\mathrm{ngb}}$), these two galaxy properties are not completely independent of each other. In Figure \ref{FOF_environment_mass_scaling}, we can clearly see that the galaxy with lower masses tend to have higher number of neighboring halos. This is not surprising given the precise definition of $N_{\mathrm{ngb}}$ %~(see section 
described in Section \ref{Subhalo-based stochastic seeding}, which only counts neighboring halos that exceed the host halo mass of the galaxy. In other words, higher mass galaxies are typically hosted by higher mass halos. Therefore, for a higher mass galaxy, there are going to be fewer neighboring halos that have enough mass to be counted in the $N_{\mathrm{ngb}}$ calculation. Notably, galaxies of a fixed mass tend to have higher $N_{\mathrm{ngb}}$ at lower redshifts; this is simply due to higher number of halos at lower redshifts in general.

Due to this negative correlation between galaxy mass and galaxy environment, applying a \textit{galaxy environment criterion}~(that favors seeding in richer environments) can cause the ESDs to form more favorably in lower mass galaxies. This can alter our desired calibration for the \textit{galaxy mass criterion} that we apply prior to the \textit{galaxy environment criterion}. To prevent this from happening, we impose the environment based seeding probabilities $p_0$ and $p_1$ to linearly increase with the galaxy mass with a slope $\gamma > 0$~(see Equation \ref{environment_based_seed_probability}). Depending on the gas based seed parameters, $\gamma$ values of $\sim1.2 - 1.6$~(quoted in Table \ref{double_power_law_table}) are the ones found to maintain the calibration of the \textit{galaxy mass criterion}. For values significantly higher or lower than $\sim1.2 - 1.6$, the \textit{galaxy environment criterion} starts to skew the galaxy mass distributions~(wherein ESDs are formed) towards higher or lower masses respectively, compared to our desired calibration. Lastly, incorporating this linear dependence with $\gamma > 0$ is also physically motivated. This is because it captures the notion that, for a given value of $N_{\mathrm{ngb}}$, seeding should be favored in a galaxy with higher mass because it exists in a more extreme environment compared to a lower mass galaxy with the same $N_{\mathrm{ngb}}$.  
\section{Evolution of star forming ~\&~ metal poor gas in halos}
\label{Evolution of star forming metal poor gas in halos}
\begin{figure*}
\includegraphics[width=16 cm]{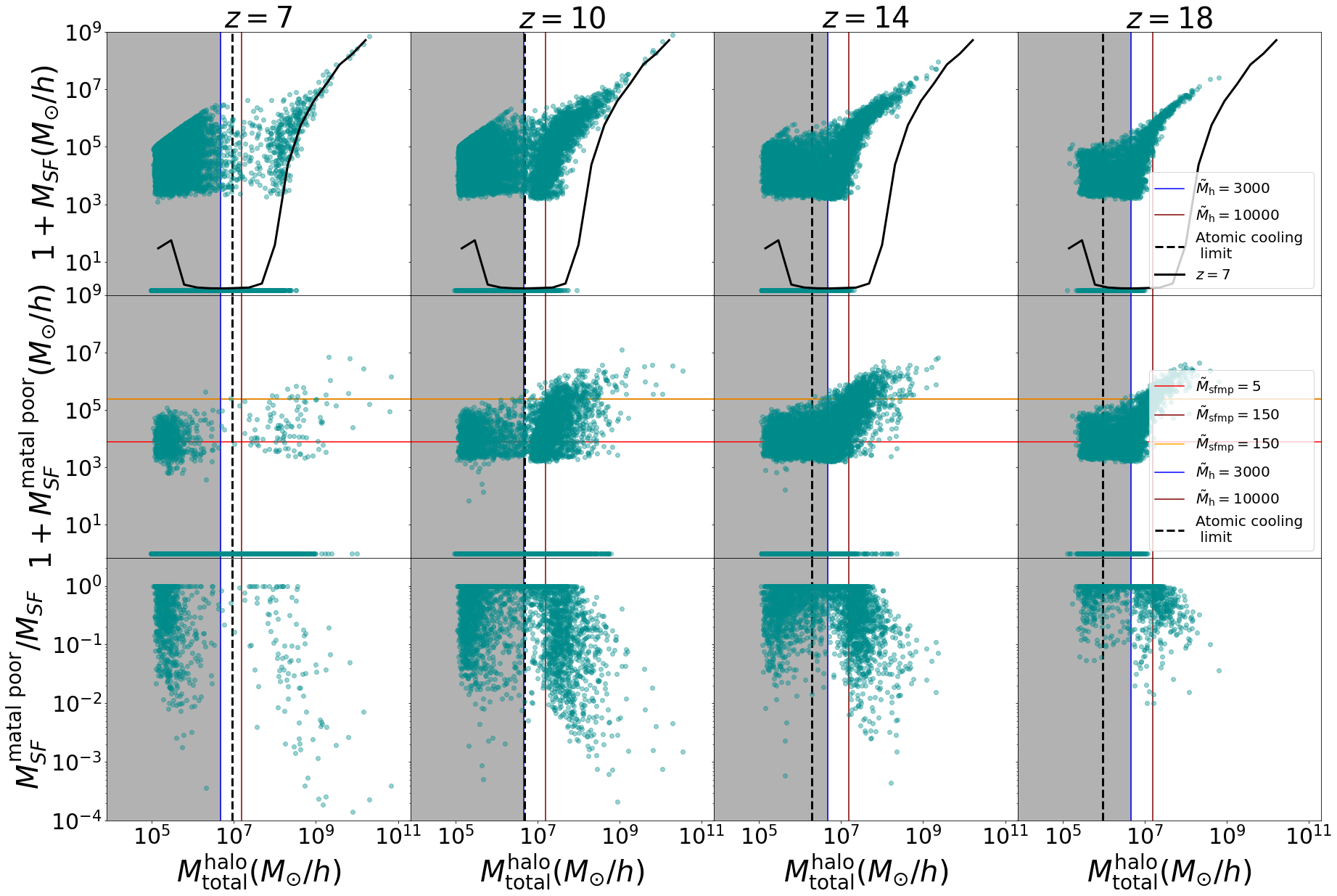}
\caption{Star forming gas masses~($M_{\mathrm{SF}}$, top panels), star forming \& metal poor gas masses~($M^{\mathrm{metal \ poor}}_{\mathrm{SF}}$, middle panels) and their ratios~(bottom panels) are plotted versus the total mass~($M_{\mathrm{total}}$) for halos in different snapshots within the \texttt{GAS_BASED} suite that explicitly resolves the $1.56\times10^3~M_{\odot}/h$ DGBs. The different columns show different redshift snapshots; however, the mean trend at $z=7$ is plotted as solid black line in all the top panels to clearly see the redshift evolution. We added 1 to the y-axis in order to include halos with no star forming gas on the log-scale. The black dashed vertical lines correspond to the atomic cooling limit~(halo virial temperature $T_{\mathrm{vir}}=10^4$ K).  The red and orange horizontal lines correspond to the seeding thresholds of $\msfmp = 5~\&~150$ respectively. The blue and brown vertical lines correspond to the seeding thresholds of $\mh=3000~\&~10000$ respectively. Shaded regions correspond to the lowest mass objects below the $\mh=3000$ limit, which are also below the atomic cooling threshold. We avoid seeding in these halos since they are impacted by the limited mass resolution and lack of sufficient physics~(absence of $H_2$ cooling). Top panels show that at fixed halo mass, star forming gas mass decreases with time due to cosmological expansion. Middle and bottom panels show that despite stronger metal enrichment in more massive halos, the star forming \& metal poor gas mass is still positively correlated with halo mass. As a result, DGB formation is favored in more massive halos when the primary driver is metal enrichment.}
\label{halo_evolution_star_formation}
\end{figure*}

In Figure \ref{halo_evolution_star_formation}, we show scatter plots of the star forming gas mass~($M_{\mathrm{SF}}$) and star forming~\&~metal poor gas mass~($M^{\mathrm{metal \ poor}}_{\mathrm{SF}}$) versus the total halo mass~($M_{\mathrm{total}}^{\mathrm{halo}}$) at different redshifts. The top row shows that there is a straightforward positive correlation between the halo mass and star forming gas mass, except at the lowest halo masses wherein the results are likely impacted by the finite simulation resolution. Notably, several of these lowest mass objects are spuriously identified gas clumps with very little DM mass. In addition, these halos are also significantly below the atomic cooling threshold~(virial Temperature of $10^4$ K, dashed black vertical lines), which we do not self-consistently simulate due to the absence of $H_2$ cooling. With our adopted halo mass thresholds~($\mh=3000~\&~10000$), we avoid seeding in these lowest mass halos~(marked as shaded grey region). Hereafter, we shall focus only on halos with reasonably well converged stellar and gas properties~(outside the grey region). The top row also shows that at fixed halo mass, the star forming gas mass~(top row) steadily decreases with time~(green circles vs. black solid line). This is a simple consequence of cosmological expansion, which increases the atomic cooling threshold with time. As a result, at later times, halos of a given mass have lower ability to contain gas and collapse it to high enough densities to form stars. This is overall responsible for the steady increase in DGB forming halo masses with time in epochs where star formation is the primary driver of DGB formation~(seen in Section \ref{where descendants assemble}). In the bottom row, the fraction of star forming gas mass that is also metal poor~($<10^{-4}~Z_{\odot}$), sharply decreases with halo mass at fixed redshift. This is not surprising because metal enrichment is expected to be more prevalent in massive halos. Regardless, the middle row shows that the overall star forming \& metal poor gas mass continues to be positively correlated with halo mass. This is simply due to more massive halos having higher overall star forming gas mass. As a result, whenever metal enrichment becomes the primary driver of DGB formation, it leads to a more rapid increase in the DGB forming halo mass with time, compared to that of simple cosmological expansion~(see again Section \ref{where descendants assemble}).

\end{document}